\documentclass[twoside,twocolumn,9pt]{article}
\usepackage{extsizes}
\usepackage[super,sort&compress,comma]{natbib} 
\usepackage[version=3]{mhchem}
\usepackage[left=1.5cm, right=1.5cm, top=1.785cm, bottom=2.0cm]{geometry}
\usepackage{balance}
\usepackage{mathptmx}
\usepackage{sectsty}
\usepackage{graphicx} 
\usepackage{lastpage}
\usepackage[format=plain,justification=justified,singlelinecheck=false,font={stretch=1.125,small,sf},labelfont=bf,labelsep=space]{caption}
\usepackage{float}
\usepackage{fancyhdr}
\usepackage{fnpos}
\usepackage[english]{babel}
\addto{\captionsenglish}{%
  
}
\usepackage{array}
\usepackage{droidsans}
\usepackage{charter}
\usepackage[T1]{fontenc}
\usepackage[usenames,dvipsnames]{xcolor}
\usepackage{setspace}
\usepackage[compact]{titlesec}
\usepackage{hyperref}

\usepackage{epstopdf}

\definecolor{cream}{RGB}{222,217,201}

\usepackage{mathtools}
\usepackage{multirow}
\usepackage{tabularx}
\usepackage{amsmath}
\usepackage{amssymb}
\usepackage{booktabs}
\usepackage{soul}

\usepackage{color,ulem}

\begin{document}

\pagestyle{fancy}
\thispagestyle{plain}
\fancypagestyle{plain}{
\renewcommand{\headrulewidth}{0pt}
}

\makeFNbottom
\makeatletter
\renewcommand\LARGE{\@setfontsize\LARGE{15pt}{17}}
\renewcommand\Large{\@setfontsize\Large{12pt}{14}}
\renewcommand\large{\@setfontsize\large{10pt}{12}}
\renewcommand\footnotesize{\@setfontsize\footnotesize{7pt}{10}}
\makeatother

\renewcommand{\thefootnote}{\fnsymbol{footnote}}
\renewcommand\footnoterule{\vspace*{1pt}%
\color{cream}\hrule width 3.5in height 0.4pt \color{black}\vspace*{5pt}} 
\setcounter{secnumdepth}{5}

\makeatletter 
\renewcommand\@biblabel[1]{#1}            
\renewcommand\@makefntext[1]%
{\noindent\makebox[0pt][r]{\@thefnmark\,}#1}
\makeatother 
\renewcommand{\figurename}{\small{Fig.}~}
\sectionfont{\sffamily\Large}
\subsectionfont{\normalsize}
\subsubsectionfont{\bf}
\setstretch{1.125} 
\setlength{\skip\footins}{0.8cm}
\setlength{\footnotesep}{0.25cm}
\setlength{\jot}{10pt}
\titlespacing*{\section}{0pt}{4pt}{4pt}
\titlespacing*{\subsection}{0pt}{15pt}{1pt}

\fancyfoot{}
\fancyfoot[LO,RE]{\vspace{-7.1pt}\includegraphics[height=9pt]{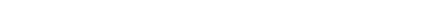}}
\fancyfoot[CO]{\vspace{-7.1pt}\hspace{13.2cm}\includegraphics{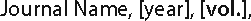}}
\fancyfoot[CE]{\vspace{-7.2pt}\hspace{-14.2cm}\includegraphics{head_foot/RF.png}}
\fancyfoot[RO]{\footnotesize{\sffamily{1--\pageref{LastPage} ~\textbar  \hspace{2pt}\thepage}}}
\fancyfoot[LE]{\footnotesize{\sffamily{\thepage~\textbar\hspace{3.45cm} 1--\pageref{LastPage}}}}
\fancyhead{}
\renewcommand{\headrulewidth}{0pt} 
\renewcommand{\footrulewidth}{0pt}
\setlength{\arrayrulewidth}{1pt}
\setlength{\columnsep}{6.5mm}
\setlength\bibsep{1pt}

\makeatletter 
\newlength{\figrulesep} 
\setlength{\figrulesep}{0.5\textfloatsep} 

\newcommand{\topfigrule}{\vspace*{-1pt}%
\noindent{\color{cream}\rule[-\figrulesep]{\columnwidth}{1.5pt}} }

\newcommand{\botfigrule}{\vspace*{-2pt}%
\noindent{\color{cream}\rule[\figrulesep]{\columnwidth}{1.5pt}} }

\newcommand{\dblfigrule}{\vspace*{-1pt}%
\noindent{\color{cream}\rule[-\figrulesep]{\textwidth}{1.5pt}} }

\makeatother

\twocolumn[
  \begin{@twocolumnfalse}
{\includegraphics[height=30pt]{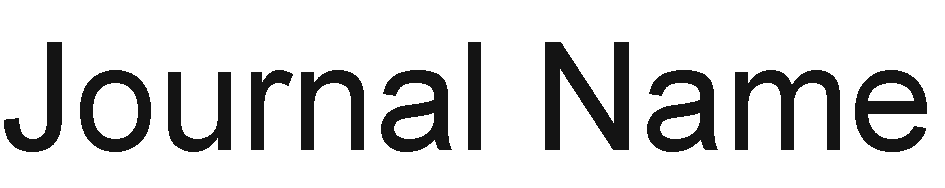}\hfill\raisebox{0pt}[0pt][0pt]{\includegraphics[height=55pt]{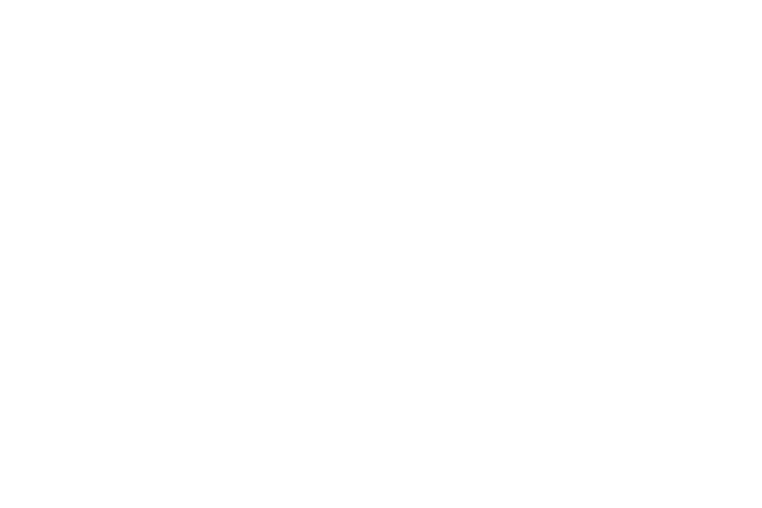}}\\[1ex]
\includegraphics[width=18.5cm]{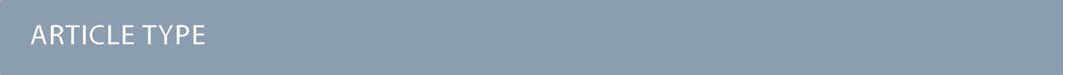}}\par
\vspace{1em}
\sffamily
\begin{tabular}{m{4.5cm} p{13.5cm} }

\includegraphics{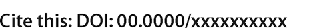} & \noindent\LARGE{\textbf{Dismai-Bench: Benchmarking and designing generative models using disordered materials and interfaces
}} \\
\vspace{0.3cm} & \vspace{0.3cm} \\

 & \noindent\large{
 Adrian Xiao Bin Yong,$^{\ast}$\textit{$^{ab}$}
 Tianyu Su,\textit{$^{ab}$} and
 Elif Ertekin$^{\ast}$\textit{$^{bc}$}} \\

\includegraphics{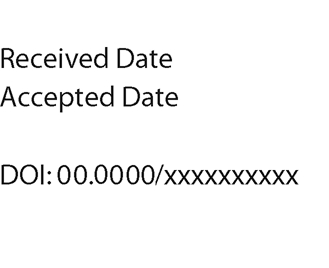} & \noindent\normalsize{Generative models have received significant attention in recent years for materials science applications, particularly in the area of inverse design for materials discovery. 
However, these models are usually assessed based on newly generated, unverified materials, using heuristic metrics such as charge neutrality, which provide a narrow evaluation of a model's performance.
Also, current efforts for inorganic materials have predominantly focused on small, periodic crystals ($\leq$ 20 atoms), even though the capability to generate large, more intricate and disordered structures would expand the applicability of generative modeling to a broader spectrum of materials.
In this work, we present the Disordered Materials \& Interfaces Benchmark (Dismai-Bench), a generative model benchmark that uses datasets of disordered alloys, interfaces, and amorphous silicon (256-264 atoms per structure). 
Models are trained on each dataset independently, and evaluated through direct structural comparisons between training and generated structures. 
Such comparisons are only possible because the material system of each training dataset is fixed. 
Benchmarking was performed on two graph diffusion models and two (coordinate-based) U-Net diffusion models. 
The graph models were found to significantly outperform the U-Net models due to the higher expressive power of graphs. 
While noise in the less expressive models can assist in discovering materials by facilitating exploration beyond the training distribution, these models face significant challenges when confronted with more complex structures.
To further demonstrate the benefits of this benchmarking in the development process of a generative model, we considered the case of developing a point-cloud-based generative adversarial network (GAN) to generate low-energy disordered interfaces. 
We tested different GAN architectures and identified reasons for good/poor performance. 
We show that the best performing architecture, CryinGAN, outperforms the U-Net models, and is competitive against the graph models despite its lack of invariances and weaker expressive power.
This work provides a new framework and insights to guide the development of future generative models, whether for ordered or disordered materials.
} \\

\end{tabular}

 \end{@twocolumnfalse} \vspace{0.6cm}

  ]

\renewcommand*\rmdefault{bch}\normalfont\upshape
\rmfamily
\section*{}
\vspace{-1cm}


\footnotetext{\textit{$^{a}$~Department of Materials Science and Engineering, University of Illinois Urbana-Champaign, Urbana, Illinois, USA. E-mail: axyong2@illinois.edu}}
\footnotetext{\textit{$^{b}$~Materials Research Laboratory, University of Illinois Urbana-Champaign, Urbana, Illinois, USA.}}
\footnotetext{\textit{$^{c}$~Department of Mechanical Science and Engineering, University of Illinois Urbana-Champaign, Urbana, Illinois, USA. Email: ertekin@illinois.edu}}

\footnotetext{\dag~Electronic Supplementary Information (ESI) available: [details of any supplementary information available should be included here]. See DOI: 00.0000/00000000.}



\section{Introduction}
Generative modeling has emerged as a powerful tool for tackling problems in materials science~\cite{RN292, RN293}. Initially limited to simpler molecules~\cite{RN289, RN290} and proteins~\cite{sinai2018variational}, generative modeling has since advanced to include inorganic materials~\cite{hoffmann2019datadriven, RN291, nouira2019crystalgan} as well. 
The primary interest in generative modeling has been its promise for inverse materials design~\cite{RN294, RN295, RN296}, where the primary objective is to create new materials tailored to specific properties rather than screening known materials for desired characteristics.
Generative models are distinguished from discriminative models, as the latter learns the conditional probability $p (y \mid x)$ of observing a property ($y$) given a material representation ($x$). Instead, a generative model learns the joint probability distribution $p (x,y)$ of the data that it was trained on, and samples from the distribution of structures. 

While generative modeling efforts for inorganic materials~\cite{RN294, RN296, RN300} have primarily centered around simpler bulk crystals, there has been comparatively less emphasis on disordered systems, despite their relevance across a wide spectrum of applications~\cite{RN297, RN298, RN299}.
Disordered systems usually have complex and irregular structures, necessitating large atomic representations and requiring more powerful generative models than those developed for simple crystals. 
They include structures that completely lack crystal lattices such as amorphous materials, as well as structures beyond bulk materials such as surfaces and interfaces. 
In direct physical modeling, disordered materials are most typically represented by large so-called ``supercells'' that (spuriously) introduce periodicity at larger length scales. 
Databases of disordered materials are growing~\cite{RN301, RN302, zheng2024ab}, offering compelling prospects for the inverse design of metal-organic frameworks, porous amorphous materials, amorphous battery materials, and more.
Beyond materials discovery, generative modeling can be used 
to generate amorphous structures of arbitrarily large sizes upon training on smaller samples that capture material correlation lengths~\cite{RN303}. 
This capability enables more thorough investigations into properties that are influenced by size effects, such as thermal conductivity and mechanical properties.
Generative modeling can also be used to refine atomic structures to align with experimental observations~\cite{aps2023fantastx}, typically focusing on the refinement of disordered structures~\cite{RN304, RN305}.
Yet, the application of generative modeling to disordered systems remains limited, such as  generating 2D morphology rather than precise atomic structures~\cite{RN303}. 
Moreover, generative models have been reported to fail when applied to large systems~\cite{RN296, RN337}.
To reap the benefits of generative modeling for disordered systems, better generative models need to be developed and evaluated on disordered systems. 

When building a generative model, two major design decisions are the type of generative model and the material representation (i.e., the input used to describe the material). 
The compatibility of these two choices is important as well.
The types of generative models that have been used for materials include variational autoencoders (VAEs)~\cite{RN291, RN296, RN294, hoffmann2019datadriven}, generative adversarial networks (GANs)~\cite{RN306, nouira2019crystalgan, RN307, RN295}, diffusion models~\cite{xie2022crystal, jiao2024diffcsp, RN353, yang2023unimat, zeni2024mattergen}, and language models~\cite{flamshepherd2023language, antunes2024crystallm, gruver2024llama}. 
Generative models were initially developed with two main types of material representations: (1) voxels ~\cite{hoffmann2019datadriven, RN291, RN296, RN300} and (2) point clouds~\cite{nouira2019crystalgan, RN307, RN294, RN295}.
Voxel representation is memory intensive, resulting in limitations in the voxelization resolution and thus number of atoms (e.g., Court et al.~\cite{RN296} restricted the number of atoms to 	$<$ 40 atoms per cell). Reconstruction issues were also reported~\cite{RN296} for non-cubic cells. 
On the other hand, point clouds directly represent structures using their atomic coordinates and lattice parameters, making them highly scalable with the number of atoms. 
However, the design of point cloud architectures that are symmetry-invariant is not trivial, as the commonly used PointNet architecture~\cite{qi2017pointnet} does not include the desired symmetry invariances. 
More recently, other representations such as graphs~\cite{xie2022crystal, zeni2024mattergen}, coordinate-based representations (e.g., UniMat~\cite{yang2023unimat}, CrysTens~\cite{RN353}), and text-based representations~\cite{flamshepherd2023language, antunes2024crystallm, gruver2024llama} have also been explored.
Graphs are particularly attractive due to their symmetry invariances and strong expressive power, capturing both geometrical features and neighbor information. However, graph convolutions become computationally and memory intensive as the number of atoms increases.

\begin{figure*}[t!]
\centering
\includegraphics[width=\textwidth]{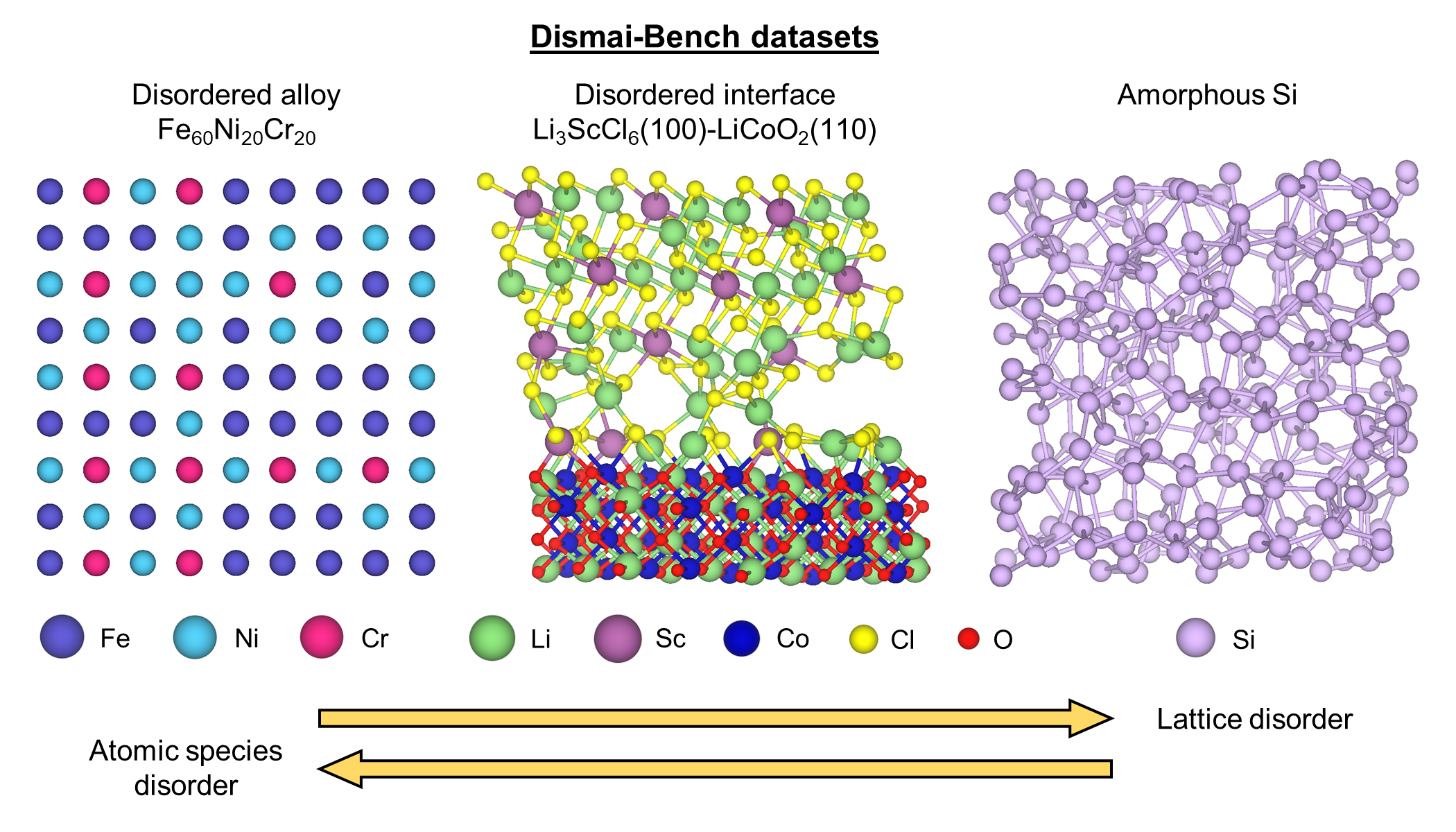}
\caption{Datasets used in Dismai-Bench, consisting of a disordered Fe$_{60}$Ni$_{20}$Cr$_{20}$ austenitic stainless steel system, a disordered Li$_3$ScCl$_6$(100)-LiCoO$_{2}$(110) battery interface system, and an amorphous silicon system.}
\label{fig_datasets}
\end{figure*}

To compare different generative models and make design choices, it is necessary to sufficiently evaluate the generated structures for their validity. 
Training generative models for materials discovery inherently makes the evaluation of the models' performance difficult. 
In more conventional problems such as image generation or speech synthesis, it is relatively easy to discern if the model has learned to generate realistic images or speech from the training data. 
However, it is much more difficult to determine if a newly generated material is realistic, or if the model is simply generating noise. 
Recent generative models~\cite{xie2022crystal, jiao2024diffcsp, yang2023unimat, gruver2024llama} have relied on limited and heuristic metrics (e.g., charge neutrality, material space coverage) to evaluate and compare between models, making it difficult to meaningfully assess model performance. 
One approach to circumventing the issues of evaluating on new, unknown materials is to instead train the model on a fixed set of materials (e.g., perovskites). Restricting the material space allows for easier determination of whether the correct structure is being generated, and direct comparison of the properties between the generated and training structures can be performed. 
In this scenario, however, the materials on which the model is trained should be sufficiently challenging to provide meaningful evaluation of the model's performance. 
In this regard, disordered materials are good candidates for the task, given that generative models can fail to generate even a single type of disordered material (e.g., amorphous silicon~\cite{RN337}).

In this work, we present the Disordered Materials \& Interfaces Benchmark (Dismai-Bench), a generative model benchmark that uses datasets of an Fe$_{60}$Ni$_{20}$Cr$_{20}$ austenitic stainless steel, a disordered Li$_3$ScCl$_6$(100)-LiCoO$_{2}$(110) battery interface, and amorphous silicon. 
Dismai-Bench evaluates generative models on a wide range of material disorder ranging from structural to configurational (see Fig. \ref{fig_datasets}).  
Structural disorder increases from left to right, and configurational disorder increases from right to left, in Fig. \ref{fig_datasets}. 
The composition of each dataset is fixed, and each structure has 256-264 atoms. 
We selected four recent diffusion models to be benchmarked on Dismai-Bench, including two models that use graph representations (CDVAE~\cite{xie2022crystal} \& DiffCSP~\cite{jiao2024diffcsp}) and two models that use coordinate-based representations (CrysTens~\cite{RN353} \& UniMat~\cite{yang2023unimat}). The models were trained on one dataset at a time, and the generated structures were compared with the training structures to obtain structural similarity metrics. 
These metrics quantify the model's ability to learn complex structural patterns found in disordered materials. 
We show that the graph models outperform the coordinate-based models due to the higher expressive power of graphs. 
The success of the less expressive models in materials discovery~\cite{yang2023unimat, gruver2024llama} suggest that noisy models are better for discovering small crystals, but face challenges when tasked with generating larger, more complex structures.

To demonstrate the application of Dismai-Bench in the development of a generative model, we further considered the design of a GAN to generate low-interface-energy Li$_3$ScCl$_6$(100)-LiCoO$_{2}$(110) interface structures. 
We chose the simple point cloud representation, and tested multiple different GAN architectures for which we included bond distance information explicitly in the GANs, instead of just atomic coordinates. 
Direct comparison between the generated and training structures identified the architecture that best achieved the intended goal, along with explanations for why the other architectures were less successful.
We demonstrate that the best architecture, Crystal Interface Generative Adversarial Network (CryinGAN), can generate the disordered interfaces with low interface energy, and similar structural features to the training structures. 
Despite its design simplicity, CryinGAN outperforms the more recent coordinate-based diffusion models on Dismai-Bench. 
It does not outperform the graph diffusion models across all datasets, however, possibly as a result of its weaker expressive power and lack of invariances. 
Through this work, we present a novel framework for conducting meaningful comparisons between models, providing valuable insights into model weaknesses and failures to inform the design of future generative models.

\section{Results and discussion}
\subsection{Datasets and interatomic potentials}
An overview of the dataset curation is outlined here; further details are found in the Methods section. A total of six datasets are used in Dismai-Bench, consisting of four alloy datasets, one interface dataset, and one amorphous silicon dataset. Each dataset contains a total of 1,500 structures, split into 80 \% training and 20 \% validation data. Test sets are not needed since model performance is measured using the benchmark metrics. 

\subsubsection{\texorpdfstring{Fe$_{60}$Ni$_{20}$Cr$_{20}$}{Fe60Cr20Ni20} austenitic stainless steel} 
The stainless steel datasets consist of face-centered cubic (FCC) crystals that are structurally simple, but configurationally complex (refer to Fig. \ref{fig_datasets}). Atoms of various species occupy the lattice sites with different ordering tendencies. The generative models are challenged with generating structures that not only have well-defined FCC lattices, but also the correct degrees of ordering.
The stainless steel datasets were created using a cluster expansion Monte Carlo (CEMC) approach~\cite{kim2022multisublattice, RN358}. The cluster expansion (CE) potential was adapted from ref. \citenum{RN358} , where pair interactions up to the 7$^{\textrm{th}}$ neighbor shell were included. The composition of the structures is Fe$_{60}$Ni$_{20}$Cr$_{20}$, and each structure contains 256 atoms. Monte Carlo (MC) simulations were carried out in the canonical ensemble at 300 K and 1,500 K, such as to obtain datasets with different degrees of short-range order (SRO). The SRO is quantified using the Warren-Cowley SRO parameter~\cite{RN354},
\begin{equation}
    \alpha_l^{AB} = 1 - \frac{P_l^{AB}}{C_AC_B} = 1 - \frac{p_{l,A}^B}{C_B} \hspace{0.5em}, 
\label{WC_SRO}
\end{equation}
where $P_l^{AB}$ is the probability of finding $AB$ pairs in the $l$-th neighbor shell, and $p_{l,A}^B$ = $P_l^{AB}$/$C_A$ is the conditional probability of finding atom $B$ in the $l$-th coordination shell of atom $A$. 
$C_A$ and $C_B$ are the concentration of $A$ and $B$ atoms respectively. 
SRO parameter $\alpha = 0$ indicates zero correlation between atoms (as in a random solution), while $\alpha < 0$ indicates an attractive interaction and $\alpha > 0$ indicates a repulsive interaction. Note that the $A$ and $B$ atoms can be of the same atomic species.

The SRO distributions of the 300 K and 1500 K alloy training datasets are shown in ESI Fig. S1 for the 1st and 2nd nearest neighbor interactions. The 300 K dataset shows more prominent SRO than the 1500 K dataset. The SRO parameter tends to distribute away from zero at 300 K, and consistently distribute near zero at 1500 K. We also created two additional datasets by filtering the CEMC-generated structures such that the SRO parameters have narrow distributions within $\pm$ 0.1 of the average values (see ESI Fig. S2). Sufficiently large number of structures were generated to obtain 1,500 structures for each dataset. 
We refer to the unfiltered and filtered datasets as the wide SRO and narrow SRO datasets respectively.
This SRO filtering was performed to enable comparison of generative model performance when trained on structures with noisy SRO distribution (wide SRO dataset) and structures with less noisy SRO distribution (narrow SRO dataset). 

\subsubsection{Amorphous silicon} 
Amorphous silicon can be thought of as the polar opposite of the FCC alloys. Amorphous silicon consists of a single atomic species only, but completely lack ordering in the form of a crystalline lattice. The generative models are not assessed on any ability to learn ordering relationships between different atomic species. Instead, they are assessed on their abilities to learn the complex structural patterns found in amorphous silicon, such as near-tetrahedral local environments and pair distribution functions.  
The amorphous silicon dataset was adapted from ref. \citenum{RN355}. The original data consists of a 100,000-atom amorphous silicon structure generated through melt-quench molecular dynamics simulation~\cite{RN355}. The structure was sliced into smaller blocks with lattice parameters corresponding to 256-atom amorphous silicon structures. The blocks were sliced at different locations to obtain a total of 1,500 blocks. Blocks with < 256 atoms had atoms added at random to low density regions, and blocks with > 256 atoms had atoms removed at random from high density regions, so that all blocks have 256 atoms. The 1,500 structures were relaxed using a pre-trained SOAP-GAP~\cite{RN356} machine learning interatomic potential for Si. 
The resulting structures have a higher concentration of defects compared to the original 100,000-atom structure (refer to ESI Fig. S3) due to the slicing and atom addition/removal, but the Si coordination geometry remains predominantly tetrahedral.

\subsubsection{\texorpdfstring{Li$_3$ScCl$_6$(100)-LiCoO$_{2}$(110)}{Li3ScCl6(100)-LiCoO2(110)} battery interface}
The disordered interface dataset assesses the generative models on structures that exhibit a mixture of structural and configurational disorder (refer to Fig. \ref{fig_datasets}). Atoms in the disordered interface region are not arranged in well-defined lattices, and coordinate with each other in a range of motifs. The models have to learn to generate these complex heterogeneous interfaces, where they need to construct the crystalline slabs and disordered interface region correctly. 
The interface considered in this work is a solid-state battery interface between the LiCoO$_{2}$ (LCO) cathode and the Li$_{3}$ScCl$_{6}$ (LSC) solid electrolyte. 
LCO is one of the most commonly used cathode materials in commercial Li-ion batteries~\cite{RN328}. LSC~\cite{RN313} belongs to the class of halide solid electrolytes, which can achieve both high ionic conductivity and high-voltage stability, enabling the use of high-voltage cathodes in all-solid-state batteries~\cite{RN329,RN330}. In previous work~\cite{RN258}, a similar disordered interface structure was observed between LCO and a different halide solid electrolyte, Li$_{3}$YCl$_{6}$, despite the different compositions and crystal structures of the halide solid electrolytes. 
Disordered interfaces have been observed experimentally across diverse material systems. 
They arise for a variety of reasons including elemental segregation to the interface~\cite{RN347,RN348}, intermixing across the interface~\cite{RN349}, and ion irradiation~\cite{RN350}. 
For the oxide-chloride interface system, we found that chlorides have an innate tendency to form disordered interfaces with oxides; further details are documented in ESI Supplementary Note 1.  

The Li$_3$ScCl$_6$(100)-LiCoO$_{2}$(110) interface dataset was created by generating random interface structures and relaxing them. Each structure was first constructed by randomly generating 3 formula units of LSC atoms in the interface region between the LSC and LCO slabs. For each structure, the thickness of the interface region was randomly selected between 4 and 6 Å, and a random lateral displacement was applied to the LSC slab (translation allowed along the full range of both lateral directions). The randomly generated structures were relaxed using density functional theory (DFT) calculations. 

To perform relaxations faster, we trained from scratch a machine learning interatomic potential, M3GNet~\cite{RN322}, for the LSC-LCO interfaces. 
A total of 15,484 training structures consisting of optimized structures and intermediate ionic steps of the DFT relaxations were used to train the M3GNet interatomic potential. The M3GNet model achieved low test set mean absolute errors (MAEs) of 2.70 mev/atom, 20.9 meV/Å, and 0.0146 GPa for energy, force, and stress respectively. Machine learning interatomic potentials with similar (or higher) MAEs showed good performance in relaxations and molecular dynamics simulations when applied to other Li-ion conductors~\cite{RN322,RN351}.
We then relaxed randomly generated interface structures using the M3GNet interatomic potential. 
The M3GNet-relaxed structures were found to be near DFT convergence (refer to Table \ref{table_m3gnet_error} in the Methods section).
The interface energies of the relaxed structures were distributed across a wide range (approximately 1.4 J/m$^2$) as shown in ESI Fig. S4.
We define structures with interface energies no higher than 0.4 J/m$^2$ relative to the lowest energy structure to be low-interface-energy structures. 
For reference, the observation frequency of a grain boundary in aluminum metal decreases by 95 \% when the grain boundary energy increases by around 0.35 J/m$^2$~\cite{RN327}. We assembled 1,500 low-interface-energy structures (all relaxed by M3GNet only) as the disordered interface dataset.

\subsection{Generative models}
Five generative models were benchmarked on Dismai-Bench. Four of these (CDVAE~\cite{xie2022crystal}, DiffCSP~\cite{jiao2024diffcsp}, CrysTens~\cite{RN353}, UniMat~\cite{yang2023unimat}) are existing models, and one  (CryinGAN) was developed as part of this work to demonstrate the application of Dismai-Bench in model development. 
An overview of the first four models is outlined here, whereas CryinGAN will be presented in detail in Section \ref{sec_CryinGAN}.

CDVAE~\cite{xie2022crystal} is a VAE that uses equivariant graph neural networks for its encoder and decoder. The encoder is a graph convolutional network that encodes material structures into latent representations ($z$). Three multilayer perceptrons (MLPs) are used to predict the composition, lattice parameters, and number of atoms from $z$. These predictions are used to initialize structures (corresponding to the sampled $z$), where the atoms are initialized at random positions. The decoder is a graph diffusion model that denoises both the atomic coordinates and atomic species of the atoms. 

In our early tests of training CDVAE on Dismai-Bench datasets, CDVAE was found to fail in generating structures with the correct compositions (see ESI Fig. S5). 
Although CDVAE was able to predict the compositions correctly from $z$, the compositions became incorrect after the atomic species of the atoms were denoised. 
Therefore, we modified CDVAE such that the atomic species denoising becomes an optional feature, and all CDVAE benchmarking was performed without atomic species denoising. 
We also tested the effect of atomic species denoising when CDVAE is trained on the MP-20 dataset~\cite{xie2022crystal}, which includes structures from the Materials Project~\cite{RN357} of various compositions. 
Here, denoising the atomic species was found to increase the composition accuracy of reconstructed structures from around 24 \% to 54 \%. 
However, denoising the atomic species appears to be detrimental for larger structures (such as those in the Dismai-Bench datasets), and fails even when all structures have the same composition.

DiffCSP~\cite{jiao2024diffcsp} is another graph diffusion model.
The main feature introduced in DiffCSP is the ability to jointly denoise the atomic coordinates and lattice parameters. In contrast, CDVAE predicts the lattice parameters first, and they remain fixed throughout the diffusion steps. 
Note that DiffCSP does not denoise the atomic species. To better capture periodicity, DiffCSP also uses fractional instead of Cartesian coordinates (as in CDVAE), and uses periodic translation invariant Fourier transformations in its message passing. 
However, DiffCSP does not include bond angle information in its graphs, whereas CDVAE does. 

When DiffCSP was trained on the disordered interface dataset allowing joint atomic coordinate and lattice parameter diffusion, the generated structures were found to be of poor quality (refer to ESI Fig. S6a). 
We modified DiffCSP to allow teacher forcing of lattice parameters during the initial training epochs, where the ground truth lattice parameters are used as input, and lattice cost is not used to update the model (only coordinate cost is used). The quality of the generated structures improved when trained with teacher forcing (see ESI Fig. S6b). 
However, structures with the best quality were still obtained when DiffCSP was trained without any lattice diffusion, using the ground truth lattice parameters as input (see ESI Fig. S6c). Therefore, all benchmarking of DiffCSP presented in Section \ref{sec_benchmark} was performed without  lattice denoising. 
This choice also provides a consistent comparison between DiffCSP and CDVAE, since CDVAE also does not perform lattice denoising. For reference, the Dismai-Bench metrics of DiffCSP trained with lattice denoising and teacher forcing are listed in ESI Table S1. DiffCSP performance drops with lattice denoising across the metrics (see Section \ref{sec_benchmark} for benchmarking details).

CrysTens~\cite{RN353} is an image-like representation for materials. 
The pixel values of a CrysTens image are filled with information of the structure such as lattice parameters, fractional coordinates, and atomic number. 
Each CrysTens image has four channels, analogous to the RGB color channels of an image. The first channel includes a pairwise distance matrix between all atom pairs, and the remaining three channels include the pairwise $\Delta x$, $\Delta y$, and $\Delta z$ matrices respectively. CrysTens is used with a 2D U-Net diffusion model for image generation, and the generated CrysTens images are reconstructed back into material structures. 
We made no major modifications to the baseline implementation. The dimensions of the CrysTens images were increased to fit Dismai-Bench structures (original implementation only allowed up to 52 atoms). 
Refer to the Methods section for more details.

UniMat~\cite{yang2023unimat} is a video-like representation for materials. 
Each frame of a UniMat video represents a single atom in the structure, and has three channels corresponding to the $x$, $y$, and $z$ coordinates. The atomic species of the atom is indicated by the pixel location in the frame (e.g., top left pixel is H), where those pixel values correspond to the coordinates of the atom, and all other pixel values are set to -1. 
UniMat is used with a 3D U-Net diffusion model for video generation, and the generated UniMat videos are reconstructed back into materials. 
Despite the lack of any geometrical information beyond atomic coordinates, UniMat was shown to outperform CDVAE in discovering materials with lower formation energy (upon DFT relaxation of the generated structures). 
As the UniMat code is currently not openly available, we used an open-access implementation~\cite{wang2024imagen} of the 3D U-Net model~\cite{ho2022videodiff} that the UniMat model was repurposed from. 
Besides this change, no major modifications were made to UniMat. The dimensions of the UniMat frames were decreased, since each Dismai-Bench structure has only five atomic species at most.

\subsection{Dismai-Bench} \label{sec_benchmark} 
Benchmarking was performed by training all generative models on each dataset separately from scratch, such that the models only generate one type of structure at a time. 
1,000 structures were generated using each model, and post-processed as described in the Methods section. 
The disordered interface structures were relaxed using the M3GNet~\cite{RN322} interatomic potential, and the amorphous Si structures were relaxed using the SOAP-GAP~\cite{RN356} interatomic potential. No relaxations were performed on the alloy structures.  
These final structures were used to calculate the benchmark metrics.
For each generative model architecture, three separate models were trained for each dataset, and the metrics were averaged across the three models.

Although the benchmark metrics for all five generative model architectures are listed in this section, only the results for CDVAE, DiffCSP, CrysTens, and UniMat will be discussed here, since the CryinGAN architecture has not been presented yet (see Section \ref{sec_CryinGAN}). 
The benchmark results for CryinGAN will be discussed in Section \ref{sec_comparisons}, along with an overall comparison of the general performance of all architectures.

\subsubsection{Disordered LSC-LCO interface}
\begin{figure}[h!]
\centering
\includegraphics[width=\columnwidth]{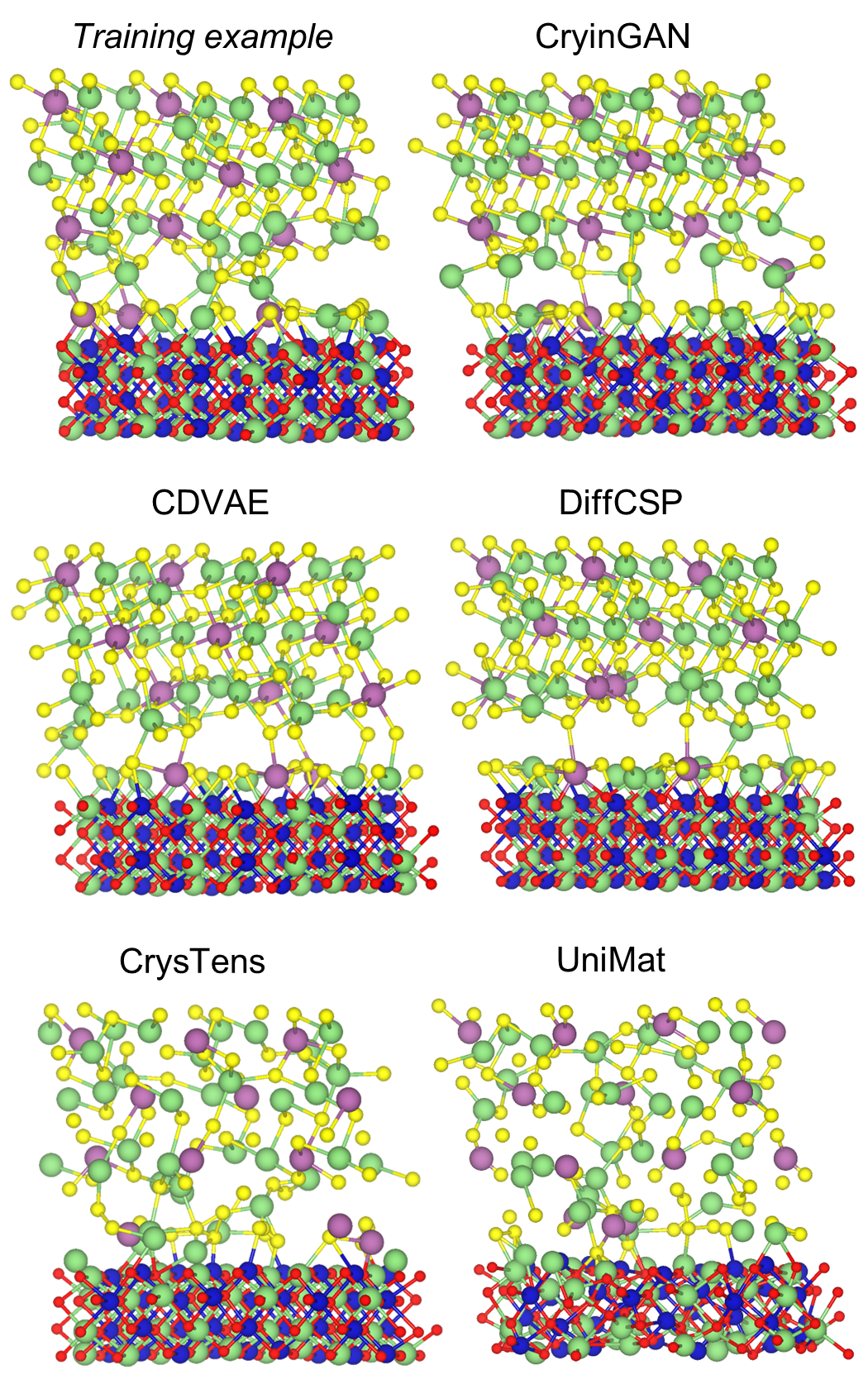}
\caption{Example disordered LSC-LCO interface structures generated by the generative models. The structures shown are as-generated by the models (i.e., no post-processing or relaxation). An example structure from the training dataset is also included for reference.}
\label{fig_int_examples}
\end{figure}

Examples of interface structures generated by the models are shown in Fig. \ref{fig_int_examples}. The structures shown are as-generated without any post-processing or relaxation. Visually, the structures generated by CryinGAN, CDVAE, and DiffCSP are similar to the training structures. 
The coordinate-based U-Net diffusion models generated the most noisy structures.

\begin{table*}[t!]
\small
\caption{Dismai-Bench metrics for the disordered LSC-LCO interfaces. Each metric is represented by the average value over 3 separately trained models. The minimum and maximum values are shown in brackets. For all metrics, lower is better.}
    \begin{tabularx}{\textwidth}{XXXXXX}
        \toprule
            Model &
            \hspace{3.7mm} $d_{\textrm{Li}}$ \newline (min, max) &
            \hspace{3.7mm} $d_{\textrm{Co}}$ \newline (min, max) &
            \hspace{3.7mm} $d_{\textrm{Sc}}$ \newline (min, max) & 
            \hspace{3.7mm} $d_{\textrm{all}}$ \newline (min, max) & 
            \% struc failed \newline (min, max)  
            \\
        \midrule
            CDVAE &
            \hspace{5mm} 0.0664 \newline (0.0601, 0.0697) &
            \hspace{5mm} 0.0427 \newline (0.0372, 0.0504) & 
            \hspace{5mm} 0.201 \newline (0.186, 0.214) & 
            \hspace{5mm} 0.0547 \newline (0.0508, 0.0589) &
            \hspace{3.7mm} 0.00  \newline (0.00, 0.00)
            \\
            DiffCSP &
            \hspace{5mm} 0.0439 \newline (0.0414, 0.0474) &
            \hspace{5mm} 0.0272 \newline (0.0240, 0.0298) & 
            \hspace{5mm} 0.0954 \newline (0.0909, 0.102) & 
            \hspace{5mm} 0.0370 \newline (0.0353, 0.0388) &
            \hspace{3.7mm} 7.17  \newline (6.00, 8.90)
            \\
            CrysTens &
            \hspace{5mm} 0.0202 \newline (0.0160, 0.0244) &
            \hspace{5mm} 0.0141 \newline (0.00707, 0.0239) & 
            \hspace{5mm} 0.0972 \newline (0.0888, 0.106) & 
            \hspace{5mm} 0.0213 \newline (0.0164, 0.0257) &
            \hspace{3.7mm} 55.3  \newline (50.6, 61.7)
            \\
            UniMat &
            \hspace{5mm} 0.131 \newline (0.0966, 0.154) &
            \hspace{5mm} 0.235 \newline (0.165, 0.292) & 
            \hspace{5mm} 0.101 \newline (0.0891, 0.111) & 
            \hspace{5mm} 0.151 \newline (0.111, 0.181) &
            \hspace{3.7mm} 64.6  \newline (60.6, 68.3)
            \\
            CryinGAN &
            \hspace{5mm} 0.0538 \newline (0.0466, 0.0600) &
            \hspace{5mm} 0.0274 \newline (0.0210, 0.0359) & 
            \hspace{5mm} 0.0888 \newline (0.0838, 0.0963) & 
            \hspace{5mm} 0.0426 \newline (0.0379, 0.0451) &
            \hspace{3.7mm} 9.00  \newline (7.20, 10.2)
            \\
        \bottomrule
    \end{tabularx} 
    \label{table_int_metrics}
\end{table*}

The generated structures were relaxed, and the benchmark metrics were calculated by analyzing the local coordination environment (motifs) of the atoms. The CrystalNNFingerprint~\cite{RN263}, which contains the coordination likelihoods and local structure order parameters of a given atom, was averaged across the structures generated by each model, and separately obtained for the training dataset as well. 
We calculated the fingerprints of the cations (Li, Co, and Sc) coordinated to the anions (Cl and O). We also appended the fraction of Cl and O neighbors in each motif to the fingerprints, so that the fingerprints contain both chemical and coordination information. 
For each model, the Euclidean distance between the average fingerprint of the training structures and the generated structures was calculated.

The benchmark metrics for the disordered interfaces are shown in Table \ref{table_int_metrics}. Parameters $d_{\textrm{Li}}$, $d_{\textrm{Co}}$, $d_{\textrm{Sc}}$, and $d_{\textrm{all}}$ represent the fingerprint Euclidean distance of Li, Co, Sc, and all cations respectively. The percentages of structures that failed to be post-processed or relaxed are also listed. Notably, CDVAE is the only model that generates structures close to being fully relaxed, requiring only an average of 10 relaxation steps per structure (see ESI Table S2). All other models required at least 68 relaxation steps on average, and the U-Net models have $>50$ \% failed structures. UniMat exhibited the highest \% failed structures and $d_{\textrm{all}}$, likely because it uses a less expressive material representation (atomic coordinates only). 
CrysTens, despite the inclusion of bond distances, also has high \% failed structures. Although its Euclidean distance metrics are small, the low values are only achieved after relaxation (refer to Fig. \ref{fig_int_examples} and Table S2 for unrelaxed structures). On the other hand, the graph diffusion models perform significantly better than the U-Net diffusion models due to the higher expressive power of graphs. Comparing between CDVAE and DiffCSP, CDVAE generates interface structures closer to convergence than DiffCSP, but upon relaxation, DiffCSP achieves lower distance metrics than CDVAE.  

\subsubsection{Amorphous silicon}

\begin{figure}[h!]
\centering
\includegraphics[width=\columnwidth]{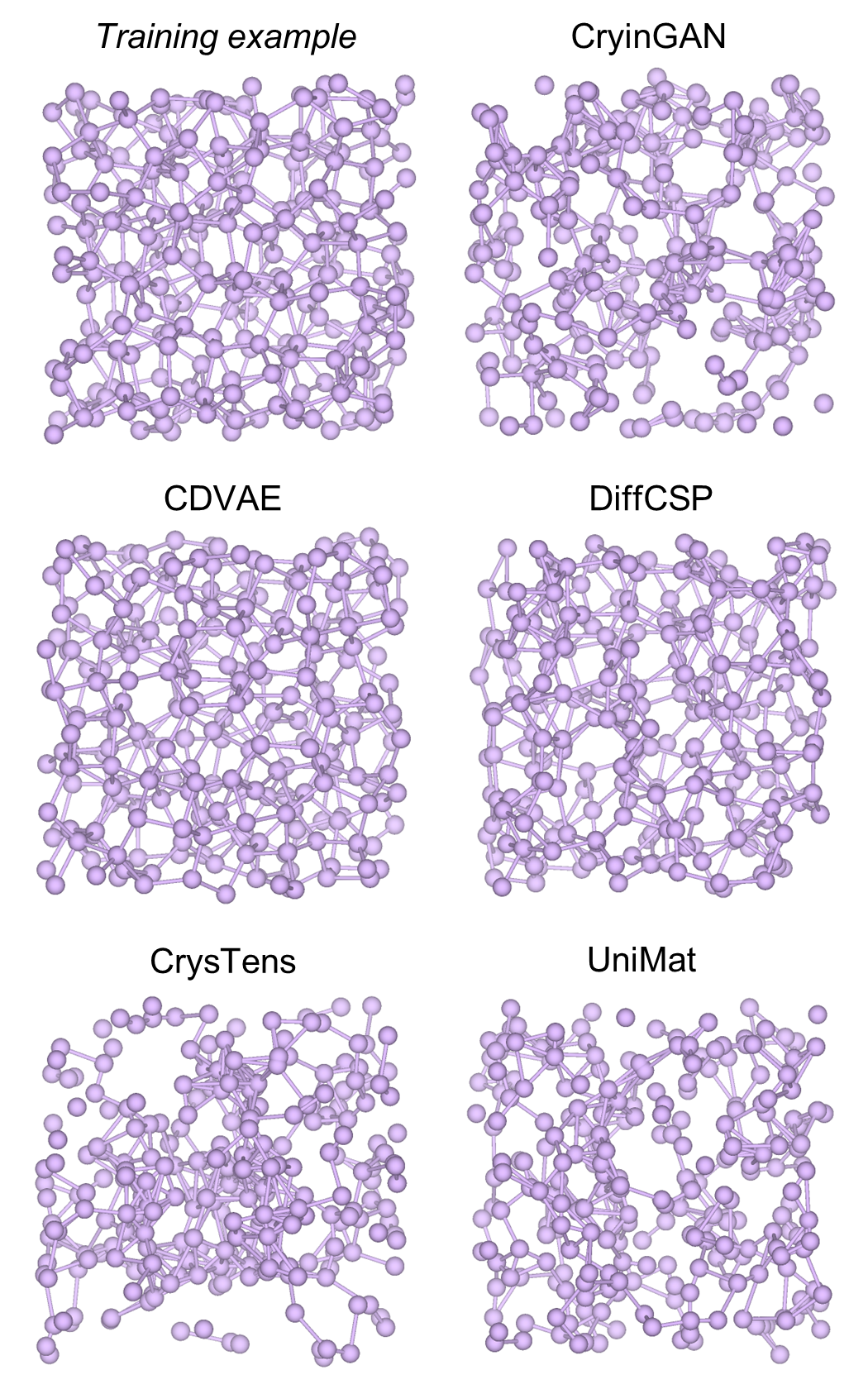}
\caption{Example amorphous Si structures generated by the generative models. The structures shown are as-generated by the models (i.e., no post-processing or relaxation). An example structure from the training dataset is also included for reference.}
\label{fig_a-Si_examples}
\end{figure}

Examples of amorphous Si structures generated by the models are shown in Fig. \ref{fig_a-Si_examples}. 
Only the graph diffusion models (CDVAE \& DiffCSP) were able to generate the structures successfully.
The other models generated random-looking structures with large voids and little bonding between atoms. Therefore, further detailed benchmarking of amorphous Si was only performed for CDVAE and DiffCSP. Similar to the disordered interfaces, we determined the coordination motif fingerprints of Si, and  the Euclidean distance between the average fingerprint of the training and generated structures. We
also determined the radial distribution functions (RDFs) and bond angle distributions of the structures, and compared them to that of the training set. 

\begin{table*}[t!]
\small
\caption{Dismai-Bench metrics for amorphous Si. Each metric is represented by the average value over 3 separately trained models. The minimum and maximum values are shown in brackets. For all metrics, lower is better.}
    \begin{tabularx}{\textwidth}{XXXXX}
        \toprule
            Model &
            \hspace{3.7mm} $d_{\textrm{motif}}$ \newline (min, max) &
            \hspace{3.7mm} $d_{\textrm{rdf}}$ \newline (min, max) &
            \hspace{3.7mm} $d_{\textrm{angle}}$ \newline (min, max) & 
            \% struc failed \newline (min, max)  
            \\
        \midrule
            CDVAE &
            \hspace{5mm} 0.0402 \newline (0.0396, 0.0412) &
            \hspace{5mm} 0.392 \newline (0.381, 0.408) & 
            \hspace{5mm} 0.00332 \newline (0.00312, 0.00353) & 
            \hspace{3.7mm} 0.00  \newline (0.00, 0.00)
            \\
            DiffCSP &
            \hspace{5mm} 0.0647 \newline (0.0462, 0.0908) &
            \hspace{5mm} 1.39 \newline (1.03, 1.69) & 
            \hspace{5mm} 0.0103 \newline (0.00797, 0.0133) & 
            \hspace{3.7mm} 0.00  \newline (0.00, 0.00)
            \\
        \bottomrule
    \end{tabularx} 
    \label{table_a-Si_metrics}
\end{table*}

The benchmark metrics for amorphous Si are shown in Table \ref{table_a-Si_metrics}. Parameters $d_{\textrm{motif}}$, $d_{\textrm{rdf}}$, and $d_{\textrm{angle}}$ represent the Euclidean distance for motif fingerprint, RDF, and bond angle distribution respectively. Both CDVAE and DiffCSP had 0 \% failed structures, but CDVAE-generated structures were significantly closer to being fully relaxed than DiffCSP-generated structures. 
CDVAE only required an average of 23 relaxation steps per structure, whereas DiffCSP required an average of 140 steps per structure (see ESI Table S3). Similarly, all distance metrics of CDVAE are lower than DiffCSP.
The lack of bond angle information in DiffCSP's graphs is likely one main contributor to these trends. 
The RDFs and bond angle distributions of CDVAE and DiffCSP are shown in ESI Fig. S7. DiffCSP and CDVAE encode distance information in their graphs, and both show similar RDFs to the training dataset. However, there is a clear difference when comparing the bond angle distributions, where DiffCSP shows a larger discrepancy with the training dataset than CDVAE. These results indicate that bond angle information is particularly beneficial for helping generative models learn amorphous structures, since amorphous structures have more complicated arrangements of atoms that do not locate on lattice sites.

\subsubsection{Disordered stainless steel alloy}
\begin{figure}[h!]
\centering
\includegraphics[width=\columnwidth]{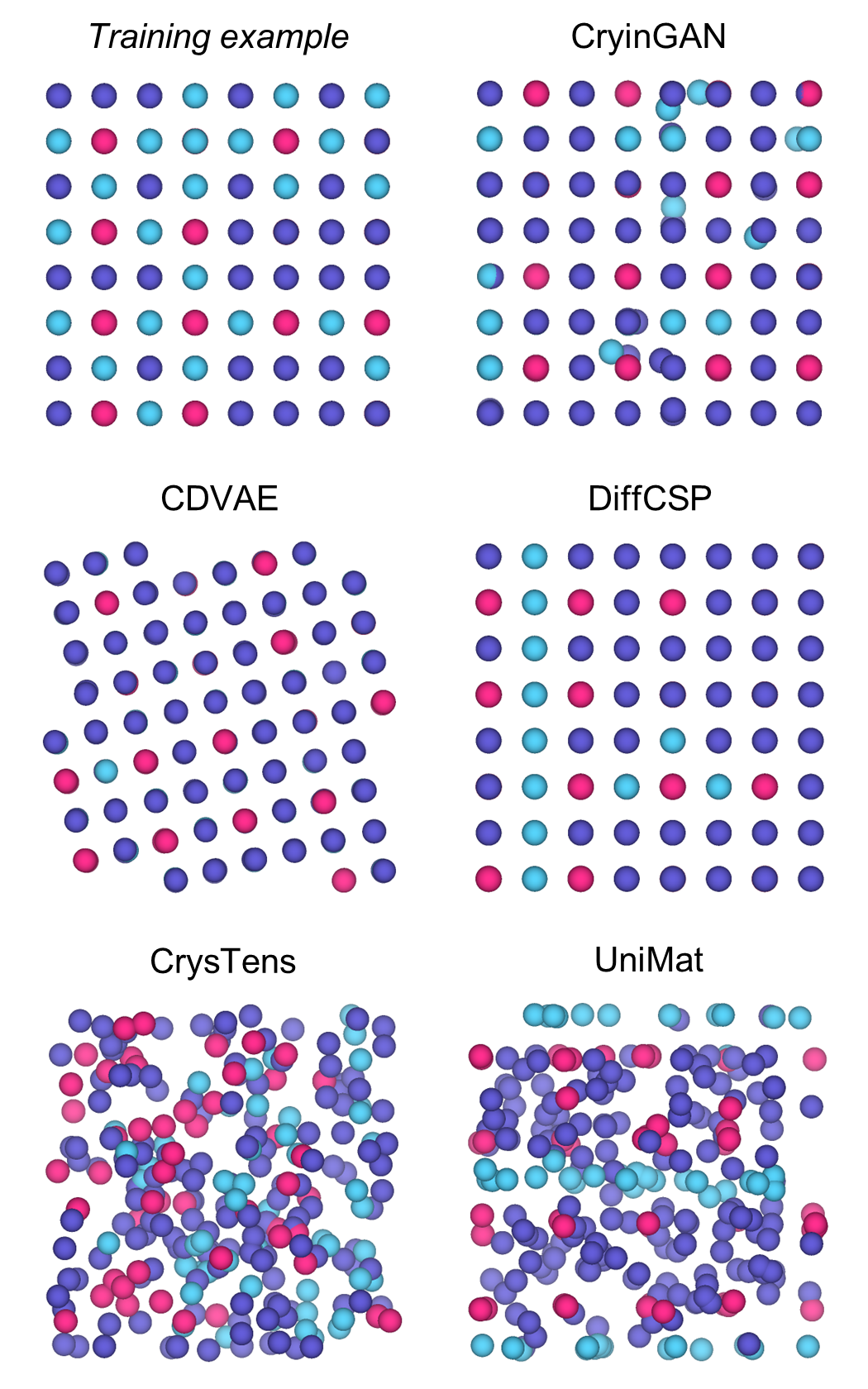}
\caption{Example alloy structures (300 K, narrow SRO) generated by the generative models. The structures shown are as-generated by the models (i.e., no post-processing). An example structure from the training dataset is also included for reference.}
\label{fig_alloy_examples}
\end{figure}

Examples of disordered alloy structures generated by the models are shown in Fig. \ref{fig_alloy_examples}. 
CryinGAN, CDVAE, and DiffCSP were able to generate the FCC alloy structures successfully, whereas CrysTens and UniMat 
were unable to reproduce the underlying FCC lattice. 
Therefore, detailed alloy benchmarking was only performed for CryinGAN, CDVAE, and DiffCSP. Interestingly, CDVAE was prone to generating rotated structures (since graphs are rotationally invariant), but all DiffCSP-generated structures were unrotated. 
Some CDVAE-generated structures had noisy lattices, where atoms exhibited relatively large deviations from pristine FCC lattice sites (see ESI Fig. S8a), whereas all DiffCSP-generated structures had well-defined lattices. 
A fraction of CDVAE-generated structures had slightly shorter lattice spacing, creating additional sites that were vacant since the total number of atoms was fixed (see ESI Fig. S8b). 
DiffCSP's stronger ability to learn the FCC lattice is likely due to the use of fractional coordinates and Fourier transformations in its message passing to capture periodicity~\cite{jiao2024diffcsp}. 

\begin{table}[h!]
\small
\caption{Dismai-Bench metrics for the disordered stainless steel alloy. Each metric is represented by the average value over 3 separately trained models. The minimum and maximum values are shown in brackets. For all metrics, lower is better.}
    \begin{tabularx}{\columnwidth}{lp{22mm}XX}
        \toprule
            Model &
            \hspace{2.5mm} $d_{\textrm{cluster}}$ \newline (min, max) &
            \% struc w/ vac \newline (min, max) &
            \% struc failed \newline (min, max)  
            \\
        \midrule
            \multicolumn{4}{l}{\textbf{300 K, narrow SRO}}
            \\
        \midrule
            CDVAE &
            \hspace{5mm} 0.0604 \newline (0.0540, 0.0675) &
            \hspace{3.7mm} 35.8 \newline (32.0, 41.2) & 
            \hspace{3.7mm} 0.13  \newline (0.00, 0.40)
            \\
            DiffCSP &
            \hspace{5mm} 0.0645 \newline (0.0566, 0.0792) &
            \hspace{3.7mm} 94.3 \newline (91.8, 96.5) & 
            \hspace{3.7mm} 0.00  \newline (0.00, 0.00)
            \\
            CryinGAN &
            \hspace{5mm} 0.117 \newline (0.106, 0.126) &
            \hspace{3.7mm} 100 \newline (100, 100) & 
            \hspace{3.7mm} 0.00  \newline (0.00, 0.00)
            \\
        \midrule
            \multicolumn{4}{l}{\textbf{300 K, wide SRO}}
            \\
        \midrule
            CDVAE &
            \hspace{5mm} 0.0658 \newline (0.0598, 0.0742) &
            \hspace{3.7mm} 37.4 \newline (34.0, 43.4) & 
            \hspace{3.7mm} 0.07  \newline (0.00, 0.10)
            \\
            DiffCSP &
            \hspace{5mm} 0.103 \newline (0.0831, 0.121) &
            \hspace{3.7mm} 95.8 \newline (95.4, 96.2) & 
            \hspace{3.7mm} 0.00  \newline (0.00, 0.00)
            \\
            CryinGAN &
            \hspace{5mm} 0.125 \newline (0.119, 0.129) &
            \hspace{3.7mm} 100 \newline (100, 100) & 
            \hspace{3.7mm} 0.00  \newline (0.00, 0.00)
            \\
        \midrule
            \multicolumn{4}{l}{\textbf{1500 K, narrow SRO}}
            \\
        \midrule
            CDVAE &
            \hspace{5mm} 0.0621 \newline (0.0589, 0.0665) &
            \hspace{3.7mm} 54.8 \newline (51.9, 56.8) & 
            \hspace{3.7mm} 0.37  \newline (0.20, 0.50)
            \\
            DiffCSP &
            \hspace{5mm} 0.0308 \newline (0.0304, 0.0313) &
            \hspace{3.7mm} 91.7 \newline (91.5, 91.8) & 
            \hspace{3.7mm} 0.00  \newline (0.00, 0.00)
            \\
            CryinGAN &
            \hspace{5mm} 0.0643 \newline (0.0629, 0.0659) &
            \hspace{3.7mm} 100 \newline (100, 100) & 
            \hspace{3.7mm} 0.00  \newline (0.00, 0.00)
            \\
        \midrule
            \multicolumn{4}{l}{\textbf{1500 K, wide SRO}}
            \\
        \midrule
            CDVAE &
            \hspace{5mm} 0.0618 \newline (0.0549, 0.0694) &
            \hspace{3.7mm} 61.3 \newline (55.9 68.2) & 
            \hspace{3.7mm} 0.27  \newline (0.20, 0.30)
            \\
            DiffCSP &
            \hspace{5mm} 0.0338 \newline (0.0332, 0.0342) &
            \hspace{3.7mm} 92.3 \newline (89.2, 94.4) & 
            \hspace{3.7mm} 0.00  \newline (0.00, 0.00)
            \\
            CryinGAN &
            \hspace{5mm} 0.0650 \newline (0.0609, 0.0691) &
            \hspace{3.7mm} 100 \newline (100, 100) & 
            \hspace{3.7mm} 0.00  \newline (0.00, 0.00)
            \\
        \bottomrule
    \end{tabularx} 
    \label{table_alloy_metrics}
\end{table}

The generated structures were post-processed to remove atoms not on lattice sites. Any structure with > 50 atoms removed ($\sim$20 \% of all atoms) was considered a failed structure and rejected. Then, the fingerprint of each structure was calculated using a vector of conditional probabilities of observing each cluster (monomer/dimer) in the structure, up to the 7$^{\textrm{th}}$ neighbor shell. The Euclidean distance between the average cluster fingerprint of the training and generated structures, $d_{\textrm{cluster}}$, was calculated for each model. 

Benchmark metrics for the disordered alloys are shown in Table \ref{table_alloy_metrics}. The percentage of failed structures is 0 \% for DiffCSP and CryinGAN, while CDVAE exhibits a tiny percentage of failed structures. However, the percentage of structures with site vacancies is 90-100 \% for DiffCSP and CryinGAN. In comparison, CDVAE has significantly lower percentages of structures with vacancies, around roughly 40 \% for 300 K structures and 60 \% for 1500 K structures. Although DiffCSP-generated structures had more well-defined lattices than CDVAE-generated structures, almost all structures had overlapping atoms on lattice sites, resulting in site vacancies in the structures. 

For the 300 K structures with narrow SRO, the $d_{\textrm{cluster}}$ values of CDVAE and DiffCSP are similar. However, for the 300 K structures with wide SRO, CDVAE has lower $d_{\textrm{cluster}}$ than DiffCSP. The wide SRO structures have less consistent SRO distributions, so DiffCSP had more difficulty in learning the wide SRO structures than the narrow SRO structures. On the other hand, $d_{\textrm{cluster}}$ only increased slightly for CDVAE between the wide and narrow SRO structures, indicating CDVAE's stronger ability to learn different degrees of SRO. 
For the 1500 K structures, which more resemble random solid solutions, there is little difference in $d_{\textrm{cluster}}$ between the narrow and wide SRO structures. 
Here, $d_{\textrm{cluster}}$ is lower for DiffCSP than CDVAE. Some factors contributing to DiffCSP's better performance in generating random solid solutions may be its stronger ability in learning the FCC lattice, weaker ability in learning SRO patterns, and low number of site vacancies ($\sim$1.5 vacancies per structure). 
The Warren-Cowley SRO parameter distributions for the 1st and 2nd nearest neighbor interactions are shown in ESI Fig. S9-11 for reference. Overall, the models were able to generate structures with similar SRO distributions to the training structures.

\subsection{CryinGAN development} \label{sec_CryinGAN}
We present here a case study of developing a generative model with the help of Dismai-Bench, demonstrating how meaningful feedback about model performance is obtained through direct comparisons between generated and training structures. 
We considered the case of a point-cloud-based GAN to generate low-interface-energy LSC-LCO interface structures. 
We chose to focus on the interface structures during development for simplicity, and subsequently benchmarked CryinGAN on other datasets to evaluate its generalizability. The disordered interfaces were also the only structures where all four diffusion models were able to generate successfully, hence providing the most informative model comparison.
We show that, given the correct architecture, a coordinate-based representation can still perform well on Dismai-Bench (unlike CrysTens and UniMat), and that more complicated architectures do not necessarily outperform simpler architectures.

\begin{figure*}[t!]
\centering
\includegraphics[width=\textwidth]{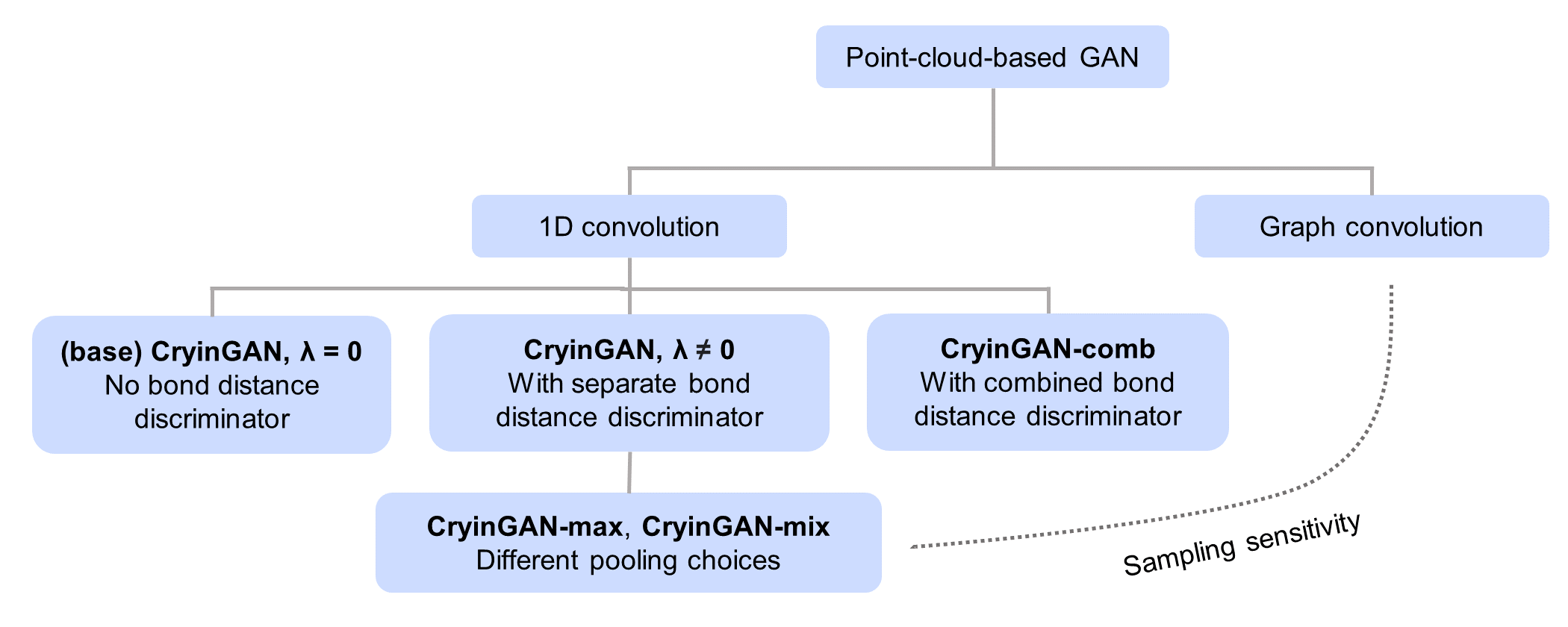}
\caption{Schematic of GAN architectures tested. The discriminator either uses 1D convolutions (PointNet) or graph convolutions (CGCNN). For the 1D-convolution-based discriminators, the primary CryinGAN design consists of a fractional coordinate discriminator and a separate bond distance discriminator, where the output of the latter is weighted by $\lambda$. Alternate pooling choices were tested (CryinGAN-max and CryinGAN-mix). Graph convolution and different types of pooling affect the sampling sensitivity of the discriminator. CryinGAN-comb combines the two discriminators into a single discriminator.}
\label{fig_GAN_chart}
\end{figure*}

A typical GAN consists of two neural networks: a generator and a discriminator. The role of the generator is to generate material structures from input noise, whereas the role of the discriminator is to distinguish between the real (training) structures and the fake (generated) structures. The generator and discriminator compete with each other during training to progressively improve the quality of the generated structures. The point cloud representation was used for the GANs, and since all of the structures in the dataset share the same lattice parameters, each structure was represented by the fractional coordinates of its atoms only. We tested a couple of different GAN architectures as summarized in Fig. \ref{fig_GAN_chart}. 

The base GAN model (CryinGAN) was adapted from the Composition-Conditioned Crystal GAN (CCCGAN) presented by Kim et al.~\cite{RN307}, which was used to generate Mg-Mn-O ternary materials. One-dimensional (1D) convolutions were used in the discriminator to extract the latent features of structures, an inspiration taken from PointNet~\cite{qi2017pointnet}, a 3D object classification and segmentation network. These 1D convolutions have been similarly implemented in other point-cloud-based crystal generative models such as FTCP-VAE~\cite{RN294} and CubicGAN~\cite{RN295}. We simplified and generalized CCCGAN to be used for any periodic system with fixed lattice, composition, and number of atoms. We did not include any conditional generation capability in CryinGAN for simplicity. The discriminator of the original CCCGAN relied solely on atomic coordinates and lattice parameters to distinguish between real and fake structures. To further provide explicit bond distance information to the discriminator, we added a second discriminator to the CryinGAN model that has the same architecture, but accepts bond distances as input instead of coordinates. 

\begin{figure*}[t!]
\centering
\includegraphics[width=\textwidth]{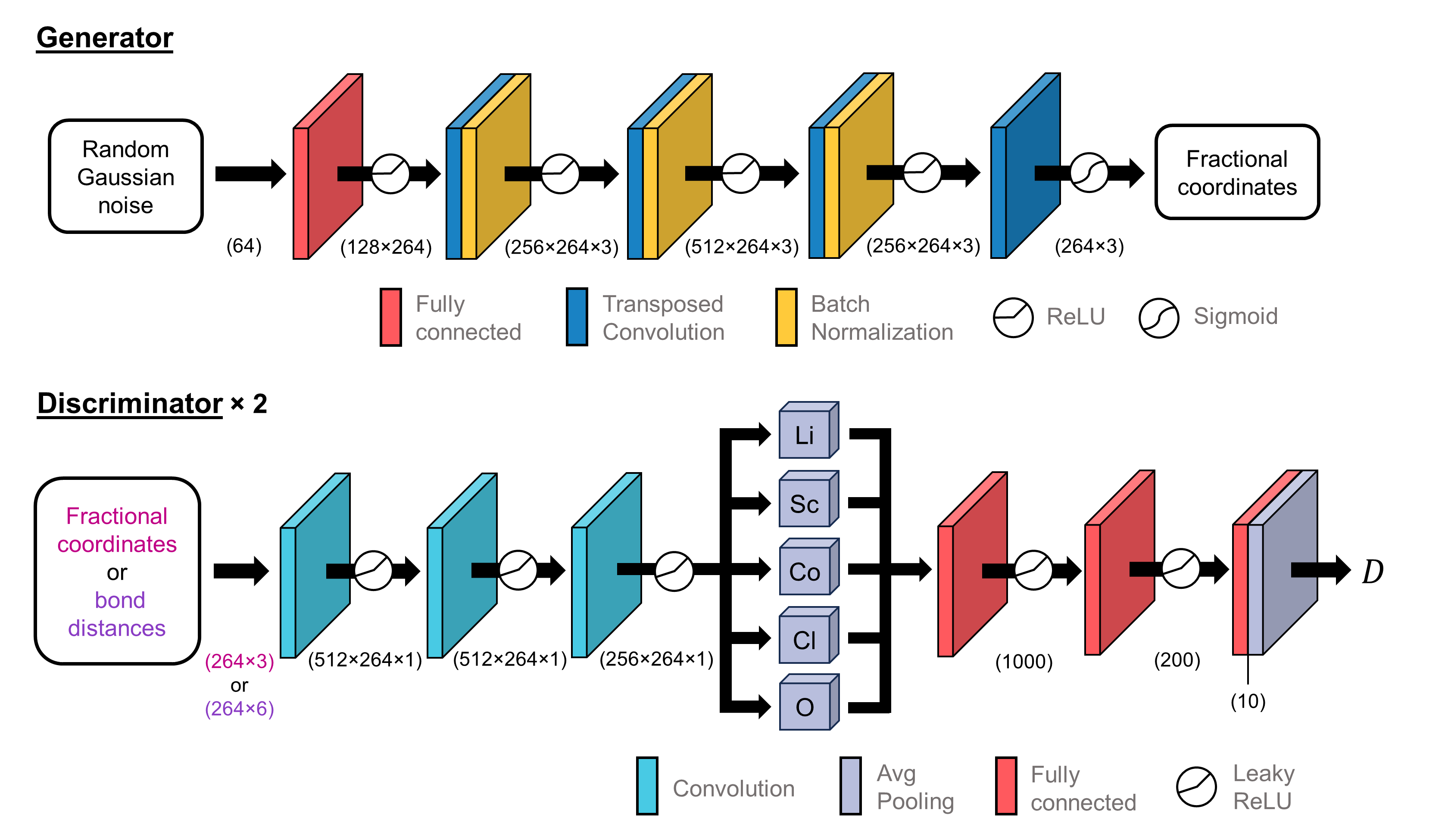}
\caption{CryinGAN architecture. The generator takes in random Gaussian noise as input and produces fractional coordinates as output. There are two discriminators with the same architecture, but with different inputs (fractional coordinates or bond distances). $D$ is the discriminator output used to calculate the losses. The numbers in brackets represent the dimensions of tensors before/after a layer, where the batch dimension is omitted. Note that 264 corresponds to the number of atoms in each interface structure. The size of the layers shown in the figure do not reflect the tensor dimensions.}
\label{fig_GAN_arc}
\end{figure*}

The CryinGAN model architecture is shown in Fig. \ref{fig_GAN_arc}. The generator accepts random gaussian noise as input, and produces fractional coordinates of structures as output. 
Of the two discriminators, one  accepts fractional coordinates as input, and the other accepts bond distances (6 nearest-neighbors of each atom) as input. In the first convolutional layer of the discriminators, the fractional coordinates/bond distances are convoluted along each row separately (i.e., separate 1D convolutions for each atom). Note that CryinGAN is permutationally invariant to atom ordering within each atomic species block (e.g., order of the 72 Li atoms does not matter). CryinGAN implements the Wasserstein loss, which was shown to provide more stable training by preventing the vanishing gradients of the traditional GAN~\cite{arjovsky2017wgan, gulrajani2017gradpenalty}. The discriminator loss function with a gradient penalty term for improved stability~\cite{gulrajani2017gradpenalty} is:
\begin{equation}
L_{\textrm{disc}} = \underset{\tilde{x} \sim \mathbb{P}_g}{\mathbb{E}}[D(\tilde{x})] - \underset{x \sim \mathbb{P}_r}{\mathbb{E}}[D(x)] + \mu \underset{\hat{x} \sim \mathbb{P}_{\hat{x}}}{\mathbb{E}}[(\lVert \nabla_{\hat{x}}D(\hat{x}) \rVert_2 - 1)^2] \hspace{0.5em},
\end{equation}
where $D$ is the discriminator output, $\tilde{x}$ and $x$ are the inputs for generated and real structures respectively. 
The distribution 
$\mathbb{P}_{\hat{x}}$ is taken over interpolated samples between the distribution of real structures $\mathbb{P}_r$, and the distribution of generated structures $\mathbb{P}_g$. 
In the code implementation, $\hat{x}$ is obtained by interpolating between the fractional coordinates/bond distances of training structures and generated structures. The interpolation point between any two data points is chosen randomly. 
The parameter $\mu$ is the gradient penalty coefficient set to 10 similar to past Wasserstein GANs~\cite{gulrajani2017gradpenalty,RN295,RN307}. 
The total discriminator loss, to be minimized, is a weighted sum of the losses from both discriminators 
\begin{equation}
L_{\textrm{disc,total}} = L_{\textrm{disc,coord}} + \lambda L_{\textrm{disc,bond}} \hspace{0.5em}.
\end{equation}
Here, $L_{\textrm{disc,coord}}$ and $L_{\textrm{disc,bond}}$ are the losses from the fractional coordinate discriminator and bond distance discriminator respectively, and $\lambda$ is the weight of the bond distance discriminator loss. The total generator loss, to be maximized, is computed similarly according to:
\begin{equation}
L_{\textrm{gen}} = \underset{\tilde{x} \sim \mathbb{P}_g}{\mathbb{E}}[D(\tilde{x})] \hspace{0.5em}, 
\end{equation}
\begin{equation}
L_{\textrm{gen,total}} = L_{\textrm{gen,coord}} + \lambda L_{\textrm{gen,bond}} \hspace{0.5em}. 
\end{equation}
Note that the base CryinGAN model uses only the fractional coordinate discriminator ($\lambda=0$), which we use as the baseline reference model. 

We developed the GAN models with the intention of evaluating the best performing model by comparing between DFT-relaxed training and generated structures. 
The training dataset was curated from structures relaxed with the M3GNet interatomic potential followed by DFT calculations (refer to Methods section). 
Note that the CryinGAN Dismai-Bench metrics shown in Section \ref{sec_benchmark} were calculated using the same procedure as described previously. 
To study the effect of $\lambda$ on CryinGAN model performance, we trained CryinGAN models on the (DFT-relaxed) dataset of interface structures with varying $\lambda$. 
A visualization of the training process is shown in ESI Movie S1, where CryinGAN progressively learns to generate low interface energy structures over training epochs. 
Due to noise in the models, the generated structures often have a small fraction of atoms that are too close to each other. 
As $\lambda$ increases, we found that the number of pairs of atoms generated too close together quickly decreases and then levels off around $\lambda$ = 0.05-0.1 (see ESI Fig. S12). 
This observation is an indicator that including the bond distance discriminator is helpful for training the generator to create structures with more reasonable atom-atom distances. 
The generated structures were then relaxed using M3GNet, refer to ESI Fig. S13 for examples of structures before and after relaxation.

\begin{figure}[t!]
\centering
\includegraphics[width=\columnwidth]{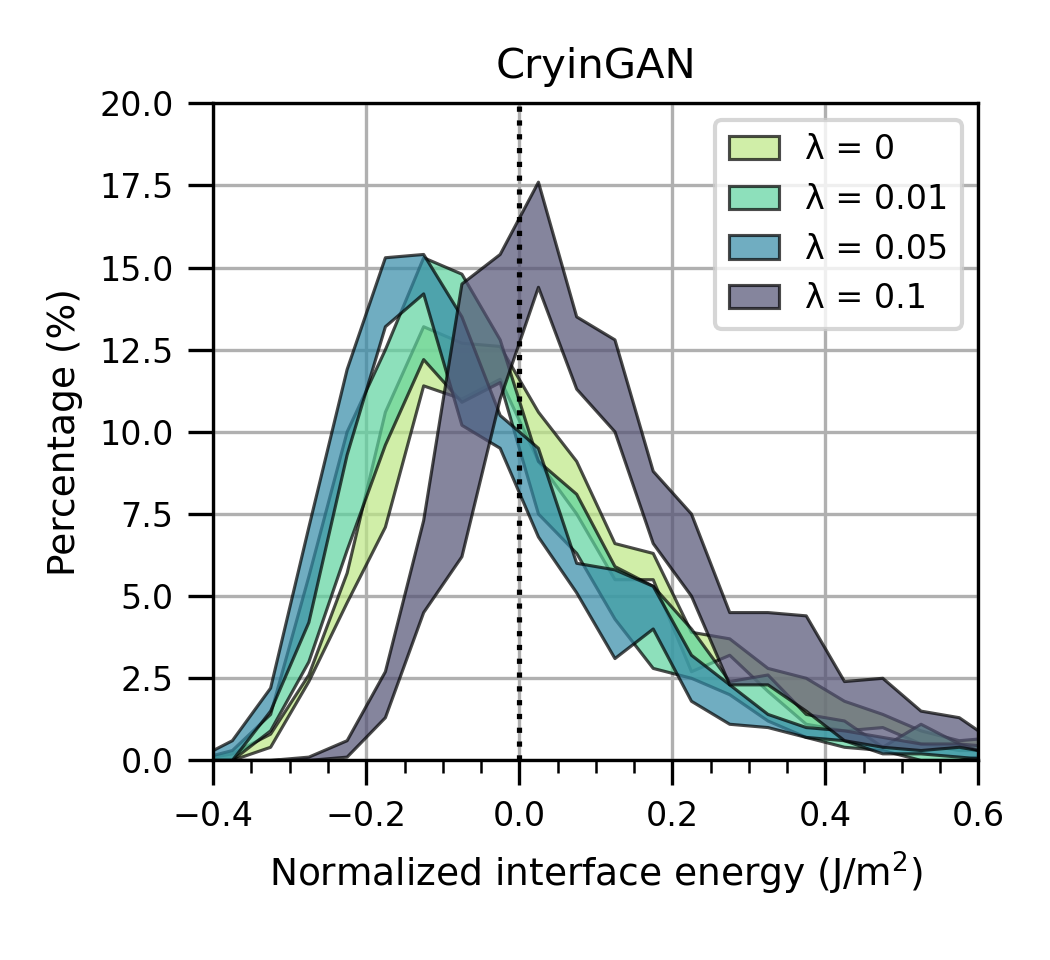}
\caption{Normalized interface energy distributions of structures generated using CryinGAN trained with different $\lambda$ values (0, 0.01, 0.05, and 0.1). The generated structures were relaxed using M3GNet, and the interface energies shown are based on M3GNet-calculated energies. For each $\lambda$ value, three separate models were trained, and the spread of the interface energy distributions is indicated by the shading. As $\lambda$ increases, the interface energy distribution initially shifts to lower energies, then shifts to higher energies.}
\label{fig_lambda}
\end{figure}

The objective was to obtain a model with the lowest interface energy distribution. 
Fig. \ref{fig_lambda} shows the interface energy distribution for the models trained with different $\lambda$. 
For each  $\lambda$, three separate models were trained and the spread of the interface energy distributions is indicated by the shading (the distribution of all trained models without shading is provided in ESI Fig. S14). 
The results indicate that as $\lambda$ increases from 0 to 0.05, the interface energy distribution shifts to lower energies. 
However, as $\lambda$ further increases to 0.1, the interface energy distribution shifts to higher energies. 
The coordination motif fingerprint distance between the generated and training structures also shows the same trend (see ESI Fig. S15), where it decreases between $\lambda=0$ and $\lambda=0.05$, then increases when $\lambda>0.05$.
These observations show that optimal values of $\lambda$ improve model performance, but excessive weight on the bond distance discriminator causes the generator to prioritize the bond distances too much over positioning the atoms correctly. 
The use of a second discriminator does slow down training compared to using only a single discriminator, requiring around twice as long to train the model for the same number of epochs. However, we found that two discriminators still outperform a single discriminator given the same amount of training time (see ESI Fig. S16), justifying the benefits of the bond distance discriminator. 
Whereas around 34 \% of the generated structures failed to converge during M3GNet relaxation for $\lambda = 0$, only around 11 \% did not converge for $\lambda = 0.05$. 
These results show that with an appropriately tuned $\lambda$, the bond distance discriminator improves model performance. 

We further considered a couple of alternative GAN architectures (refer to Fig. \ref{fig_GAN_chart}). CryinGAN-comb combines both discriminators into a single discriminator, circumventing the need to tune $\lambda$. CryinGAN-max uses max pooling in the discriminator, instead of average pooling as in CryinGAN. CryinGAN-mix uses the mix pooling operation proposed by Wang et al.~\cite{wang2020sampling}, where both max and average pooling operations are used together. The type of pooling operation affects the sampling sensitivity of the discriminator and the overall performance of the GAN. The sampling sensitivity describes how sensitive the discriminator is to changes in point density or the sampling pattern of the input point cloud. Max pooling was reported to cause lower sampling sensitivity than average pooling~\cite{wang2020sampling}. Overall, CryinGAN was found to outperform all of these alternative architectures, see ESI Supplementary Note 2 for further details.
 
The superior performance of CryinGAN over CryinGAN-max and CryinGAN-mix shows that higher sampling sensitivity is beneficial for learning atomic configurations.
The sampling sensitivity can be further increased through the use of graph convolutions, where each atom is convoluted with its surrounding atoms and bonds. 
We attempted a graph convolutional discriminator by adapting the commonly used Crystal Graph Convolutional Neural Networks (CGCNN)~\cite{RN215}.
However, the training losses diverged and it was not possible to train a GAN that could generate useful structures (see ESI Fig. S17). This result is consistent with the findings of Wang et al.~\cite{wang2020sampling}, whose graph convolutional GAN also failed to produce point clouds of 3D objects. 
Graph convolutions are highly sensitive to sampling, making them prone to overfocus on the sampling pattern of a point cloud instead of the overall structure. 
This overfocus bears similarity to the behavior observed here for CryinGAN at high $\lambda$ values. 
For graph convolutions to be implemented in point-cloud-based GANs, we expect that a carefully designed architecture will be required to take advantage of its sampling sensitivity without destabilizing training.

\subsection{Detailed evaluation of CryinGAN-generated interfaces}
\begin{figure}[h!]
\centering
\includegraphics[width=\columnwidth]{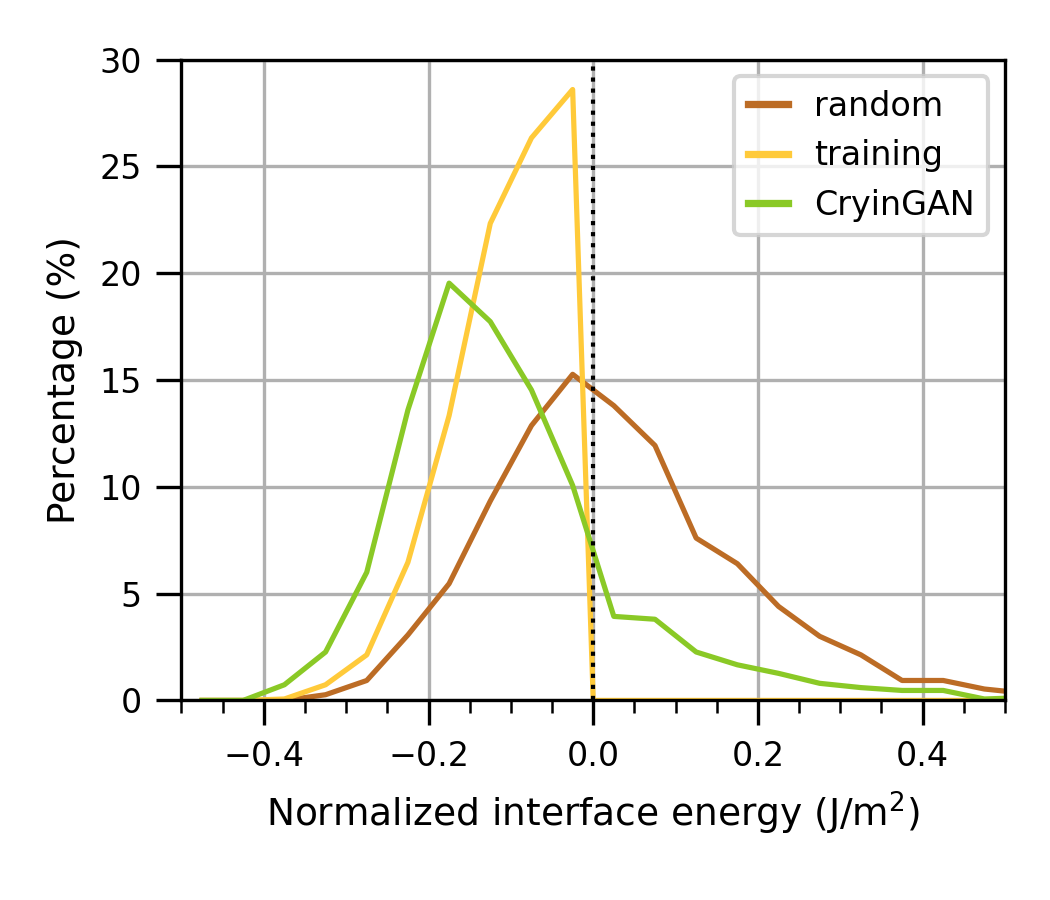}
\caption{Normalized interface energy distribution of structures generated using CryinGAN (green), compared against structures that were randomly generated (brown). All structures shown were relaxed using M3GNet followed by DFT calculations, and the energies shown are with respect to DFT-calculated energies. Randomly generated structures with (relaxed) normalized interface energy $\leq$ 0 J/m$^2$ were used as the CryinGAN training dataset (yellow).}
\label{fig_e_best}
\end{figure}

Here, we further evaluate CryinGAN to demonstrate that the generated structures are energetically and structurally similar to the training structures.
We trained a CryinGAN model ($\lambda = 0.05$) for a higher number of epochs with a shorter interval between generator trainings (see Methods section for more details). 
Fig. \ref{fig_e_best} shows the interface energy distribution of the relaxed CryinGAN structures, compared to randomly generated structures (also relaxed). 
The shift of the interface energy distribution to the left shows that CryinGAN has learnt to generate low-interface-energy structures. 
With random generation, only around 48 \% of the structures had low interface energy, whereas with CryinGAN, the percentage is around 85 \%. 

We filtered CryinGAN structures with low interface energy ($\leq$ 0 J/m$^2$) for structural comparison with the training structures.
We also compared the CryinGAN structures to a dataset of high-interface-energy structures ($>$ 0 J/m$^2$) that were randomly generated and relaxed. 
We analyzed the coordination motif fingerprints of the cations (Li and Sc) in the interface region coordinated to the anions (Cl and O), to determine the structural similarities/differences. 
The comparison for Li motifs showed smaller differences (see ESI Supplementary Note 3 for more details), so we focus the discussion here on Sc motifs. 
Table \ref{table_fingerprint_Sc} shows the Euclidean distance and cosine similarity between the average interface Sc fingerprint of the CryinGAN/high energy dataset and the training dataset.
The CryinGAN dataset has a lower Euclidean distance and higher cosine similarity than the high energy dataset. 
These results indicate that the interface Sc atoms in the CryinGAN structures are more similar to the low-interface-energy training structures, than the high-interface-energy structures. 

\begin{table}[h!]
\small
\caption{Euclidean distance and cosine similarity between the average interface Sc site fingerprint of the training structures and the CryinGAN/high-interface-energy structures. The 95 \% bootstrap confidence intervals are shown in brackets.}
    \begin{tabularx}{\columnwidth}{lXX}
    \toprule
        Dataset &  Euclidean distance \newline (95 \% CI) &  Cosine similarity \newline (95 \% CI) \\
    \midrule
         CryinGAN &  0.2791 \newline (0.2500
 to 0.3077) &  0.9905 \newline (0.9886 to 0.9926)\\
         High energy & 0.7074 \newline (0.6734 to 0.7429) & 0.9316 \newline (0.9250 to 0.9380) \\
    \bottomrule
    \end{tabularx}
    \label{table_fingerprint_Sc}
\end{table}

\begin{figure}[t!]
\centering
\includegraphics[width=\columnwidth]{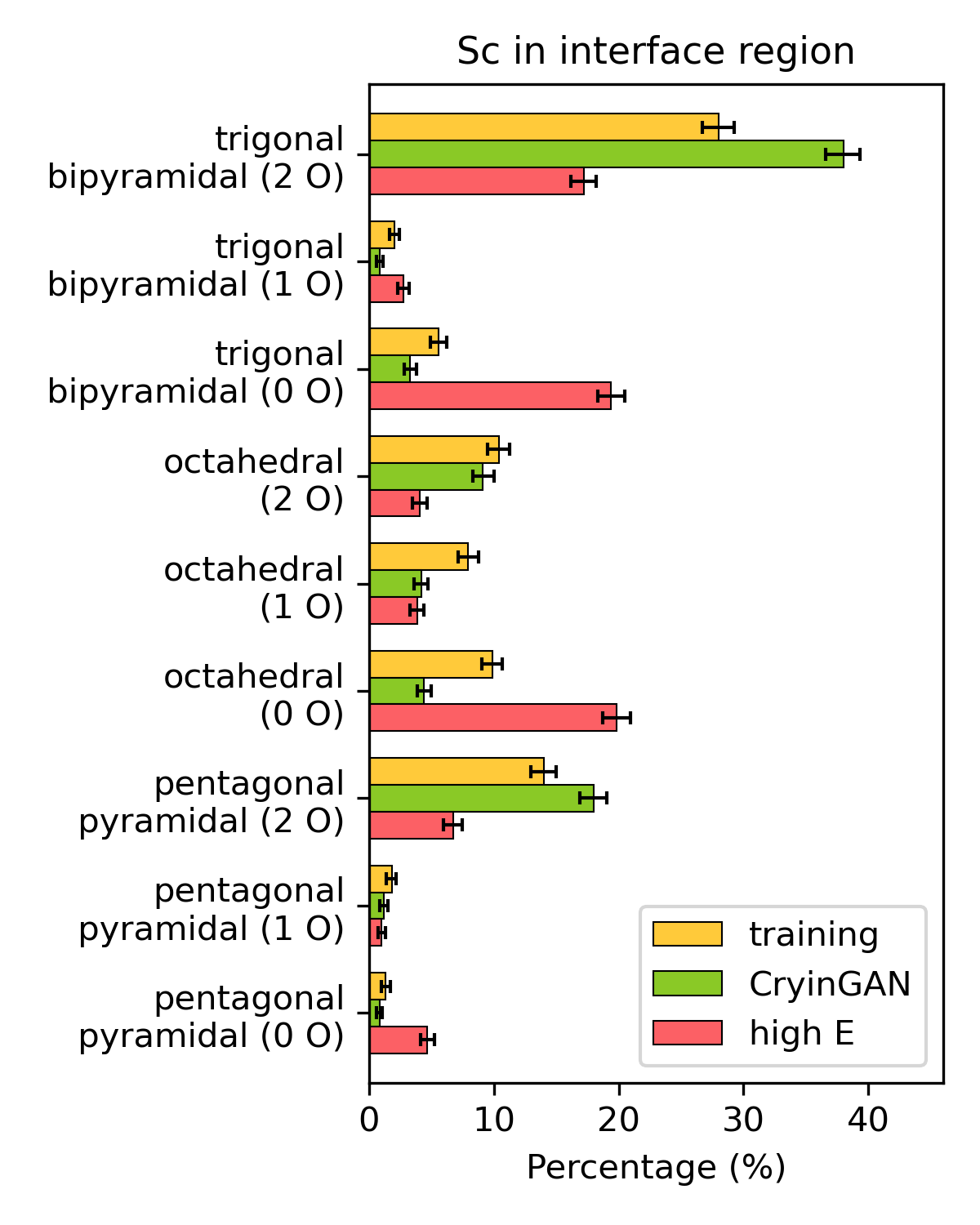}
\caption{Percentages of selected coordination motifs for Sc in the interface region. The motifs are subdivided based on the number of O atoms in the motif as indicated in brackets in the $y$-labels. The percentages of coordination motifs not shown are similar across the three datasets. Error bars represent 95 \% bootstrap confidence intervals.}
\label{fig_Sc_motifs}
\end{figure}

Fig. \ref{fig_Sc_motifs} shows the distribution of selected Sc coordination motifs, focusing on the motifs that show the largest differences across the datasets. 
Each motif is further subdivided based on the number of O atoms in the coordination shell of Sc (the distribution of all coordination motifs is provided in ESI Fig. S18b for reference). 
Compared to the training dataset, the high energy dataset shows significantly higher percentages of motifs with no O atoms present, and lower percentages of motifs with O atoms. 
This finding indicates that the lower frequency of bonding between O (in LCO) and Sc leads to weaker interface binding and higher interface energy. 
In contrast, the CryinGAN dataset shows a higher frequency of Sc-O bonding than the training dataset, with higher percentages of motifs with 2 O atoms (trigonal bypyramidal and pentagonal pyramidal), and lower percentages of motifs with no O atoms (trigonal bypyramidal and octahedral). 
The average O bond count per Sc atom provided in ESI Fig. S19b also shows that the high energy structures have fewer Sc-O bonds, whereas the CryinGAN structures have higher number of Sc-O bonds.
This analysis shows that although the Sc coordination environments of the CryinGAN structures differ from the training structures due to a higher frequency of Sc-O bonding, the Sc-O bonds are still a feature of the low-interface-energy structures.
We also analyzed the RDFs of the interface Li and Sc atoms. 
The RDFs confirm the trends revealed by the coordination motif analysis, in which the CryinGAN dataset shows higher similarity to the training dataset than the high energy dataset (see ESI Supplementary Note 4 for more details). 

\subsection{Overall generative model comparisons and insights} \label{sec_comparisons}
The Dismai-Bench metrics of CryinGAN across all datasets are listed in Section \ref{sec_benchmark}. 
For the disordered interfaces, CryinGAN outperforms both U-Net diffusion models, despite that all three models use coordinate-based representations, and diffusion models are often reported to outperform GANs in image synthesis~\cite{dhariwal2021diffvsgan, RN371, RN353}. 
CryinGAN performs competitively with the graph diffusion models for the disordered interfaces, which is unexpected considering the lack of invariances and graph convolutions in CryinGAN. Nonetheless, the limited expressive power of point clouds does introduce challenges when CryinGAN is tasked with 
generating amorphous Si structures. 
However, CryinGAN is still able to generate the disordered alloy structures and reproduce the SRO distributions, while the other coordinate-based diffusion models struggled with this task. 
Apart from the 300 K dataset with narrow SRO, CryinGAN demonstrated similar metrics to the graph diffusion models (refer to Table \ref{table_alloy_metrics}). 
The results of this simple point-cloud-based GAN highlights the benefits and importance of robust generative model evaluation. 

\begin{figure*}[h!]
\centering
\includegraphics[width=\textwidth]{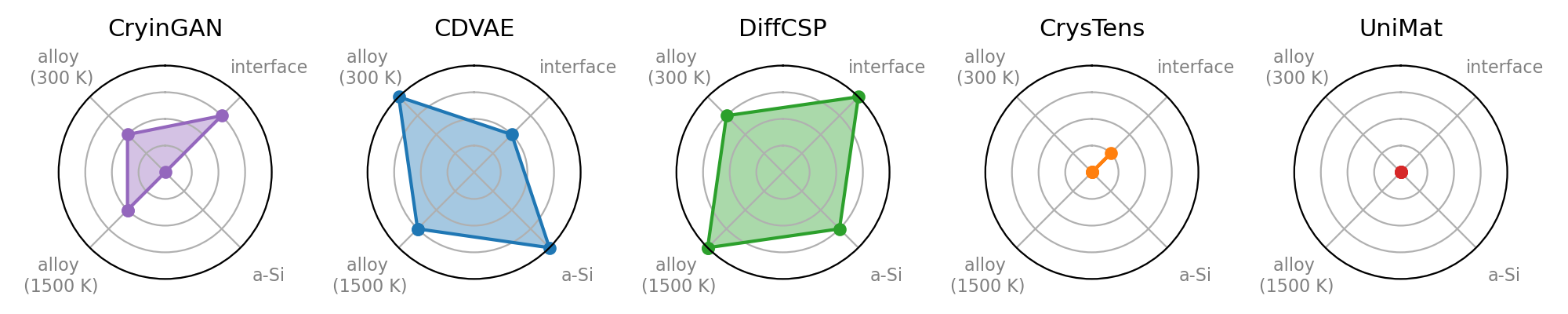}
\caption{Spider chart of generative model ranking based the models' general performance on each dataset. The models are ranked 1-5, where the outermost ring corresponds to rank 1 (best), and the center of the chart corresponds to rank 5 (worst).}
\label{fig_spider_chart}
\end{figure*}

An overall ranking of all generative models based on their general performance on each Dismai-Bench dataset is visualized using a spider chart as shown in Fig. \ref{fig_spider_chart}. 
The U-Net diffusion models struggled with most tasks in Dismai-Bench, while performing the best on the disordered interface dataset. 
Between the two, CrysTens performed marginally better on the disordered interfaces, probably as a result of the inclusion of distance matrices (albeit at a memory cost that scales with O($N^2$), $N$ being the number of atoms).
Also, both UniMat and CrysTens require data augmentation to compensate for their lack of all invariances, further increasing their memory requirements by orders of magnitude. 
We expect that the reason for the better performance on the disordered interface dataset is that the interface structures have sorted atom orderings that make learning easier for these models with image-/video-like representation. 
For example, the first 72 atoms are always Li, with the first 9 Li atoms always in the interface region, the next 9 Li atoms in the top layer of LCO, etc. 
In contrast, the other datasets do not have such well-sorted atom orderings. 
When CrysTens and UniMat were trained on disordered interface structures with randomized atom orderings, they were unable to generate the interface structures (see ESI Fig. S20). 
On the other hand, CryinGAN is permutationally invariant, so it does not rely on the sorted atom orderings to perform well on the interface dataset, and was able to generate the crystalline alloy structures.

The graph diffusion models demonstrated the best performance overall. Comparing the two models, CDVAE performed better on the amorphous Si and the 300 K alloy datasets, whereas DiffCSP performed better on the 1500 K alloy and interface datasets. 
CDVAE stands out as the only model that is able to generate structures that are nearly fully relaxed. 
In use cases where relaxation/post-processing is not feasible/practical (e.g., high-throughput materials discovery), CDVAE may show some advantages here. 
Although the graph models were successful in performing diffusion on atomic coordinates alone, they suffered when performing joint diffusion with atomic species or lattice parameters. 
When using atomic species diffusion, CDVAE was not able to generate the compositions correctly even though all Dismai-Bench datasets have fixed compositions. 
When using lattice diffusion, DiffCSP generated structures of lower quality. 
Such issues may be less severe when training on datasets with smaller number of atoms or lattice lengths, but future models can still improve upon their joint diffusion performance. 
To this end, Dismai-Bench provides a means for testing and evaluating different features/architectures (e.g., training DiffCSP with teacher forcing of the lattice parameters; see ESI Fig. S6). 

Despite the superior performance of graph models on Dismai-Bench, there are also reports that less expressive generative models (e.g., UniMat~\cite{yang2023unimat}, language models~\cite{flamshepherd2023language, antunes2024crystallm, gruver2024llama}) can show comparable or even better performance in discovering novel materials than graph models.  
These seemingly conflicting observations suggest a need to reconsider material discovery strategies. 
Materials discovery can be formulated as an exploration-exploitation problem. 
Graph models are superior at learning structural patterns, and should likely learn and exploit more of the training distribution during generation. 
On the other hand, less expressive models learn more noisy distributions, and can readily explore outside of the training distribution during generation. 
Considering the relatively small size of training datasets ($\sim 10^{4}$ for the MP-20 dataset~\cite{xie2022crystal}) vs. the material space of possible compounds ($\sim 10^{12}$ possible combinations of quaternary compounds~\cite{RN359}), it is arguable that exploration is more important that exploitation in materials discovery. 
However, good performance for materials discovery should not be confused with the ability to learn from training structures. 
While less expressive models may do well for the discovery of small structures, they may struggle when the size/complexity of the structures increases, as suggested here. 

As the number of atoms and/or structural complexity of training samples increases, the importance of symmetry invariances and the expressive capability of generative models also grows. 
Graphs serve as one type of invariant representation among alternatives like smooth overlap of atomic positions (SOAP) vectors~\cite{RN340} and atom-centered symmetry functions (ACSFs)~\cite{RN341}. 
Exploring these invariant representations further is crucial, even if reconstruction back to atomic coordinates is necessary. 
Importantly, reconstruction does not necessarily pose a fundamental barrier, as demonstrated by Fung et al.~\cite{RN342}, who achieved it through gradient-based optimization using automatic differentiation. 
The Dismai-Bench findings underscore the need for higher expressive power to model more complex structures, despite potential increases in computational and memory requirements. 
As generative modeling advances towards larger datasets and more complex structures, the development of models capable of parallelization will be a critical research direction.

In this work, we introduced Dismai-Bench as a novel method for assessing generative models and obtaining valuable insights into their performance. 
The development of generative models, and machine learning models in general, is iterative and often counterintuitive, as evidenced by the development of CryinGAN. 
Surprising findings include instances where an older generation GAN architecture has outperformed newer diffusion architectures, and that using two separate discriminators has yielded better results than using a single discriminator. 
Dismai-Bench proves effective in evaluating a generative model's capability to learn intricate structural patterns, particularly those present in disordered materials. 
Consequently, Dismai-Bench may contribute to improving generative models not only for ordered materials but also for disordered ones.
However, Dismai-Bench's scope is limited in that it primarily assesses the accuracy of atomic positions, and does not evaluate a model's proficiency in learning compositions or lattice parameters, because Dismai-Bench operates by fixing the composition and lattice while varying atomic positions. Moreover, generative models incorporating symmetry constraints~\cite{RN338, jiao2024diffcsp++} may not leverage Dismai-Bench for evaluation unless these constraints can be disabled.
Dismai-Bench marks a significant initial stride towards robustly evaluating generative models, and we anticipate that new complementary benchmarks and datasets will emerge in the future.

\section{Conclusions}
We developed a new benchmark for generative models, Dismai-Bench, which evaluates models on datasets of large, disordered materials exhibiting different degrees of structural and configurational disorder. 
Instead of training across different compositions and space groups, models are trained on datasets with fixed composition and structure type.
By fixing the material system, Dismai-Bench circumvents the challenges of evaluating models based on newly generated materials that cannot be verified. 
Graph diffusion models (CDVAE \& DiffCSP) were found to outperform coordinate-based diffusion models (CrysTens \& UniMat) on Dismai-Bench, due to the invariant nature and higher expressive power of graphs. 
Additionally, we introduced CryinGAN, a novel GAN based on point clouds, that was developed by evaluating candidate architectures through direct comparisons between training and generated structures. 
Despite its simple architecture without symmetry invariances or complex components, CryinGAN outperformed the coordinate-based diffusion models and demonstrated competitiveness with the graph diffusion models.
Dismai-Bench provides meaningful evaluation for comparing between architectures, understanding model strengths and weaknesses, and ultimately informing design choices.
Building the next generation of generative models will rely on not only developing better architectures and representations, but also adopting better evaluation methods. 
We hope that this work will help advance future generative models for both ordered and disordered materials, and inspire the development of other new innovative benchmarks. 

\section{Methods}
\subsection{Dismai-Bench datasets}
Dismai-Bench uses a total of six datasets. Each dataset contains a total of 1,500 structures, split into 80 \% training and 20 \% validation data. 

\subsubsection{\texorpdfstring{Fe$_{60}$Ni$_{20}$Cr$_{20}$}{Fe60Cr20Ni20} austenitic stainless steel} 
A cluster expansion Monte Carlo approach~\cite{kim2022multisublattice, RN358} was used to generate the datasets of FCC Fe$_{60}$Ni$_{20}$Cr$_{20}$ austenitic stainless steels. 
The CE model operates as a generalized Ising model~\cite{Sanchez1984gen} to describe the formation energy as a function of configuration.
We adapted the CE model from ref. \citenum{RN358}. 
A CE model with seven chemical dimers and three spin dimers was used to fit a dataset of FCC Fe-Ni-Cr alloys that were generated from spin-polarized DFT calculations. 
The CE model was fit using the least absolute shrinkage and selection operator (LASSO) and 10-fold cross-validation (CV) to determine the effective cluster interactions (ECI) values of the clusters. 
A comparison between the CE-calculated and DFT-calculated formation energies is shown in ESI Fig. S21a, and the ECI values obtained by the LASSO CV fit are shown in ESI Fig. S21b. 

To generate the Dismai-Bench alloy structures, only the chemical terms of the CE model were used to obtain structures in a non-magnetic state. 
A 4$\times$4$\times$4 conventional FCC supercell, containing 256 atoms (154 Fe atoms, 51 Ni atoms, 51 Cr atoms), was used.
We performed MC simulations in the canonical ensemble based on the Metropolis-Hastings algorithm~\cite{metropolis1953eq}.
Kawasaki dynamics~\cite{kawasaki1966diffusion} for atom swaps was applied to ensure that the composition of the system remained fixed.
The alloy structures were sampled at temperatures of 300 K and 1,500 K, where 10 independent MC simulations were initialized for each temperature. 
For every MC simulation, 50,000 passes were performed, and a MC snapshot was saved every 10 passes. 
For each temperature, 1,500 structures were assembled (without any restriction on SRO distribution) as the wide SRO dataset. 
Another 1,500 structures with SRO distribution within $\pm$ 0.1 of the average SRO values were filtered as the narrow SRO dataset.
Comparisons of the SRO between the Fe$_{60}$Ni$_{20}$Cr$_{20}$ alloy modelled in Dismai-Bench and the original Fe$_{56}$Cr$_{21}$Ni$_{23}$ alloy modelled in ref. \citenum{RN358} is shown in ESI Fig. S21c. 
The Dismai-Bench alloy shows qualitatively similar SRO trends to the original alloy that was modelled with magnetism and a larger set of clusters.

\subsubsection{Amorphous silicon}
We adapted the amorphous silicon dataset from ref. \citenum{RN355}. 
The original data consists of a 100,000-atom amorphous silicon structure generated through melt-quench molecular dynamics simulation~\cite{RN355}. 
We sliced the structure into smaller blocks with lattice parameters corresponding to 256-atom amorphous silicon structures. 
The lattice lengths of the blocks were calculated by linearly scaling the lattice lengths of the 100,000-atom structure to 256 atoms. 
The blocks were sliced at different locations to obtain a total of 1,500 blocks. 
Blocks with < 256 atoms had atoms added at random to low density regions, and blocks with > 256 atoms had atoms removed at random from high density regions, until all blocks had 256 atoms. 
Atoms were added/removed to the boundary of the blocks only (where they were sliced).
The density was calculated by dividing each face of a block into 2$\times$2 regions and counting the number of atoms in each region.
The 1,500 blocks were relaxed using a pre-trained SOAP-GAP~\cite{RN356} machine learning interatomic potential for Si. 
The structures were optimized using a conjugate gradient algorithm~\cite{RN323} through the atomic simulation environment (ASE) package~\cite{RN324}. 
Only the atomic positions were allowed to relax, and the relaxations were stopped when the force on each atom was below 0.05 eV/Å.

\subsubsection{\texorpdfstring{Li$_3$ScCl$_6$(100)-LiCoO$_{2}$(110)}{Li3ScCl6(100)-LiCoO2(110)} battery interface} \label{sec_int_dataset}
To construct the interface structures, we chose the orientations of LCO(110), a Li fast-diffusing plane~\cite{RN312}, and LSC(100), a representative plane. The surfaces of the LSC(100) slab are polar, so half of the Cl atoms were moved from one surface to the other to neutralize the polarity (resulting in ‘Tasker Type 2b’ surfaces~\cite{RN132, RN131}). Lattice matching between the two slabs was carried out using the MPInterfaces package~\cite{RN128}, which implements the lattice matching algorithm proposed by Zur et al.~\cite{RN129}. The configuration of the lattice-matched interface is given in ESI Table S4, where the average lattice mismatch is 2.17 \%. 
 
In a preliminary test calculation, the interface was constructed by simply placing the LCO(110) and LSC(100) slabs in contact with each other, and we obtained a DFT-relaxed structure similar to that depicted in Fig. \ref{fig_datasets}. An interface region with disordered LSC atoms formed with a thickness of around 5 Å. Using this structure as reference, we then generated random interface structures using the CALYPSO package~\cite{RN315, RN316, RN319}. We used LCO(110) and LSC(100) slabs with 4 and 7 layers respectively. For each structure, an interface region thickness was randomly chosen between 4 and 6 Å, then the region was randomly populated with 3 formula units of LSC (9 Li, 3 Sc, and 18 Cl atoms). A random lateral displacement (in-plane direction parallel to the interface) was also applied to the LSC slab. A vacuum spacing of 14 Å was included in all the interface structures. Each interface structure has a total of 264 atoms. 

We relaxed the randomly generated interface structures with the M3GNet interatomic potential. Only the atom positions were allowed to relax, where the lateral lattice vectors were fixed to the optimized values of the LCO slab (the elastic moduli of LCO~\cite{RN320} is significantly larger than LSC~\cite{RN321}). Similar to the amorphous Si dataset, the structures were optimized using a conjugate gradient algorithm through the ASE package, and the relaxations were stopped when the force on each atom was below 0.05 eV/Å. The normalized interface energies, $\tilde{\gamma}_{\textrm{int}}$, of the relaxed structures were calculated using the following equation:
\begin{equation}
\tilde{\gamma}_{\textrm{int}} = \frac{E_{\textrm{int}} - N(-4.78\ \textrm{eV/atom})}{A} \times (1.60218 \times 10^{-19} \textrm{J/eV})
\end{equation}
where $E_{\textrm{int}}$ is the total energy of the interface in eV, $N=264$ is the total number of atoms, and $A$ is the interface area. The interface energies are normalized such that structures with interface energy $\leq$ 0 J/m$^2$ are considered to be low-interface-energy structures. 1,500 low-interface-energy structures were filtered from the relaxed structures to form the dataset.

\subsection{Generative models}
\subsubsection{CDVAE}
We modified CDVAE such that atomic species denoising becomes an optional feature, since atomic species denoising caused CDVAE to generate incorrect compositions when trained on Dismai-Bench datasets. All CDVAE models were trained without atomic species denoising. The same hyperparameters, optimizer, and learning rate scheduler as those used for the MP-20 dataset~\cite{xie2022crystal} were applied. All models were trained using a batch size of 8 for 1,000 epochs. Disordered interface and alloy structures were generated by running Langevin dynamics using 100 steps per noise level. For the amorphous Si structures, we were able to match the energy distribution of generated structures to the training structures (refer to example in ESI Fig. S22), so we used 6-7 steps per noise level based on whichever setting gave the best match.

\subsubsection{DiffCSP}
We used 4 layers and 256 hidden states for all DiffCSP models. The exponential noise scheduler parameter, $\sigma_{T}$~\cite{jiao2024diffcsp}, was set to 0.05 for amorphous Si, and 0.1 for disordered interfaces and alloys. The weight of the lattice cost was set to 0 for all models (i.e., no lattice diffusion). All other hyperparameters were set to the default values, and we used the same optimizer and learning rate scheduler as the original implementation~\cite{jiao2024diffcsp}. 
All models were trained using a batch size of 4 for 1,000 epochs. All structures were generated using step size $\gamma = 1 \times 10^{-5}$ and 1000 time steps. Although we added the option to train DiffCSP with teacher forcing of the lattice parameters, this feature was not used for the DiffCSP models benchmarked in Section \ref{sec_benchmark}, which did not use any lattice diffusion.

\subsubsection{CrysTens}
CrysTens uses the Imagen~\cite{saharia2022imagen} model, a cascaded diffusion model consisting of a base U-Net that generates a lower resolution image, followed by a super-resolution U-Net that upsamples the lower resolution image to a higher resolution. We used 64 base channels for both U-Nets, where the first U-Net generates 64$\times$64 images and the second U-Net upsamples them to the full-size CrysTens images. The optimizer and all other hyperparameters were set to be the same as the original CrysTens implementation~\cite{RN353}. Each U-Net was trained using a batch size of 8 for 150,000 training steps. Structures were generated using 100 time steps, where the energy distribution of the generated structures had already converged at this setting (refer to ESI Fig. S23). The CrysTens images were reconstructed into atomic structures using the ground truth lattice parameters. Althought the generated lattice lengths and angles from the CrysTens images were not used, their MAEs were small, around 0.02 \r{A} and 0.02$^{\circ}$ respectively. The composition accuracy of the generated structures was 100~\%. The atomic coordinates of each atom was determined by constructing directional graphs using the coordinate pixels and the pairwise $\Delta x$, $\Delta y$, and $\Delta z$ pixels, as described in ref. \citenum{RN353}. Then, the coordinate predictions from the directional graphs were averaged. However, unlike the original implementation, we did not perform $k$-means clustering on the averaged atomic coordinates (and atomic species), since this was a post-processing step intended to manually "denoise" the reconstructed structures using an arbitrary choice of $k$.

For CrysTens, we augmented the disordered interface dataset by applying random permutations (shuffling) to the atoms in the interface region. CrysTens was found to perform poorly when data augmentation was performed by permutating all atoms (see ESI Fig. S20). Each structure in the interface dataset was constructed with the same LSC and LCO slabs, as well as three formula units of LSC randomly generated in the interface region. As a result, the ordering of the atoms in the slabs is consistent across all structures, making it easier for CrysTens to learn these structures. Therefore, we only shuffled the ordering of the atoms in the interface region, where each structure was augmented 49 times. For amorphous Si and the alloys, which have no consistent atom orderings in their datasets, CrysTens was unable to generate meaningful structures even with data augmentation (see Fig. \ref{fig_a-Si_examples} and Fig. \ref{fig_alloy_examples}). 

\subsubsection{UniMat}
The original UniMat~\cite{yang2023unimat} code is unfortunately not openly available, so we used an open-access implementation~\cite{wang2024imagen} of the 3D U-Net model~\cite{ho2022videodiff} that the UniMat model was repurposed from. We used the hyperparameters as listed in ESI Table S5 for the U-Net. 4$\times$4 video frames were used for the UniMat representation, which is the smallest frame size compatible with U-Net model configuration. The Adam optimizer~\cite{kingma2017adam} was used with learning rate = $1 \times 10^{-4}$, $\beta_{1} = 0.9$, and $\beta_{2} = 0.99$. The U-Net was trained using a batch size of 8 for 150,000 training steps. Structures were generated using 100 time steps, where the energy distribution of the generated structures had already converged at this setting (refer to ESI Fig. S24). We did not include the lattice parameters in the UniMat representation, and simply used the ground truth lattice parameters to reconstruct the structures (i.e., no lattice diffusion). The average composition accuracy of the generated structures was 99.5 \%. Structures with incorrect compositions were considered failed structures and rejected. Similar to CrysTens, we augmented the disordered interface dataset by shuffling the atoms in the interface region, augmenting 49 times per structure. For amorphous Si and the alloys, UniMat was unable to generate meaningful structures even with data augmentation (see Fig. \ref{fig_a-Si_examples} and Fig. \ref{fig_alloy_examples}). 

\subsubsection{CryinGAN}
The CryinGAN architecture is as shown in Fig. \ref{fig_GAN_arc}. CryinGAN was developed using a different interface dataset from the Dismai-Bench interface datatset, consisting of 1,500 low-interface-energy structures relaxed using M3GNet followed by DFT calculations. This is to facilitate evaluation of the best performing model through comparisons between DFT-relaxed training and generated structures. Benchmarking of CryinGAN was still carried out by training models on the Dismai-Bench datasets. During model development, CryinGAN models were trained with different $\lambda$ values to study the effect of $\lambda$ on model performance. The models were trained with a batch size of 32 for 100,000 epochs. Adam optimizers with learning rate of $5 \times 10^{-5}$ were used for the generator and discriminator(s). The generator was trained only once every 5 batches to help stabilize the training. CryinGAN-comb, CryinGAN-max, and CryinGAN-mix models were trained using the same procedure. CryinGAN-comb uses only a single combined discriminator, so no $\lambda$ testing was required. CryinGAN-max and CryinGAN-mix were trained using $\lambda$ = 0 and 0.05. For $\lambda$ = 0, max pooling and mix pooling were used for the fractional coordinate discriminator of CryinGAN-max and CryinGAN-mix respectively. For $\lambda$ = 0.05, max pooling and mix pooling were used for the bond distance discriminator of CryinGAN-max and CryinGAN-mix respectively, whereas average pooling was used for the fractional coordinate discriminator of both models. The various GAN models were compared using 1,000 structures generated from each model.

We also tested a graph convolutional neural network architecture by adapting CGCNN~\cite{RN215} as the discriminator. Instead of predicting material properties, the output of CGCNN was used to estimate the Wasserstein distance between the training and generated structures. Default values for the hyperparameters were used, as provided in the code repository of CGCNN. The GAN was optimized using the Adam optimizer and a batch size of 32, where we tested learning rates of $10^{-2}$, $10^{-3}$, $10^{-4}$, and $10^{-5}$. We found that the losses diverged and the GAN was unable to generate meaningful structures. We did not further pursue the development of graph convolutional neural networks as discriminators, considering similar failures reported in literature for point cloud GANs~\cite{wang2020sampling}.

The best GAN architecture was found to be CryinGAN using both discriminators. For the disordered interface, $\lambda = 0.05$ was found to be optimum (see ESI Fig. S15). For the alloys, $\lambda = 0.1$ was found to be optimum (see ESI Fig. S25). For amorphous Si, CryinGAN was unable to generate meaningful structures (see Fig. \ref{fig_a-Si_examples}), so it was not benchmarked for amorphous Si. Benchmark models were trained on the Dismai-Bench disordered interface and alloy datasets using $\lambda$ = 0.05 and 0.1 respectively. For the disordered interface dataset, we did not perform any data augmentation. CryinGAN is permutationally invariant, and the model does not need to learn to generate rotated interface structures. Each structure in the dataset was originally generated with the LCO slab fixed and the LSC slab randomly displaced laterally (parallel to the interface). Training models on a dataset with lateral translation augmentations was found to slow down CryinGAN's learning, since it had to learn to generate structures with displaced LCO (and LSC) slabs. Therefore, we did not apply any translation augmentation as well. For the alloy datasets, we performed data augmentation by translating the structures by integer multiples of the unit cell lattice constant (3.6 \r{A}). Each structure is a 4$\times$4$\times$4 supercell, giving $4^{3} - 1 = 63$ unique translations, so we augmented each structure 63 times. The disordered interface and alloy benchmark models were trained for 100,000 and 1,500 epochs respectively, and models from the last epoch were used for benchmarking. All models were trained with a batch size of 32.

\subsection{Dismai-Bench benchmarking}
Benchmarking was performed by training all generative models on each dataset separately from scratch, such that the models only generate one type of structure at a time. 
For each generative model architecture, three separate models were trained for each dataset, and the benchmark metrics were averaged.
1,000 structures were generated for each model to calculate the benchmark metrics.

\subsubsection{Disordered LSC-LCO interface}
The generated interface structures were first post-processed by moving apart atoms that were too close together using an iterative algorithm. In each iteration, the algorithm determines all unique pairs of atoms too close together, and increases the magnitude of their bond vectors. The algorithm repeats itself until all atomic distances are $>$ 1.5 Å. We set a maximum of 100 iterations, and any structure that still had atoms too close together after 100 iterations was rejected. The structures were then relaxed using the M3GNet interatomic potential, allowing only the atom positions to relax. The relaxations were stopped when the force on each atom was below 0.05 eV/Å. The percentage of failed structures was calculated.

For each successfully relaxed structure, the CrystalNNFingerprint~\cite{RN263} was calculated for the cations (Li, Co, and Sc), allowing only the anions (Cl and O) to be considered as neighbors. We also appended the fraction of Cl and O neighbors in each motif to the fingerprints, so that the fingerprints contain both chemical and coordination information. Average fingerprints were calculated by averaging the site fingerprints of each structure, then averaging across all structures. The Euclidean distances between the average fingerprint of the generated structures and the training structures were calculated.

\subsubsection{Amorphous Si}
The generated amorphous Si structures were post-processed to move apart atoms too close together, using the same procedure described for the disordered interfaces. The structures were then relaxed using the SOAP-GAP interatomic potential, allowing only the atom positions to relax. The relaxations were stopped when the force on each atom was below 0.05 eV/Å. The percentage of failed structures was calculated.

For each successfully relaxed structure, the CrystalNNFingerprint~\cite{RN263} was calculated for all Si atoms. The Euclidean distances between the average fingerprint of the generated structures and the training structures were calculated. The RDF of each structure was calculated using the vasppy~\cite{morgan2021vasppy} package. The RDFs were calculated between 0.0 \r{A} and 10.0 \r{A} using a bin width of 0.02~\r{A}. The RDFs were averaged across structures, and the Euclidean distances between the average RDF of the generated structures and the training structures were calculated. The bond angles of each structure was also calculated. The neighbors of each atom were determined using the CrystalNN~\cite{RN263} algorithm. The bond angle distribution was calculated by binning the bond angles using a bin width of 0.5$^{\circ}$. The bond angle distributions were averaged across structures, and the Euclidean distances between the average bond angle distribution of the generated structures and the training structures were calculated.

\subsubsection{Disordered stainless steel alloy}
The generated alloy structures were post-processed to remove atoms not on lattice sites. An atom was considered to be on a lattice site if it was within a 0.8 \r{A} radius of the lattice site. If a lattice site already had an atom assigned to it, no other atom can be assigned to the site. Any structure with > 50 atoms removed ($\sim$20 \% of all atoms) was considered a failed structure and rejected. The percentages of failed structures and accepted structures with site vacancies were calculated.

The clusters (1 monomer and 7 dimers) of each structure were counted, where we considered up to the 7$^{\textrm{th}}$ neighbor shell. A fingerprint was constructed for each structure using the vector of conditional probabilities of observing each cluster (e.g., the probability of observing a 1st nearest neighbor Cr-Fe dimer given two nearest neighbor sites). The Euclidean distances between the average cluster fingerprint of the generated structures and the training structures were calculated.

\subsection{Detailed evaluation of CryinGAN-generated interfaces}
We further evaluated the disordered interface structures generated by CryinGAN. We trained a CryinGAN model on the dataset of DFT-relaxed interface structures with $\lambda = 0.05$. The generator was trained once every 2 batches to speed up the training. We found the training to be stable at this frequency as shown by the Wasserstein distance plot in ESI Fig. S26a. Although the Wasserstein distance plateaued early in the training, the percentage of (relaxed) low-interface-energy structures continued to gradually increase (see ESI Fig. S26b and c). We stopped the training when the energy improvements have mostly plateaued. Structures were generated using the model from the last epoch, and relaxed using M3GNet followed by DFT calculations. Low-interface-energy structures were filtered from the relaxed structures.

Three datasets were used to perform the structural analysis: (1) CryinGAN-generated structures with low interface energy ($\leq 0$ J/m$^2$), (2) randomly generated structures with low interface energy (i.e., the training structures), and (3) randomly generated structures with high interface energy ($> 0$ J/m$^2$). Each dataset contains 1,500 structures that were relaxed using M3GNet followed by DFT calculations. The coordination environment of atoms in the interface region was compared between the datasets. The coordination motif fingerprints of Li and Sc atoms in the interface region were calculated for each dataset and averaged. The Euclidean distance and cosine similarity between the average fingerprint of the CryinGAN/high energy dataset and the training dataset were calculated. To visualize the coordination motif distribution of the datasets, the most likely coordination motif of each interface Li/Sc atom was identified by first selecting the coordination number with the highest likelihood, then selecting the coordination motif of this coordination number with the highest local structure order parameter (LoStOP)~\cite{RN263}. We only plotted coordination motifs that appeared in $>$ 1 \% of the interface Li/Sc atoms in the training structures. 95 \% bootstrap pivotal confidence intervals~\cite{RN336} were calculated using 1,000 bootstrap samples with 1,500 structures in each sample.

\subsection{DFT calculations}
All DFT calculations were performed using the Vienna Ab initio Simulation Package (VASP)~\cite{RN100, RN101, RN102, RN103}, with the projector augmented-wave (PAW) method~\cite{RN95, RN104}. The Li ($1s^2\ 2s^1$), Sc ($3s^2\ 3p^6\ 3d^2\ 4s^1$), Co ($3d^8\ 4s^1$), Cl ($3s^2\ 3p^5$), and O ($2s^2\ 2p^4$) electrons were treated as valence electrons in the pseudopotentials. The generalized gradient approximation (GGA) with the Perdew-Burke-Ernzerhof (PBE) exchange-correlation functional\cite{RN105} was used. The orbitals were expanded using a plane wave basis with cutoff energy of 520 eV for bulk structures, and 450 eV for interface structures (to lower computational cost). The DFT+$U$ approach~\cite{RN125} was used to account for the electron localization of the Co-$3d$ states, and we selected a $U$ value of 4 eV as reported in other literature~\cite{RN126}.  

We first performed structural relaxations on unit cells of LCO ($R\overline{3}m$) and LSC ($C2/m$~\cite{RN313}), allowing the cell shapes, cell volumes, and atom positions to relax, until the force on each atom was below 0.001 eV/Å. The Brillouin zone was sampled using a (9×9×1) gamma-centered k-point grid for LCO, and a (6×4×6) Monkhorst-Pack k-point grid for LSC. The relaxed unit cells were used to construct the LCO(110) and LSC(100) slabs as described in Section \ref{sec_int_dataset}.

To generate the dataset for training the M3GNet interatomic potential, LSC(100)-LCO(110) interfaces were randomly generated as described in Section \ref{sec_int_dataset}. A total of 350 structures were generated without mutual exchanges, and 600 structures were generated with mutual exchanges. Random mutual exchanges were performed between atoms in the interface region and the top layer of LCO, where we allowed up to 3 Sc $\leftrightarrow$ Co, 3 Li $\leftrightarrow$ Co, and 6 Cl $\leftrightarrow$ O exchanges per structure. Structural relaxations were performed on all interface structures with the cell shapes and cell volumes fixed. The relaxations were performed in two stages. In the first stage, a kinetic energy cutoff of 374.3 eV was used, and only the atoms in the interface region were allowed to relax for 50 ionic steps. In the second stage, a kinetic energy cutoff of 450 eV was used, and all atoms were allowed to relax until the force on each atom was below 0.1 eV/Å. The Brillouin zone was sampled at the gamma point only for both stages. 

For structures that have been pre-relaxed using M3GNet (after the M3GNet model was trained), DFT relaxations were performed without the first stage relaxation (second stage only). The interface dataset for Dismai-Bench was created by relaxing randomly generated structures with M3GNet only. The interface dataset for developing CryinGAN was created by relaxing randomly generated structures with M3GNet followed by DFT calculations. No mutual ion exchanges were performed when the structures were generated since the exchanges mostly led to high-interface-energy structures. The initial 950 structures that were relaxed using DFT calculations only were just used for training the M3GNet model, but not any generative model.

\subsection{M3GNet}
We trained the M3GNet interatomic potential~\cite{RN322} on the LSC-LCO interface structures to allow us to perform relaxations quickly. The dataset used for M3GNet training and evaluation included the 950 relaxed interface structures and 14,534 intermediate ionic steps from the DFT relaxations (second stage only). The intermediate steps were sampled starting from the 5th ionic step of every relaxation with an interval of 10 steps. The dataset was split into 80 \% training, 
10 \% validation, and 10 \% test data. We used the Adam optimizer to optimize the loss function, $L$, as follows:
\begin{equation}
L = \textrm{MSE}_{\textrm{E}} + \textrm{MSE}_{\textrm{F}} + 0.1(\textrm{MSE}_{\textrm{S}})
\end{equation}
where MSE$_\textrm{E}$, MSE$_\textrm{F}$, and MSE$_\textrm{S}$ are the mean squared error of energy, force and stress respectively. We tested different learning rates and batch sizes, and trained M3GNet models for 24 hours each. We found that the losses plateaued within the given training time, and there was little difference in errors between the different hyperparameters (see ESI Table S6). We chose the model with the smallest loss (learning rate = 0.001, batch size = 4) as the working interatomic potential of this work (see ESI Fig. S27 for the loss curves). The test set mean absolute errors for energy, force, and stress are 2.70 meV/atom, 20.9 meV/Å, and 0.0146 GPa respectively.

We compared interface structures that were pre-relaxed with M3GNet with their final structures after subsequent DFT relaxation. Table \ref{table_m3gnet_error} shows the errors between the M3GNet- and DFT-calculated total energies. The errors are generally low even when compared to the final DFT-relaxed structures, showing that the M3GNet relaxations are able to give accurate predictions of the energy, and yield structures close to DFT convergence.

\begin{table}[h!]
\small
\caption{Performance of M3GNet relaxations in reaching DFT convergence. Interface structures are first optimized using M3GNet relaxations followed by DFT relaxations. $n_{\textrm{steps}}$ is the number of DFT ionic steps required to fully relax the M3GNet-optimized structures. $\lvert \Delta E_{\textrm{M-opt}}\rvert$ is the absolute energy difference between the M3GNet and DFT energies of the M3GNet-optimized structure. $ \lvert\Delta E_{\textrm{D-opt}}\rvert$ is the absolute energy difference between the M3GNet energy of the M3GNet-optimized structure and the DFT energy of the final DFT-optimized structure. The mean and standard deviation for each quantity are listed.}
\begin{tabularx}{0.48\textwidth}{lXXXXXX}
\toprule
    \multirow{3}{*}[-2.5pt]{Structure type} & 
    \multicolumn{2}{l}{\multirow{2}{*}{$n_{\textrm{steps}}$}} &
    \multicolumn{2}{l}{$\lvert \Delta E_{\textrm{M-opt}}\rvert$} &
    \multicolumn{2}{l}{$ \lvert\Delta E_{\textrm{D-opt}}\rvert$} \\
 & 
    \multicolumn{2}{l}{} &
    \multicolumn{2}{l}{(meV/atom)} &
    \multicolumn{2}{l}{(meV/atom)} \\
    \cmidrule(lr){2-7}
    & {mean} & {std} & {mean} & {std} & {mean} & {std} \\
\midrule
low energy & 17.6 & 25.7 & 3.14 & 3.41 & 2.15 & 1.42 \\
high energy & 56.1 & 51.8 & 11.0 & 15.3 & 4.75 & 4.51 \\
all & 31.7 & 41.8 & 6.03 & 10.4 & 3.10 & 3.21 \\
\bottomrule
\end{tabularx}
\label{table_m3gnet_error}
\end{table}

\section*{Data Availability}
The datasets and interatomic potentials used are available openly at \url{https://doi.org/10.5281/zenodo.12710372}.
The Dismai-Bench benchmarking and generative model code used in this work  are available at \url{https://github.com/ertekin-research-group/Dismai-Bench}.
The CryinGAN code is available separately at \url{https://github.com/ertekin-research-group/CryinGAN}.

\section*{Author Contributions}
A.X.B.Y. and E.E conceived the idea. A.X.B.Y. designed the benchmark, wrote the code, carried out the calculations/experiments, and performed the analysis. T.S. assisted with the alloy dataset preparation and code for cluster counting. E.E. supervised and guided the project. The manuscript was prepared by A.X.B.Y. and T.S. All authors reviewed and edited the manuscript.

\section*{Conflicts of interest}
There are no conflicts to declare.

\section*{Acknowledgements}
This work was supported by the Defense Advanced Research Projects Agency (DARPA) HR00112220028. 
We also acknowledge partial financial support from the U.S. Army CERL, awarded under Grant No. W9132T-19-2-0008. 
This work made use of the Illinois Campus Cluster,  operated by the Illinois Campus Cluster Program (ICCP) in conjunction with the National Center for Supercomputing Applications (NCSA) and supported by the University of Illinois at Urbana-Champaign. This work also utilized the Hardware-Accelerated Learning (HAL) cluster~\cite{kindratenko2020hal}, supported by the National Science Foundation’s Major Research Instrumentation program, grant \#1725729, as well as the University of Illinois at Urbana-Champaign. 
This work also used Bridges-2~\cite{brown2021bridges2} at Pittsburgh Supercomputing Center through allocation MAT220011 from the Advanced Cyberinfrastructure Coordination Ecosystem: Services \& Support (ACCESS) program, which is supported by National Science Foundation grants \#2138259, \#2138286, \#2138307, \#2137603, and \#2138296.



\balance


\bibliography{rsc.bib} 
\bibliographystyle{rsc.bst} 

\end{document}


\vspace*{50pt}
    
    \begin{center}
    {\Huge Supplementary Information}
    \end{center}
    
   \vspace{0pt}
    
    \maketitle

	
\clearpage
\baselineskip=12pt
\setlength{\parindent}{0pt}

\begin{figure}[h!]
\centering
\includegraphics[width=\textwidth]{"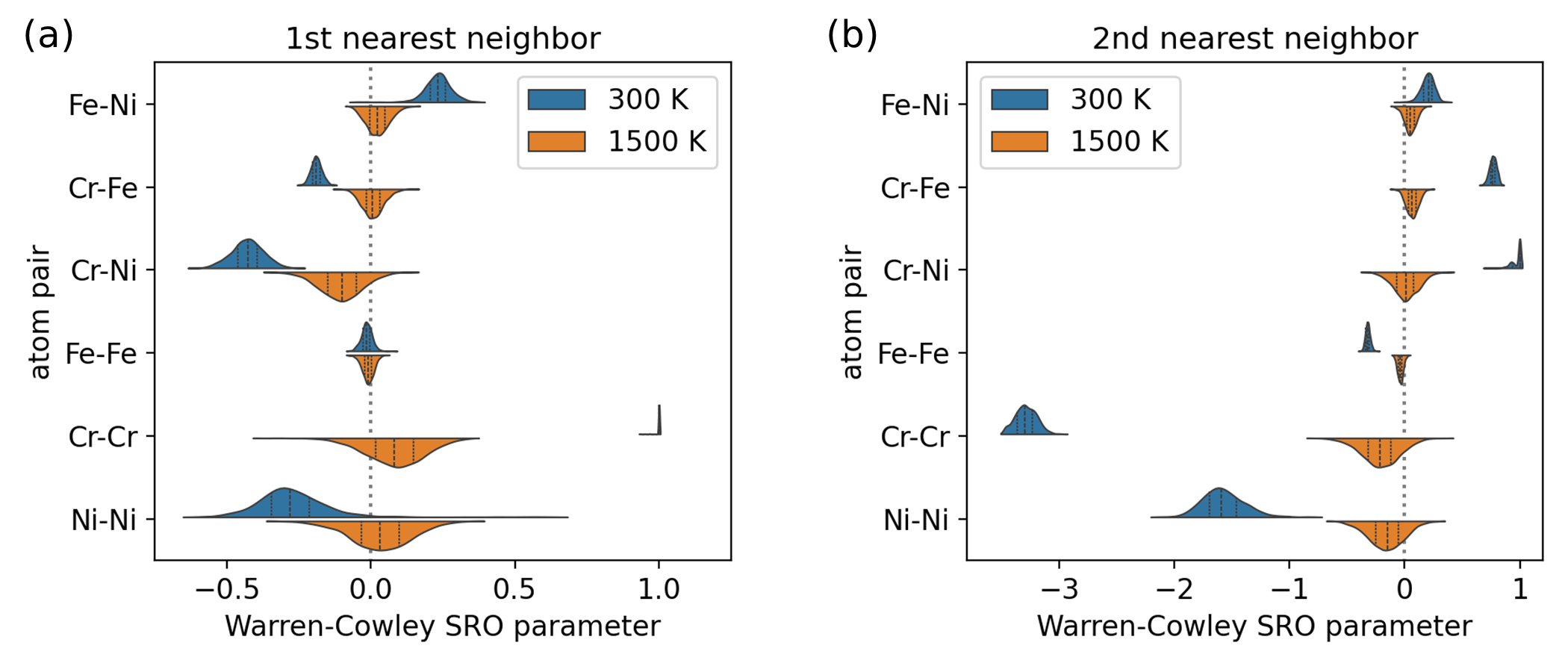"}
\caption{Warren-Cowley short-range order (SRO) parameter distributions of 300 K and 1500 K alloy training datasets (no SRO filtering). The SRO distributions are shown for the (a) 1st and (b) 2nd nearest neighbor interactions.}
\label{fig_sro_train_wide}
\end{figure}

\begin{figure}[h!]
\centering
\includegraphics[width=\textwidth]{"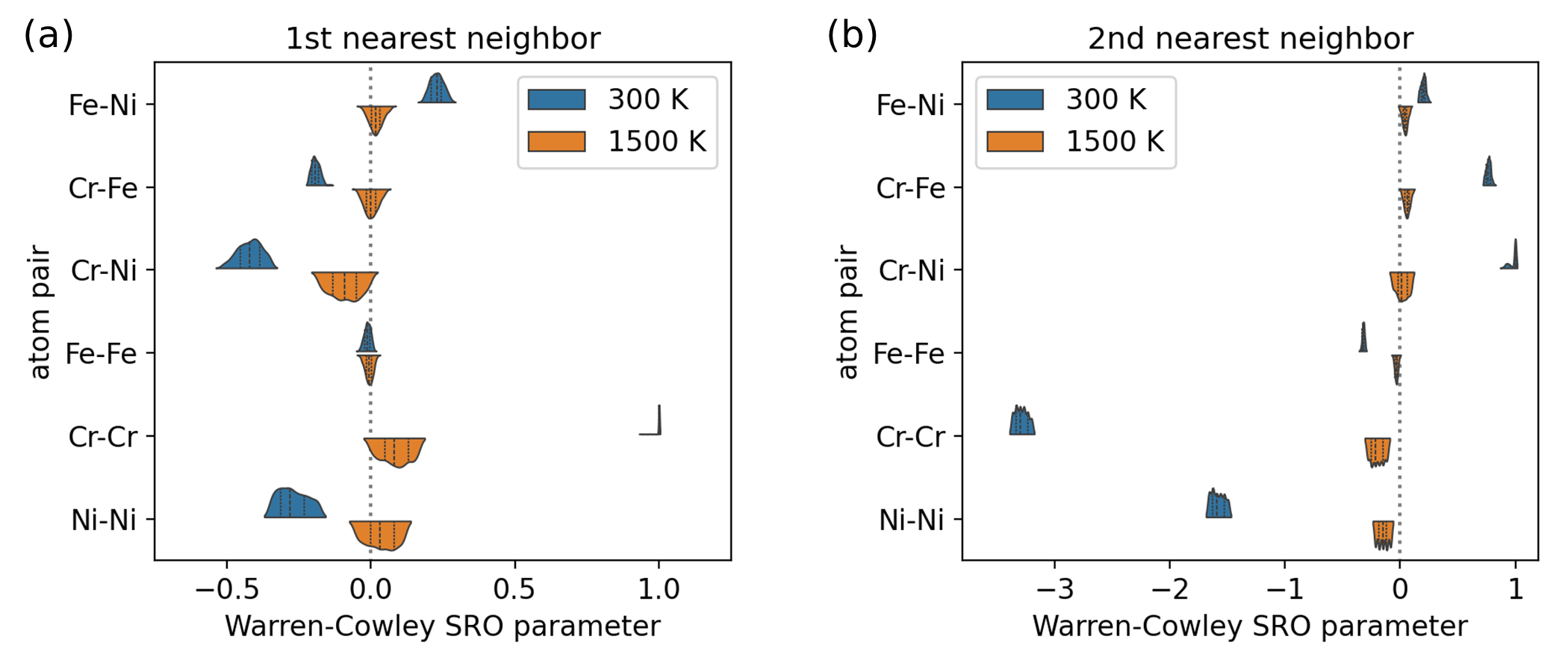"}
\caption{Warren-Cowley SRO parameter distributions of 300 K and 1500 K alloy training datasets (with SRO filtering). The SRO distributions are shown for the (a) 1st and (b) 2nd nearest neighbor interactions.}
\label{fig_sro_train_narrow}
\end{figure}

\clearpage

\begin{figure}[h!]
\centering
\includegraphics[width=5.0in]{"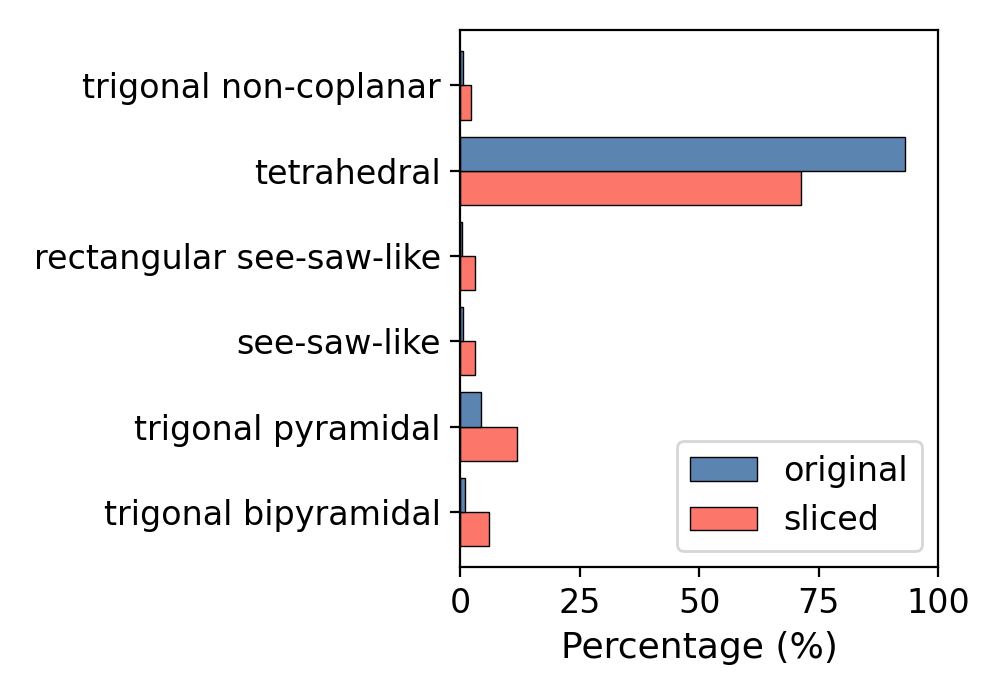"}
\caption{Coordination motif distribution of amorphous silicon, showing the comparison between the original 100,000-atom structure and the sliced 256-atom structures.}
\label{fig_a-Si_sliced}
\end{figure}

\begin{figure}[h!]
\centering
\includegraphics[width=4in]{"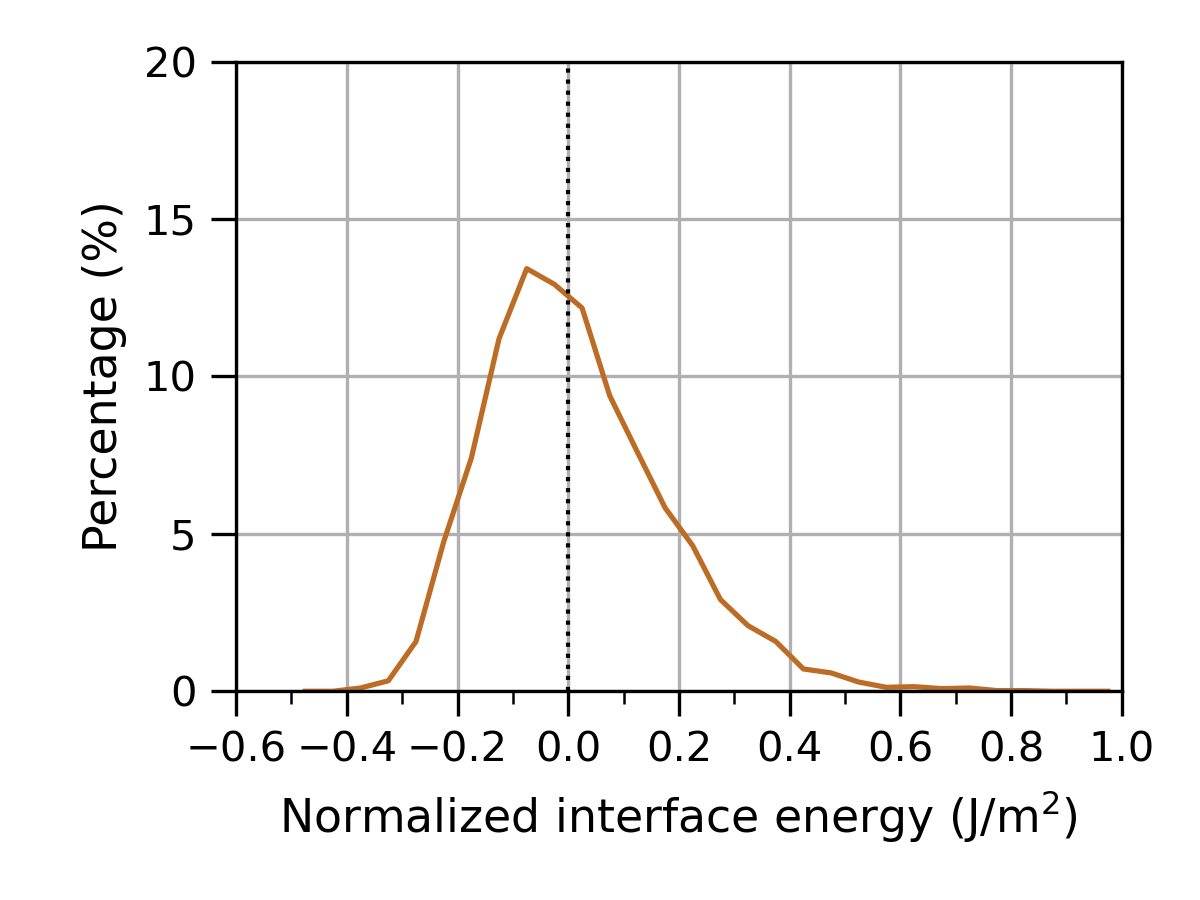"}
\caption{Normalized interface energy distribution of randomly generated interface structures. All structures were relaxed using M3GNet.}
\label{fig_e_random}
\end{figure}

\clearpage

\begin{figure}[h!]
\centering
\includegraphics[width=3.5in]{"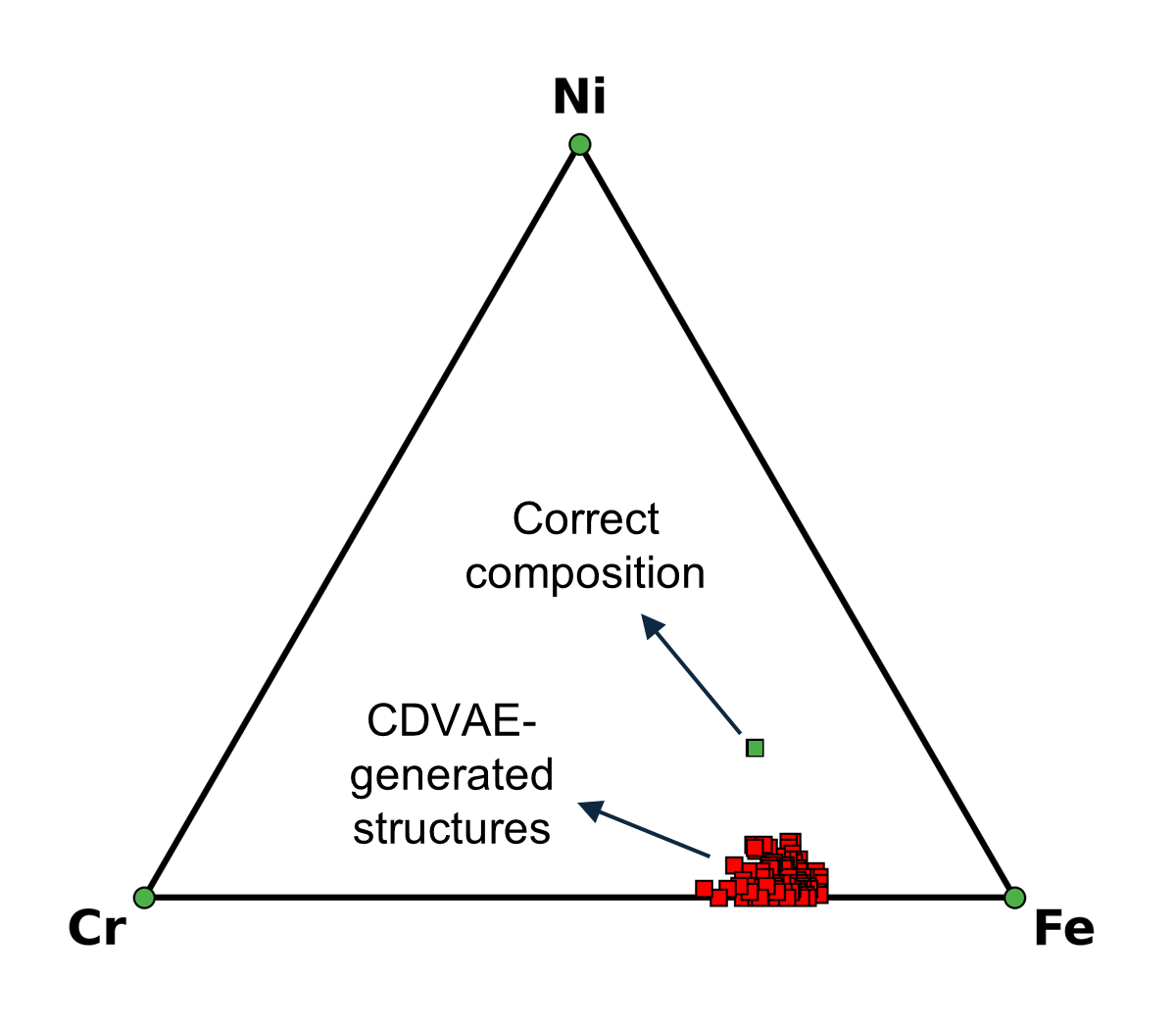"}
\caption{Phase diagram of the alloy structures generated by CDVAE (with atomic species denoising). The correct composition, Fe$_{60}$Ni$_{20}$Cr$_{20}$, is indicated by the green square. The composition of 100 CDVAE-generated structures are indicated by the red squares.}
\label{fig_cdvae_phase_diagram}
\end{figure}

\begin{figure}[h!]
\centering
\includegraphics[width=\textwidth]{"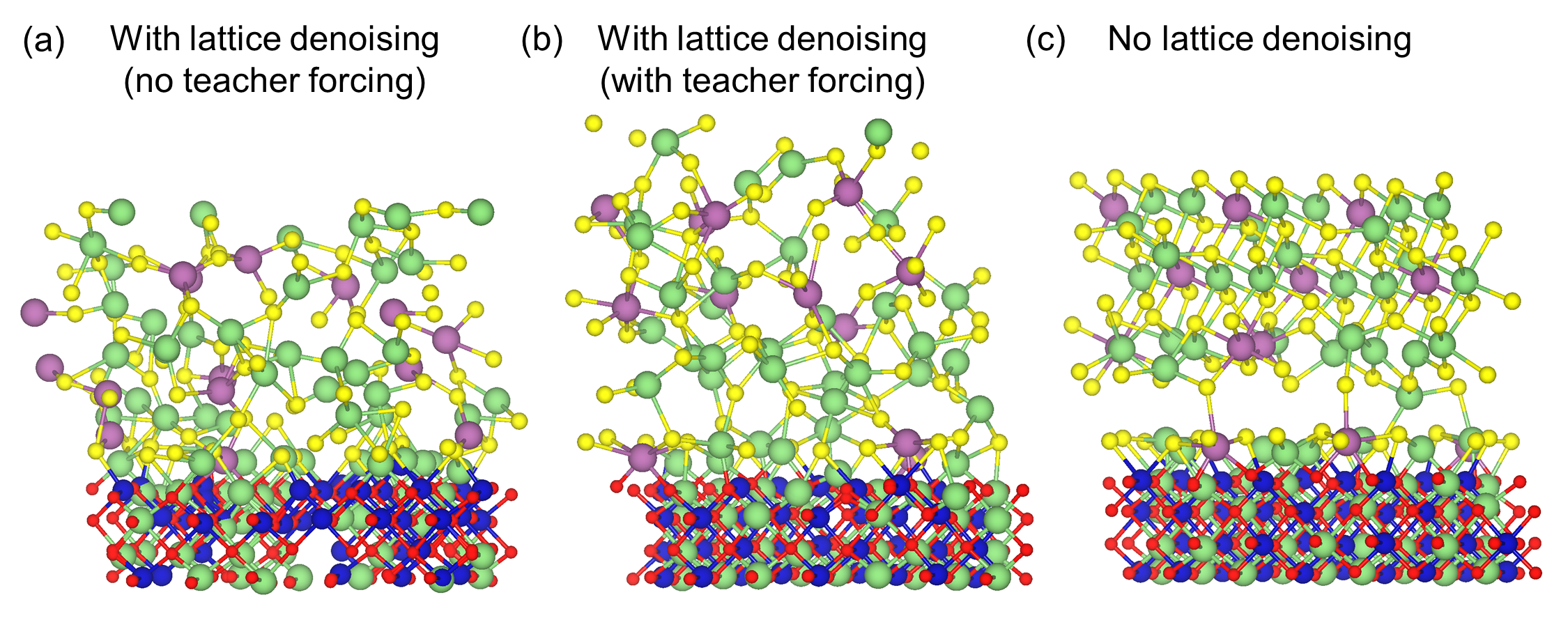"}
\caption{Example DiffCSP-generated interface structures (a) with lattice denoising but no teacher forcing, (b) with lattice denoising and teacher forcing, and (c) without lattice denoising.}
\label{fig_diffcsp_lattice_denoising}
\end{figure}

\clearpage

We trained DiffCSP models with lattice denoising (and teacher forcing) to compare with DiffCSP models without lattice denoising. For the disordered interface and alloy datasets, teacher forcing was applied for the first 150 epochs. For the amorphous Si dataset, teacher forcing was applied for the first 500 epochs. The weight of the lattice cost was set to 1 for all models. All other hyperparameters were set to be the same as the DIffCSP models without lattice denoising. 

\begin{table}[h]
\centering
\caption{Dismai-Bench metrics for DiffCSP with lattice denoising and teacher forcing. Each metric is represented by the average value over 3 separately trained models. The minimum and maximum values are shown in round brackets. The difference in the metric value between DiffCSP with lattice denoising and without lattice denoising is shown in square brackets. DiffCSP performs worse with lattice denoising than without lattice denoising.}
    \begin{tabularx}{\textwidth}{XXXXX}
        \toprule 
            \multicolumn{5}{l}{\textbf{Disordered interface}} \\
        \toprule
            \hspace{5mm} $d_{\textrm{Li}}$ \newline (min, max) &
            \hspace{5mm} $d_{\textrm{Co}}$ \newline (min, max) &
            \hspace{5mm} $d_{\textrm{Sc}}$ \newline (min, max) & 
            \hspace{5mm} $d_{\textrm{all}}$ \newline (min, max) & 
            \% struc failed \newline (min, max)  
            \\
        \midrule
            0.215 [\textbf{\texttt{+}0.171}] \newline (0.0853, 0.297) &
            0.273 [\textbf{\texttt{+}0.245}] \newline (0.0905, 0.420) & 
            0.317 [\textbf{\texttt{+}0.222}] \newline (0.240, 0.407) & 
            0.192 [\textbf{\texttt{+}0.155}] \newline (0.0671, 0.278) &
            43.8  [\textbf{\texttt{+}36.7}] \newline (11.9, 79.3)
            \\
        \bottomrule
    \end{tabularx} 
    \begin{tabularx}{\textwidth}{XXXX}
        \toprule 
            \multicolumn{4}{l}{\textbf{Amorphous Si}} \\
        \toprule
            \hspace{4mm} $d_{\textrm{motif}}$ \newline (min, max) &
            \hspace{5mm} $d_{\textrm{rdf}}$ \newline (min, max) &
            \hspace{4mm} $d_{\textrm{angle}}$ \newline (min, max) & 
            \% struc failed \newline (min, max)   
            \\
        \midrule
            0.863 [\textbf{\texttt{+}0.798}] \newline (0.405, 1.11) &
            9.70 [\textbf{\texttt{+}8.31}] \newline (8.70, 10.6) & 
            0.0492 [\textbf{\texttt{+}0.0390}] \newline (0.0260, 0.0621) & 
            8.7  [\textbf{\texttt{+}8.7}] \newline (6.0, 12.6)
            \\
        \bottomrule
    \end{tabularx}
    \begin{tabularx}{\textwidth}{XXX}
        \toprule 
            \multicolumn{3}{l}{\textbf{Alloy (300 K, narrow SRO)}} \\
        \toprule
            \hspace{3mm} $d_{\textrm{cluster}}$ \newline (min, max) &
            \% struc w/ vac \newline (min, max) & 
            \% struc failed \newline (min, max)  
            \\
        \midrule
            0.0824 [\textbf{\texttt{+}0.0179}] \newline (0.0621, 0.102) &
            99.5 [\textbf{\texttt{+}5.13}] \newline (98.4, 100) & 
            0.00 [\textbf{\texttt{+}0.00}] \newline (0.00, 0.00)
            \\
        \bottomrule
    \end{tabularx}
    \label{table_diffcsp_lattice_denoise}
\end{table}

\begin{table}[h!]
\centering
\caption{M3GNet relaxation results of disordered interfaces generated by the generative models. $n_{\textrm{steps}}$ is the number of M3GNet relaxation steps required to relax the generated structures. $E_{\textrm{initial}}$ and $E_{\textrm{final}}$ are the M3GNet-calculated energies of the unrelaxed and relaxed structures respectively. The mean and standard deviation for each quantity are listed. For each architecture, 3 separate models were trained and the results were averaged.}
\begin{tabularx}{0.6\textwidth}{lXXXX}
\toprule
    \multirow{3}{*}[-2.5pt]{Model} & 
    \multicolumn{2}{l}{\multirow{2}{*}{$n_{\textrm{steps}}$}} &
    \multicolumn{2}{l}{$E_{\textrm{initial}}$ - $E_{\textrm{final}}$} \\
    & 
    \multicolumn{2}{l}{} &
    \multicolumn{2}{l}{(meV/atom)} \\
    \cmidrule(lr){2-5}
     & {mean} & {std} & {mean} & {std}\\
\midrule
    CDVAE & 10.3 & 8.71 & 3.30 & 1.46 \\
    DiffCSP & 68.2 & 53.9 & 12.3 & 9.68 \\
    CrysTens & 171 & 74.3 & 640 & 137 \\
    UniMat & 257 & 116 & 989 & 70.0 \\
    CryinGAN & 147 & 72.4 & 234 & 67.3 \\
\bottomrule
\end{tabularx}
\label{table_int_relaxation}
\end{table}

\clearpage

\begin{table}[h!]
\centering
\caption{SOAP-GAP relaxation results of amorphous Si generated by CDVAE and DiffCSP. $n_{\textrm{steps}}$ is the number of SOAP-GAP relaxation steps required to relax the generated structures. $E_{\textrm{initial}}$ and $E_{\textrm{final}}$ are the SOAP-GAP-calculated energies of the unrelaxed and relaxed structures respectively. The mean and standard deviation for each quantity are listed. For each architecture, 3 separate models were trained and the results were averaged.}
\begin{tabularx}{0.6\textwidth}{lXXXX}
\toprule
    \multirow{3}{*}[-2.5pt]{Model} & 
    \multicolumn{2}{l}{\multirow{2}{*}{$n_{\textrm{steps}}$}} &
    \multicolumn{2}{l}{$E_{\textrm{initial}}$ - $E_{\textrm{final}}$} \\
    & 
    \multicolumn{2}{l}{} &
    \multicolumn{2}{l}{(meV/atom)} \\
    \cmidrule(lr){2-5}
     & {mean} & {std} & {mean} & {std}\\
\midrule
    CDVAE & 22.5 & 16.3 & 8.64 & 45.3 \\
    DiffCSP & 140 & 39.0 & 275 & 21.6 \\
\bottomrule
\end{tabularx}
\label{table_a-Si_relaxation}
\end{table}

\begin{figure}[h!]
\centering
\includegraphics[width=\textwidth]{"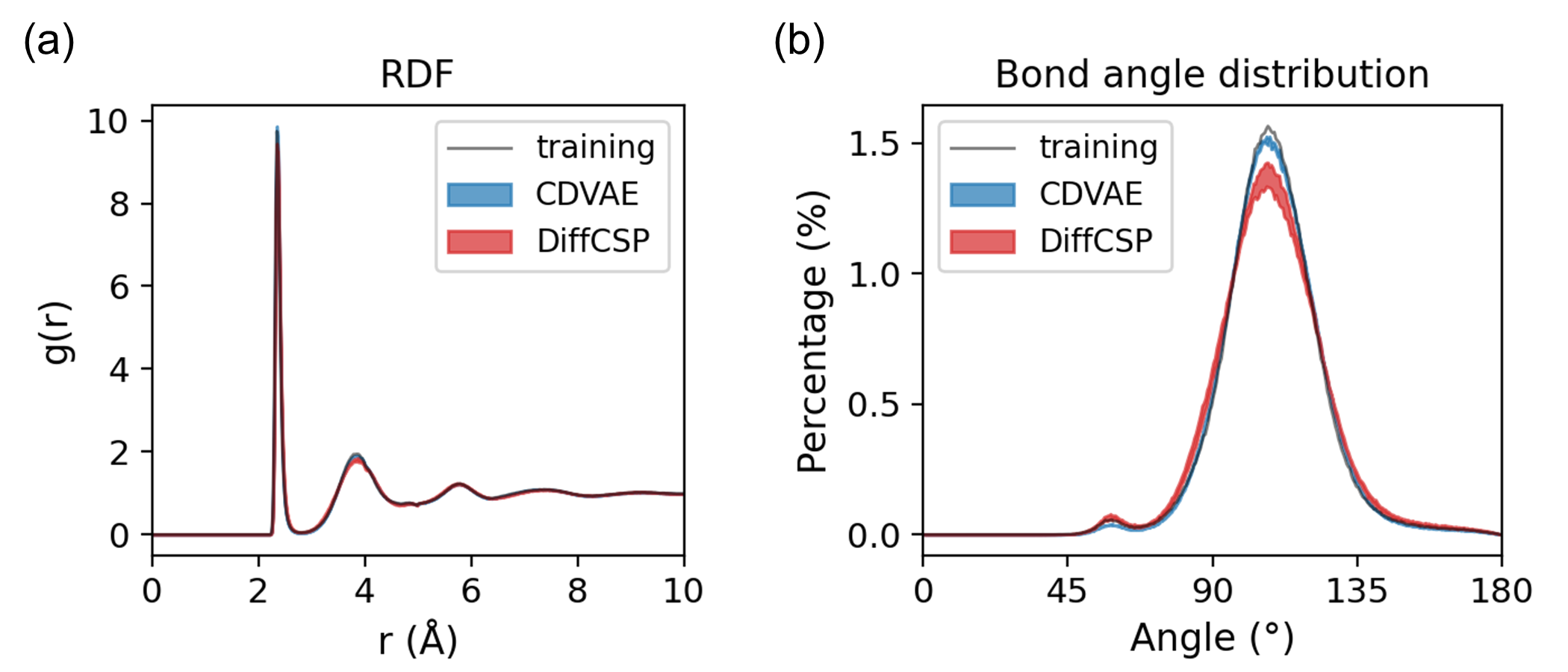"}
\caption{(a) Radial distribution functions (RDFs) and (b) bond angle distributions of amorphous Si structures generated by CDVAE and DiffCSP. The distributions of the training dataset are also shown for reference.}
\label{fig_a-Si_rdf_angle}
\end{figure}

\begin{figure}[h!]
\centering
\includegraphics[width=0.8\textwidth]{"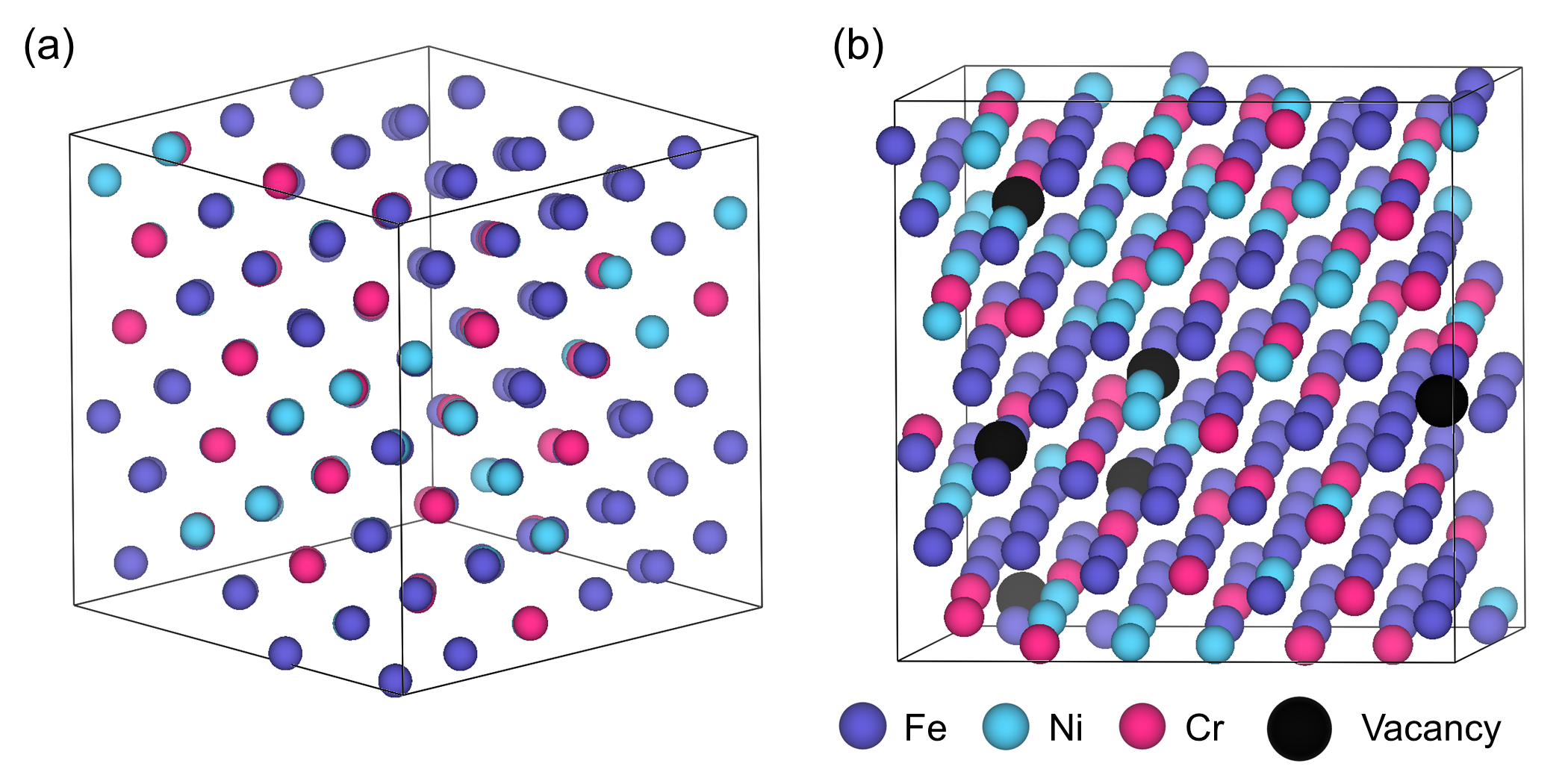"}
\caption{Examples of CDVAE-generated alloy structures with (a) noisy lattice and (b) vacancies.}
\label{fig_cdvae_nosiy_alloy}
\end{figure}

\clearpage

\begin{figure}[h!]
\centering
\includegraphics[width=\textwidth]{"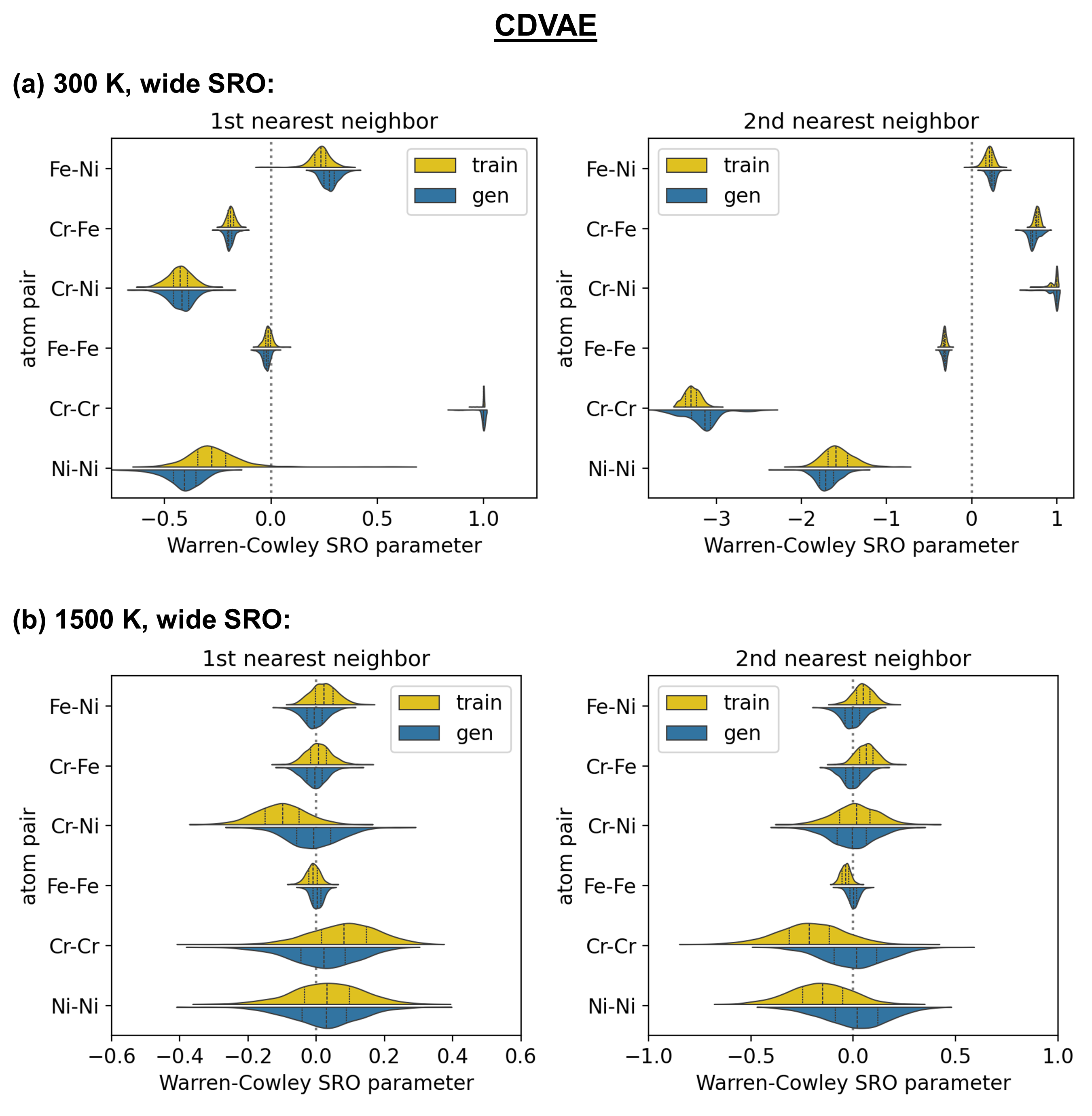"}
\caption{Warren-Cowley SRO parameter distributions of (a) 300 K and (b) 1500 K alloy structures (wide SRO) generated by CDVAE. The SRO distributions are shown for the 1st and 2nd nearest neighbor interactions.}
\label{fig_alloy_sro_cdvae}
\end{figure}

\clearpage

\begin{figure}[h!]
\centering
\includegraphics[width=\textwidth]{"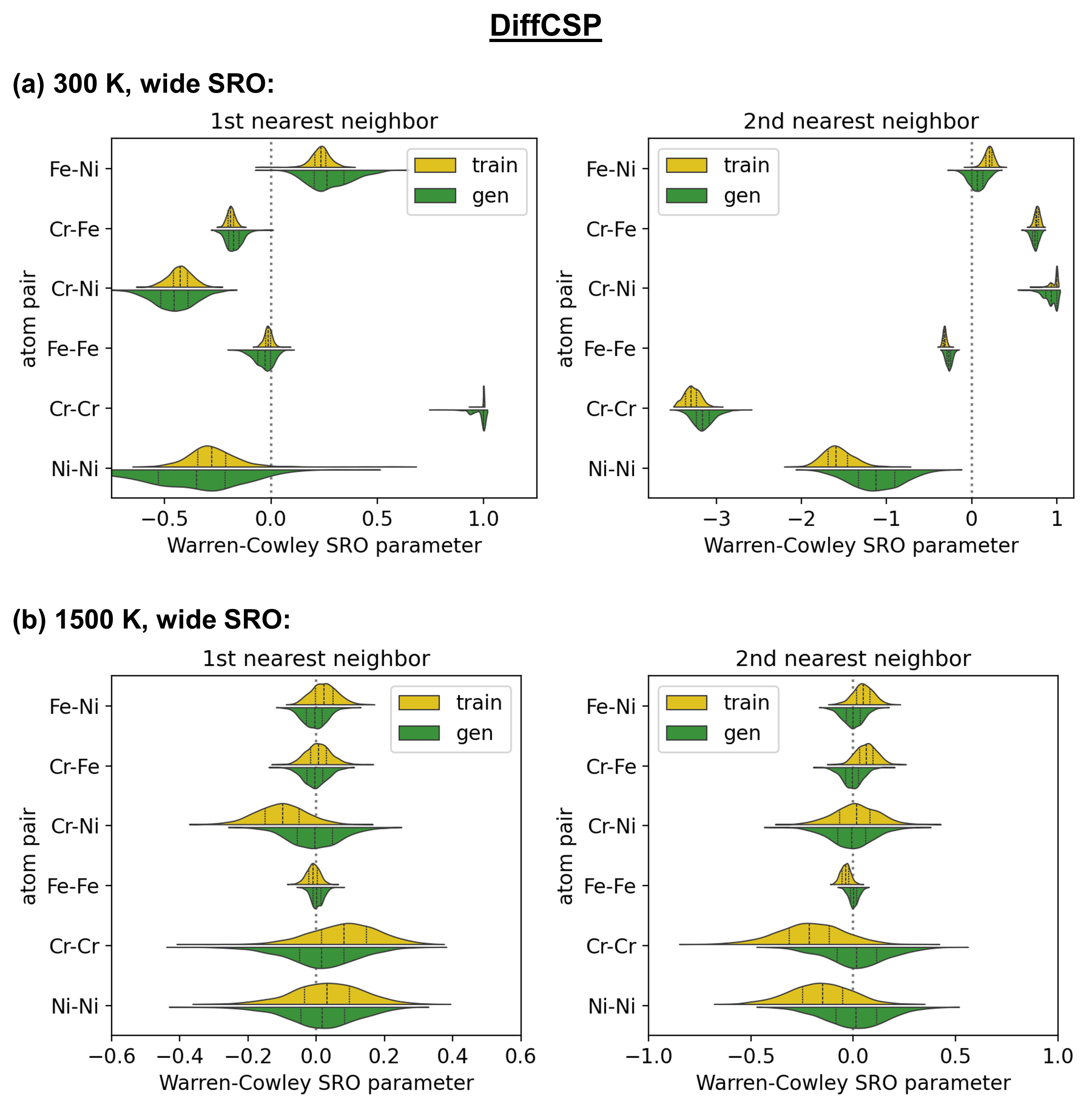"}
\caption{Warren-Cowley SRO parameter distributions of (a) 300 K and (b) 1500 K alloy structures (wide SRO) generated by DiffCSP. The SRO distributions are shown for the 1st and 2nd nearest neighbor interactions.}
\label{fig_alloy_sro_diffcsp}
\end{figure}

\clearpage

\begin{figure}[h!]
\centering
\includegraphics[width=\textwidth]{"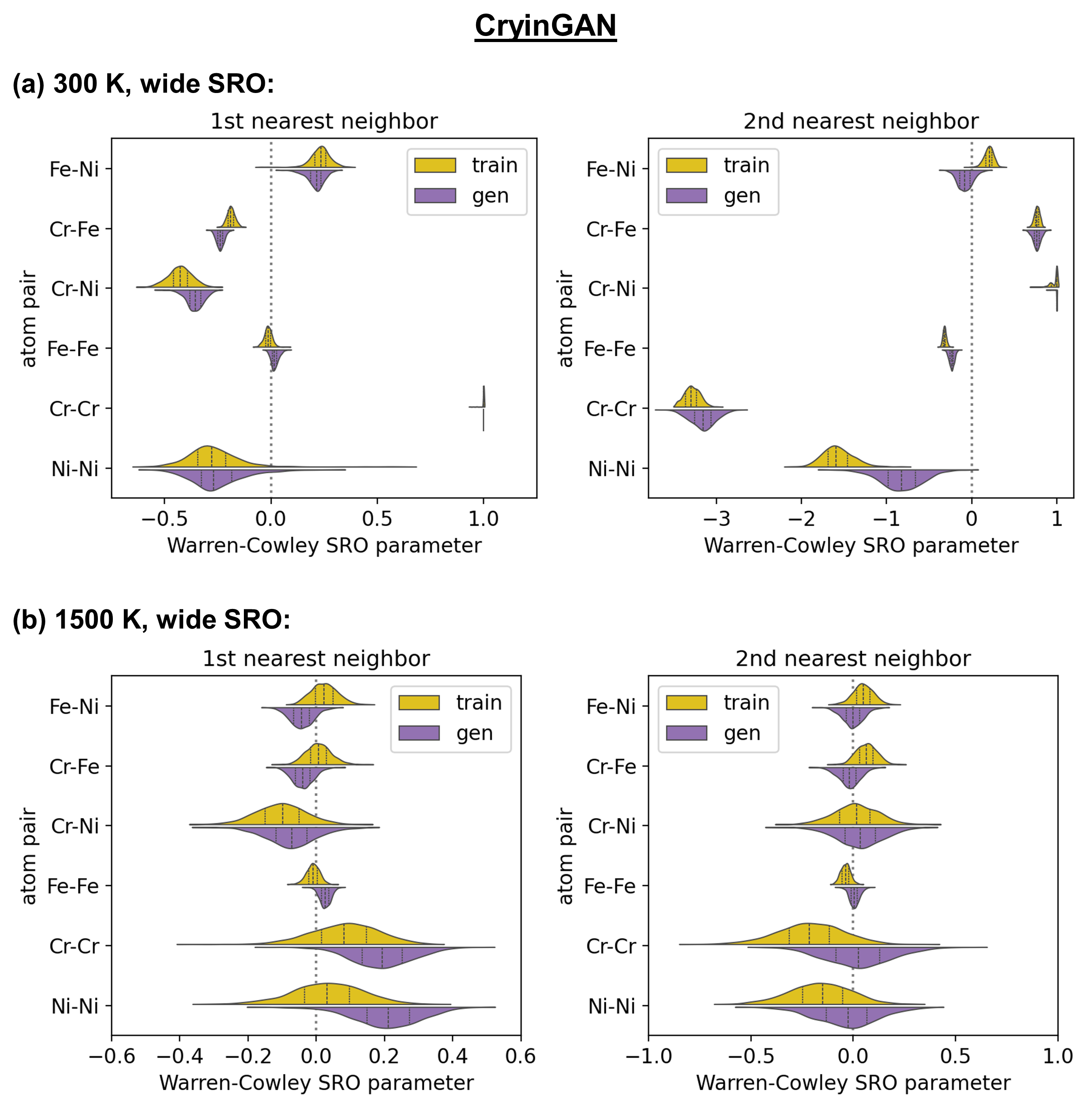"}
\caption{Warren-Cowley SRO parameter distributions of (a) 300 K and (b) 1500 K alloy structures (wide SRO) generated by CryinGAN. The SRO distributions are shown for the 1st and 2nd nearest neighbor interactions.}
\label{fig_alloy_sro_cryingan}
\end{figure}

\clearpage

\begin{figure}[h!]
\centering
\includegraphics[width=4in]{"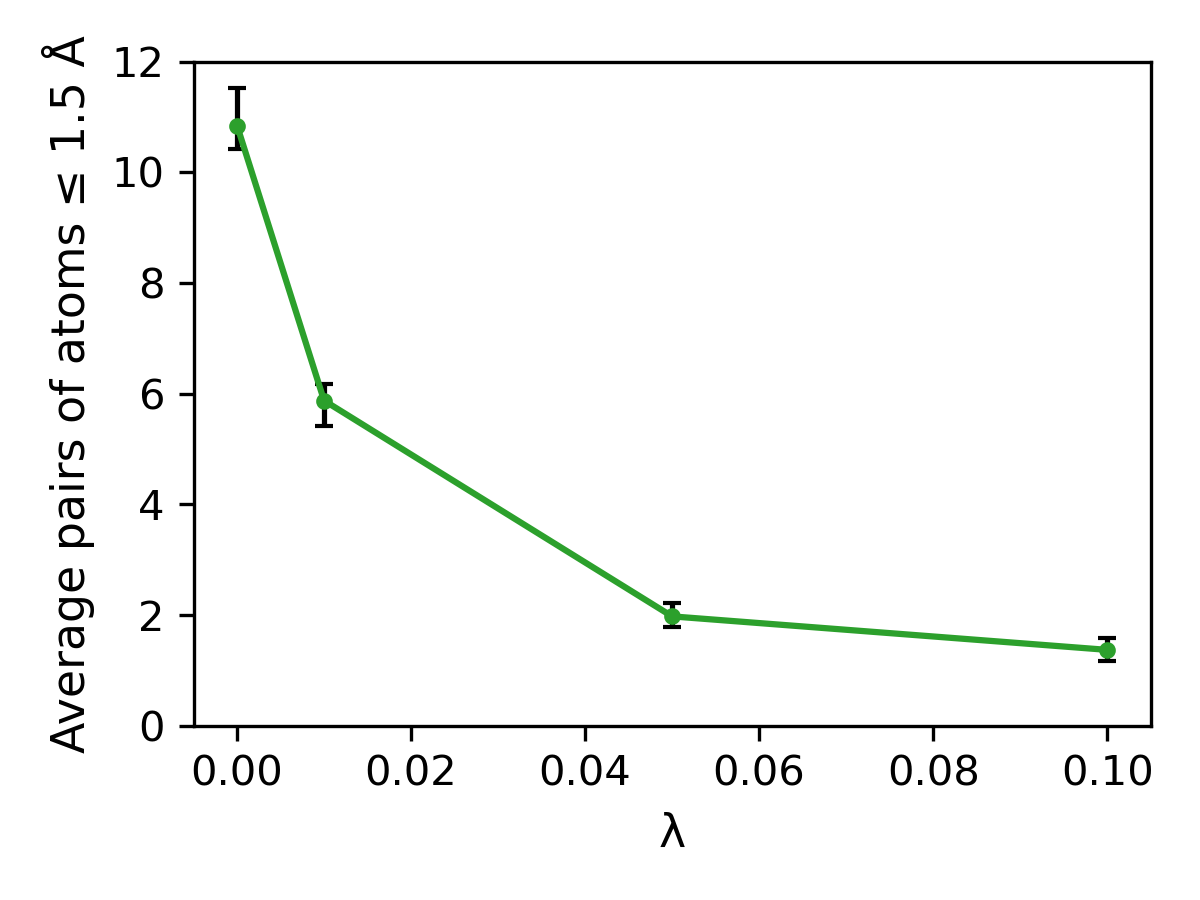"}
\caption{Average number of unique pairs of atoms with bond distance $\leq$ 1.5 \r{A} for interface structures generated using CryinGAN trained with different $\lambda$ values. As $\lambda$ increases, fewer atoms are generated too close to each other. For each $\lambda$ value, 3 separate models were trained, 1000 structures were generated using each model, and the counts were averaged across the models.}
\label{fig_too_close}
\end{figure}

\begin{figure}[h!]
\centering
\includegraphics[width=\textwidth]{"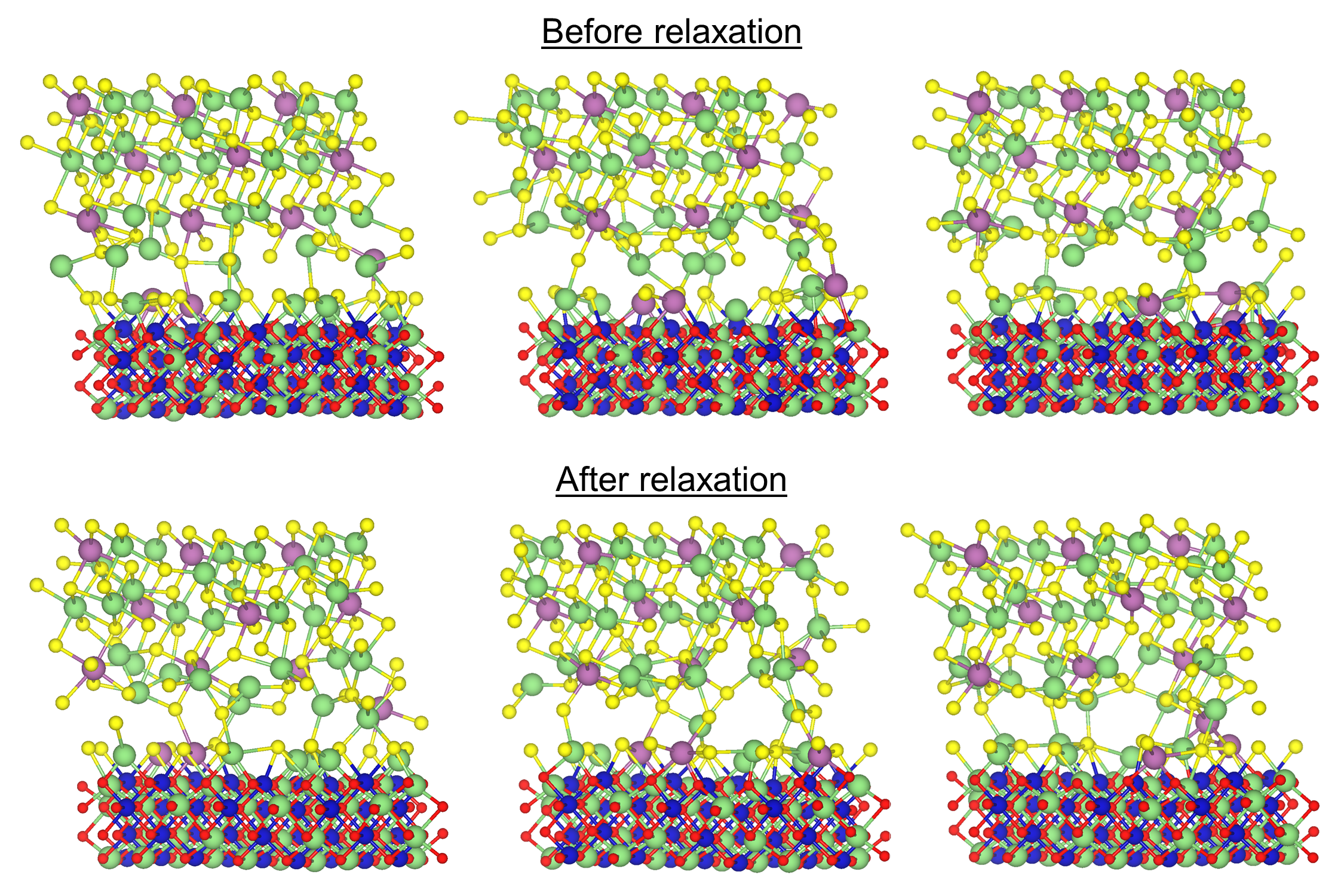"}
\caption{Examples of CryinGAN-generated structures before relaxation (as-generated) and after M3GNet relaxation.}
\label{fig_cryingan_before_after_relax}
\end{figure}

\clearpage

\begin{figure}[h!]
\centering
\includegraphics[width=5in]{"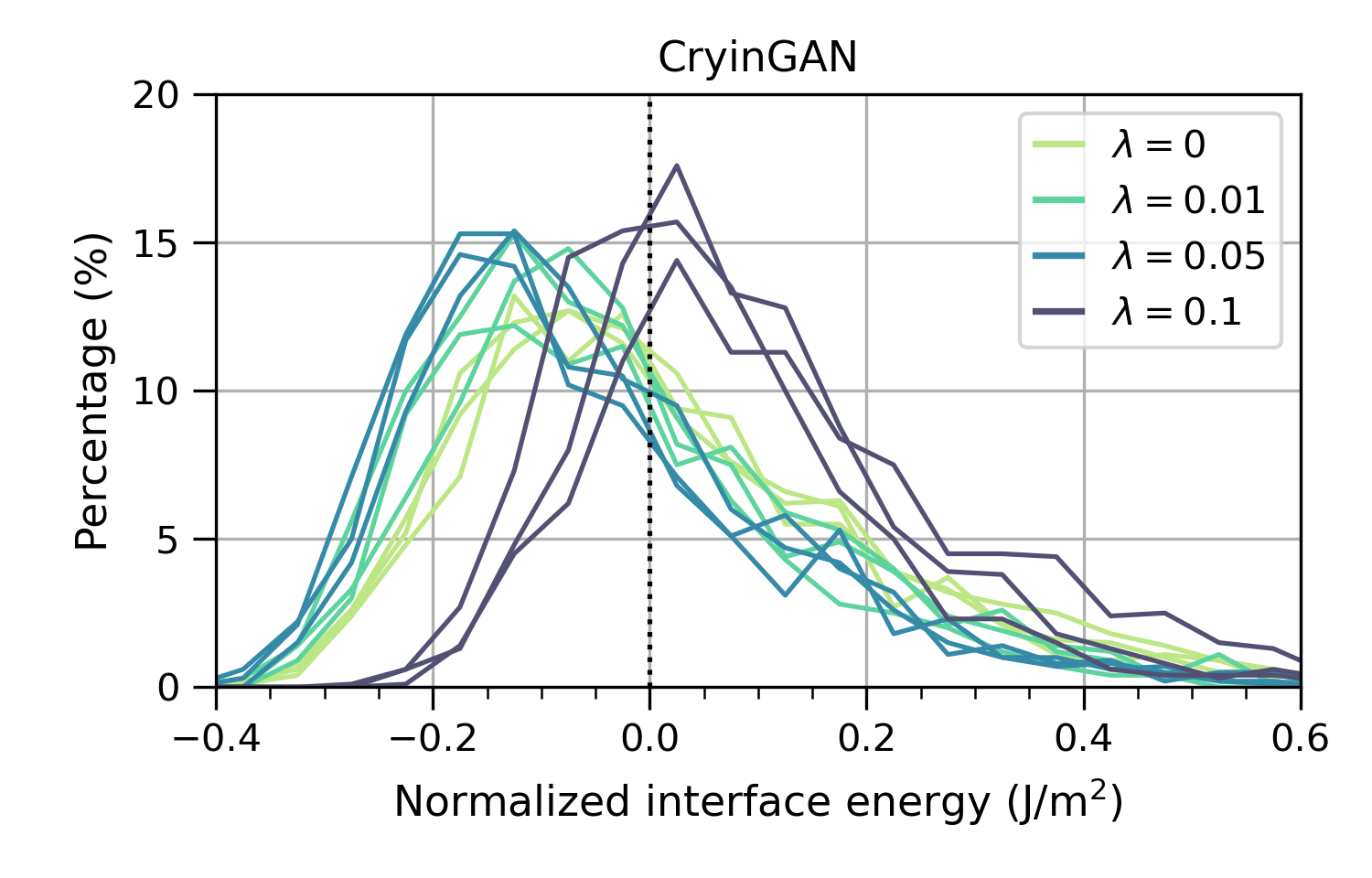"}
\caption{Normalized interface energy distributions of structures generated using CryinGAN trained with different $\lambda$ values (0, 0.01, 0.05, and 0.1). For each $\lambda$ value, 3 separate models were trained. The generated structures were relaxed using M3GNet, and the interface energies shown are based on M3GNet-calculated energies. }
\label{fig_lambda_3runs}
\end{figure}

\begin{figure}[h!]
\centering
\includegraphics[width=4in]{"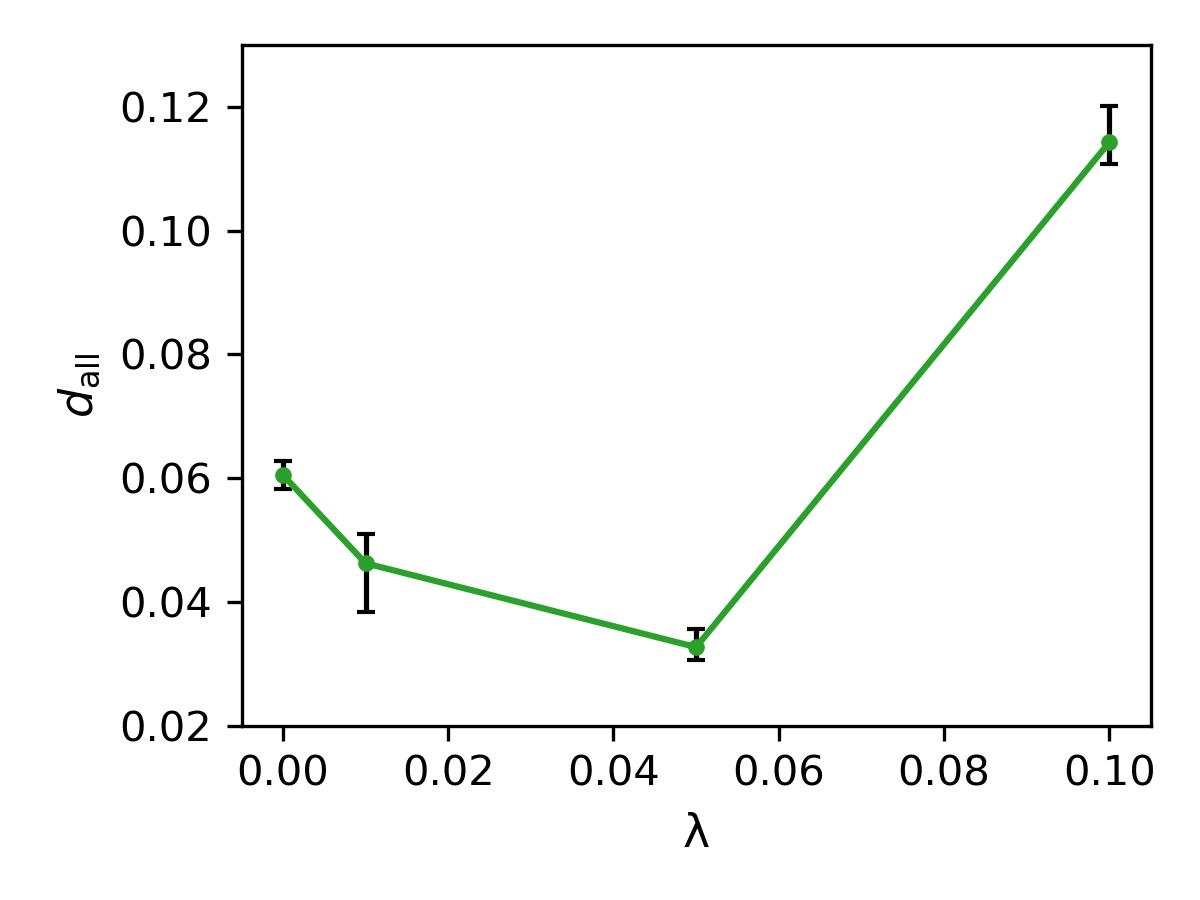"}
\caption{Euclidean distance of the coordination motif fingerprint of all cations for interface structures generated using CryinGAN trained with different $\lambda$ values.}
\label{fig_lambda_d_all}
\end{figure}

\clearpage

\begin{figure}[h!]
\centering
\includegraphics[width=5in]{"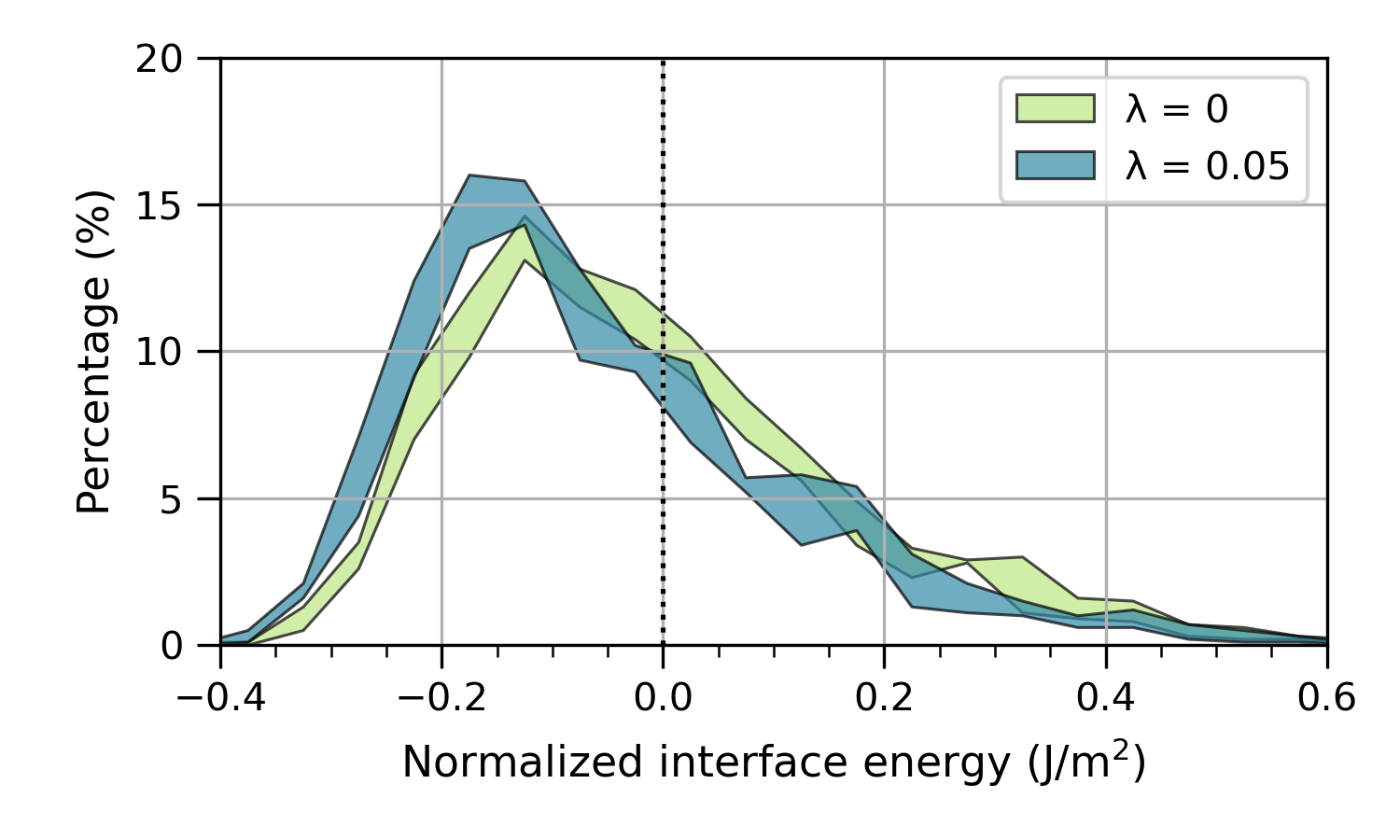"}
\caption{Normalized interface energy distributions of structures generated using CryinGAN trained with $\lambda = 0$ (200,000 epochs) and $\lambda = 0.05$ (100,000 epochs). The generated structures were relaxed using M3GNet, and the interface energies shown are based on M3GNet-calculated energies. CryinGAN runs approximately two times faster when the bond distance discriminator is not used ($\lambda = 0$), but the use of the bond distance discriminator still yields a lower interface energy distribution for the same amount of training time.}
\label{fig_epoch200000}
\end{figure}

\begin{figure}[h!]
\centering
\includegraphics[width=2.75in]{"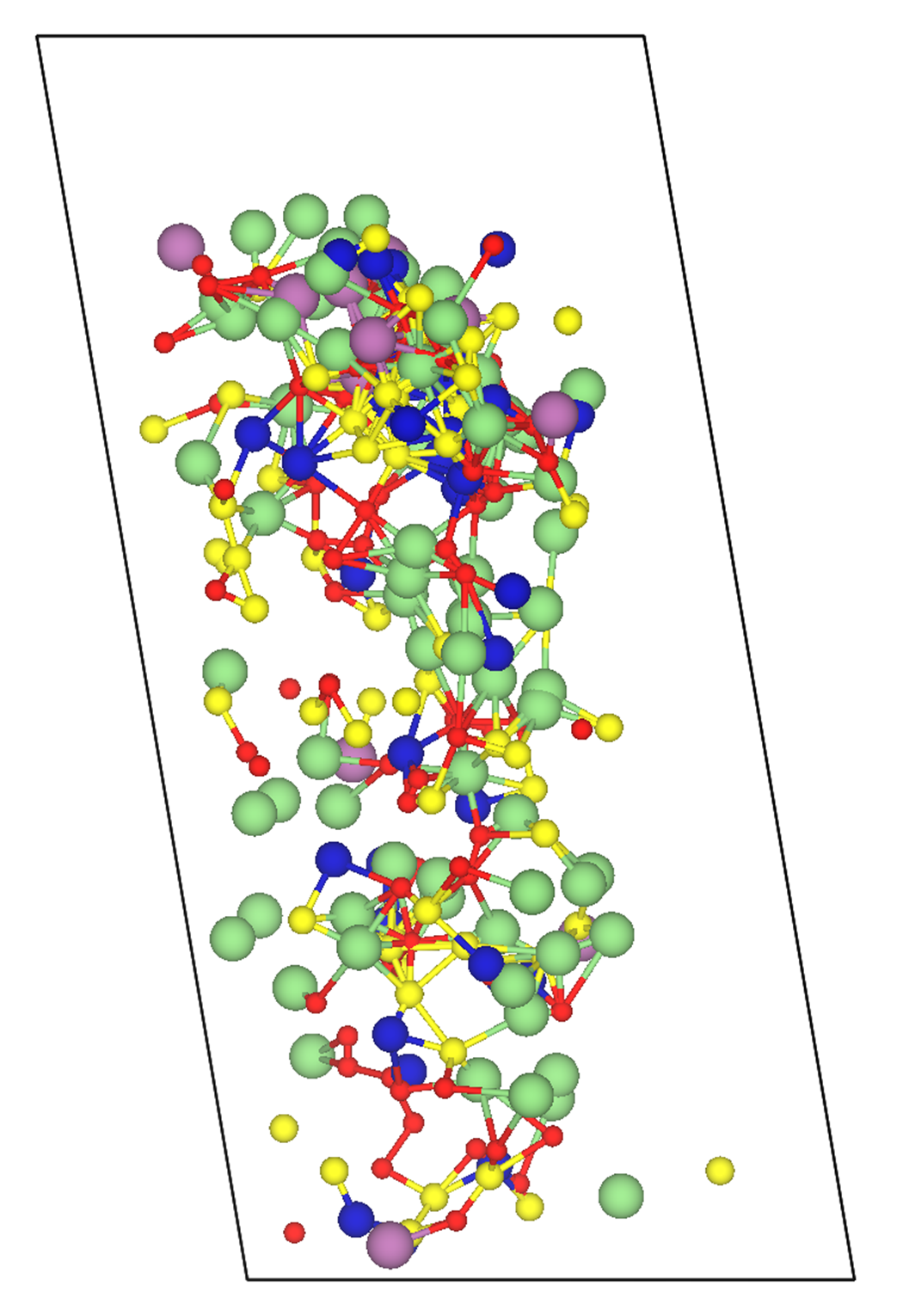"}
\caption{Example generated structure of a GAN model using Crystal Graph Convolutional Neural Networks (CGCNN) as the discriminator. The GAN was unable to learn to generate meaningful interface structures.}
\label{fig_cgcnn}
\end{figure}

\clearpage

\begin{figure}[h]
\centering
\includegraphics[width=\columnwidth]{"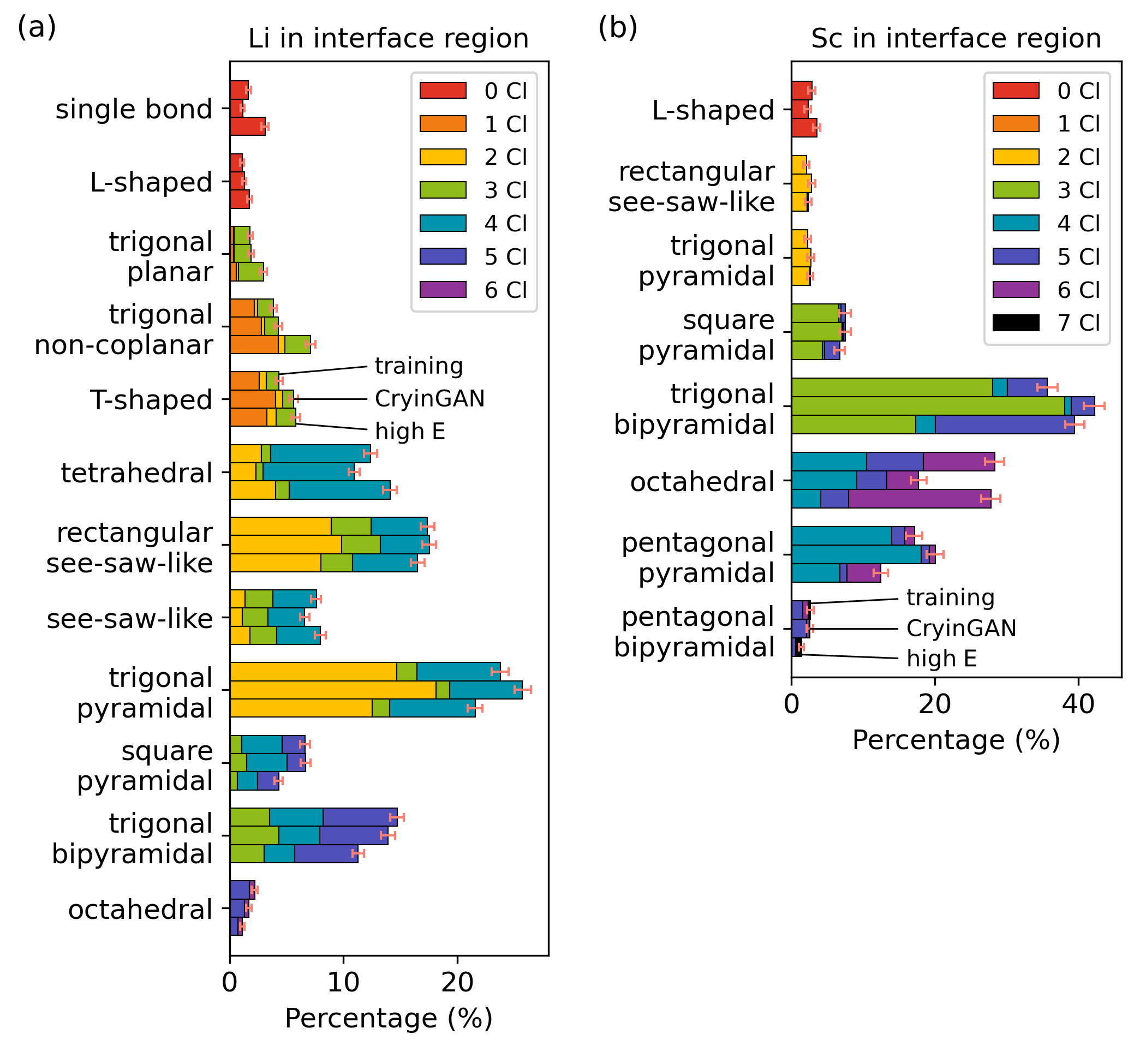"}
\caption{Coordination motif distributions of (a) Li and (b) Sc in the interface region, where each coordination motif is further subdivided based on the number of Cl bonds. The distributions of three datasets are shown: (1) training structures with low interface energy, (2) CryinGAN-generated structures, and (3) structures with high interface energy. All structures were relaxed using M3GNet followed by DFT calculations. Error bars represent 95 \% bootstrap confidence intervals.}
\label{fig_motifs_species}
\end{figure}

\clearpage

\begin{figure}[h!]
\centering
\includegraphics[width=5in]{"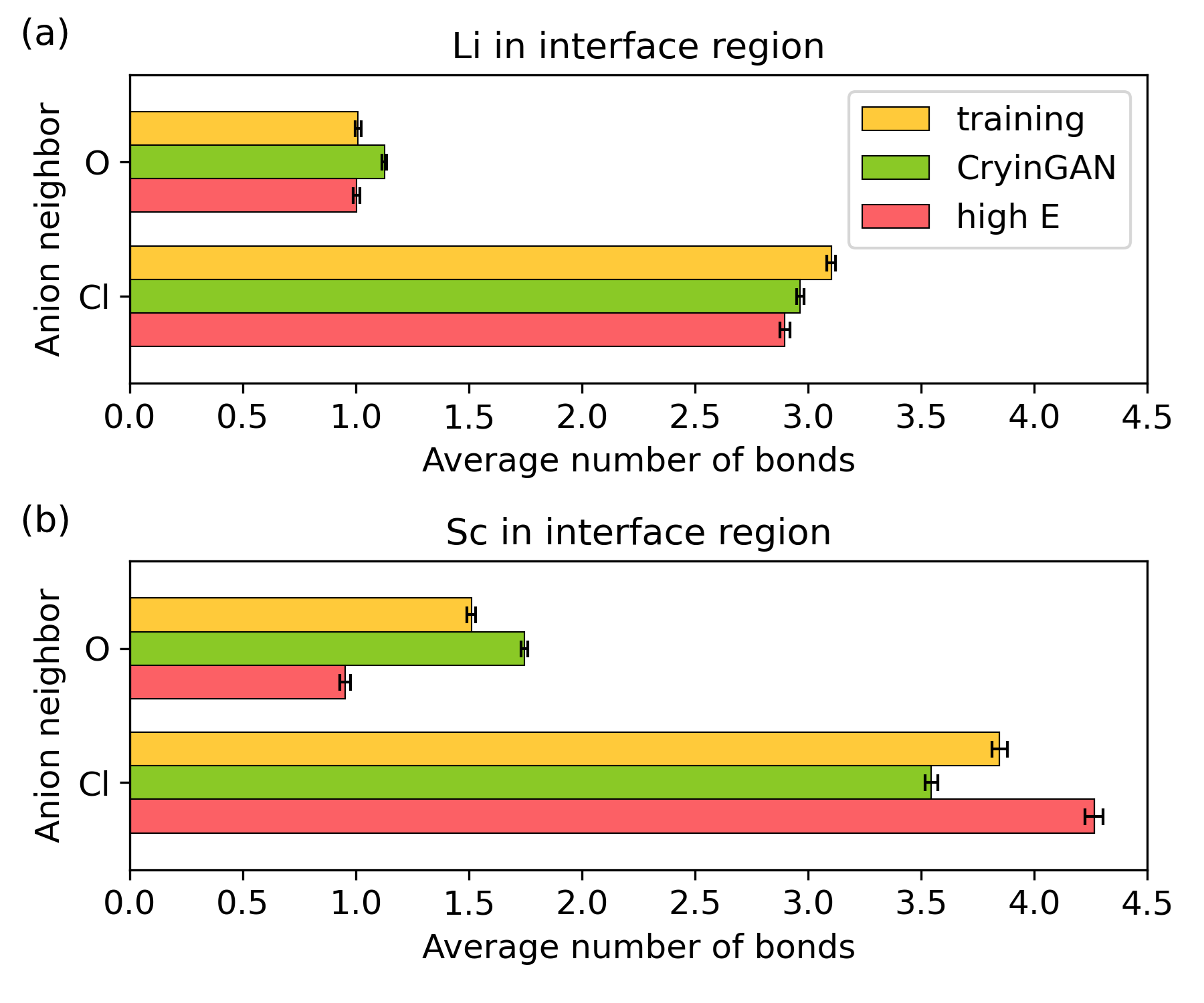"}
\caption{Histograms of the average number of O and Cl bonds for (a) Li and (b) Sc in the interface region. Error bars represent 95 \% bootstrap confidence intervals.}
\label{fig_avg_n_bonds}
\end{figure}

\begin{figure}[h!]
\centering
\includegraphics[width=6.5in]{"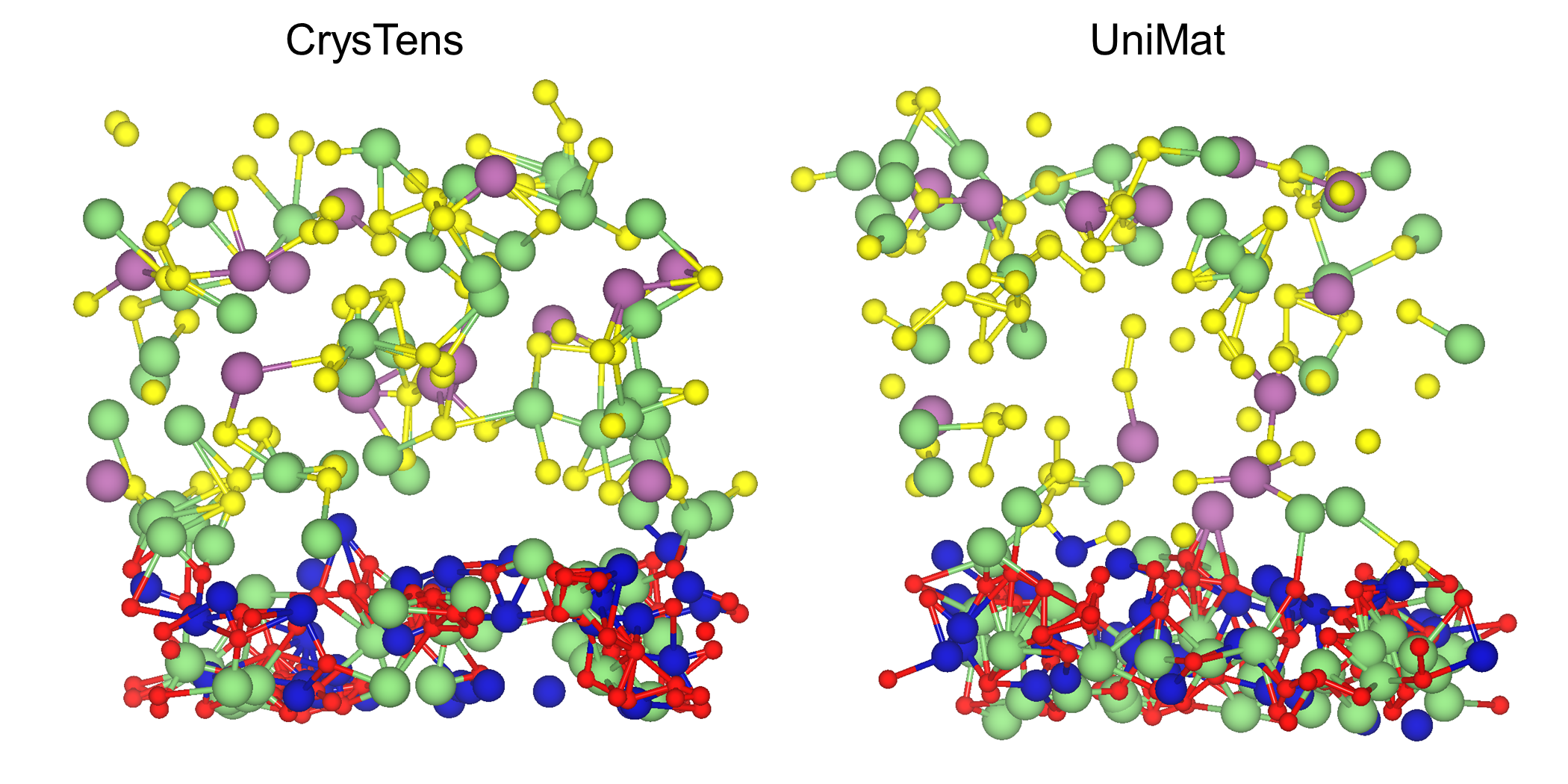"}
\caption{Example interface structures generated by CrysTens and UniMat when trained on structures with randomized atom orderings.}
\label{fig_crysten_unimat_bad_int}
\end{figure}

\clearpage

\begin{figure}[h]
\centering
\includegraphics[width=0.9\textwidth]{"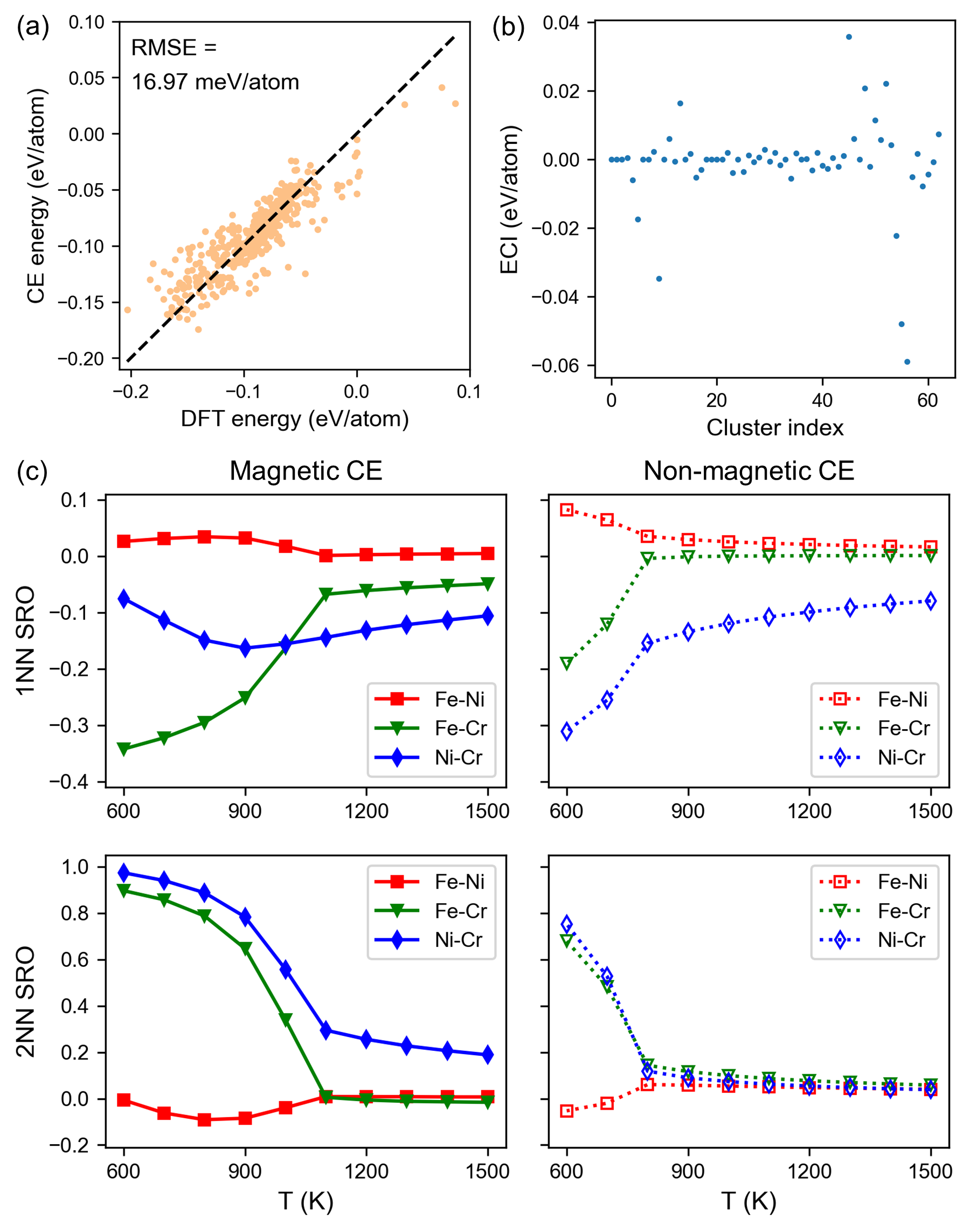"}
\caption{(a) Plot of cluster expansion (CE) energies against DFT energies. The root mean square error (RMSE) of the CE fit is 16.97 meV/atom. (b) All effective cluster interactions (ECI) values of the CE fit. (c) Plots of the Warren-Cowley SRO parameter against temperature. Plots on the left are adapted from ref. \citenum{RN358} and correspond to an alloy with composition Fe$_{56}$Cr$_{21}$Ni$_{23}$ and 4000 atoms. Plots on the right correspond to the alloy of this work, with composition Fe$_{60}$Ni$_{20}$Cr$_{20}$ and 256 atoms. Plots on the left use the full CE model with magnetic terms as described in ref. \citenum{RN358}, whereas plots on the right use only terms corresponding to the seven chemical dimers (non-magnetic). The top and bottom plots correspond to the 1st and 2nd nearest neighbor interactions respectively. The Dismai-Bench alloy shows qualitatively similar SRO trends to the larger magnetic alloy.}
\label{fig_ce_fit}
\end{figure}

\clearpage

\begin{table}[h!]
\centering
\caption{Configurations of the interfaces calculated. The slab orientation, number of layers, and average lattice mismatch between any given two slabs are listed. $M$ represents the transformation matrix used to transform the lattice vectors of a given slab surface ($\vec{u}, \vec{v}$) into the superlattice vectors of the interface  ($\vec{u}_{s}, \vec{v}_{s}$), according to the relation ($\vec{u}_{s}, \vec{v}_{s}$) = $M \cdot$($\vec{u}, \vec{v}$). Note that the number of layers of the Li$_3$ScCl$_6$(100) slab is indicated for the randomly generated LiCoO$_2$(110)-Li$_3$ScCl$_6$(100) structures, and does not include the atoms randomly generated in the interface region.}
\def\arraystretch{1.1} 
\begin{tabular}{@{}lSlSlSlSlSlSl@{}}
\toprule
Slab 1 & \makecell[l]{number \\ of layers} & \hspace{3pt} $M$ & Slab 2 & \makecell[l]{number \\ of layers} & \hspace{3pt} $M$ & \makecell[l]{average \\ mismatch (\%)} \\
\midrule
LiCoO$_2$(110)  &  4  &
    $\begin{pmatrix}      
        1  &  2  \\
        -3 &  3  \\
            \end{pmatrix}$
      & 
    Li$_3$ScCl$_6$(100)  &  9  &
    $\begin{pmatrix}      
        1  &  1  \\
        -2 &  1  \\
            \end{pmatrix}$
      &
    2.17  \\
Li$_2$O(100)  &  9  & 
    $\begin{pmatrix}      
        1  &  2  \\
        -4 &  2  \\
            \end{pmatrix}$
     & 
    LiCl(100)  &  4  &
    $\begin{pmatrix}      
        2  &  0  \\
        0  &  4  \\
            \end{pmatrix}$
      &
    0.545  \\
Li$_2$O(110)  &  4  &
    $\begin{pmatrix}      
        1  &  0  \\
        0  &  3  \\
            \end{pmatrix}$
      & 
    Li(100)  &  6  &
    $\begin{pmatrix}      
        1  &  0  \\
        0  &  3  \\
            \end{pmatrix}$
      &
    3.00  \\
Li$_2$O(110)  &  4  &
    $\begin{pmatrix}      
        -2 &  1  \\
        3  &  2  \\
            \end{pmatrix}$
      & 
    LiCl(100)  &  4  &
    $\begin{pmatrix}      
        2  &  1  \\
        -2 &  3  \\
            \end{pmatrix}$
      &
    2.13  \\
Li$_2$O(110)  &  4  &
    $\begin{pmatrix}      
        -2 &  1  \\
        3  &  2  \\
            \end{pmatrix}$
      & 
    MgS(100)  &  4  &
    $\begin{pmatrix}      
        2  &  1  \\
        -2 &  3  \\
            \end{pmatrix}$
      &
    2.17  \\
\bottomrule
\end{tabular}
\label{table_lat_mismatch}
\end{table}

\begin{figure}[h!]
\centering
\includegraphics[width=4.5in]{"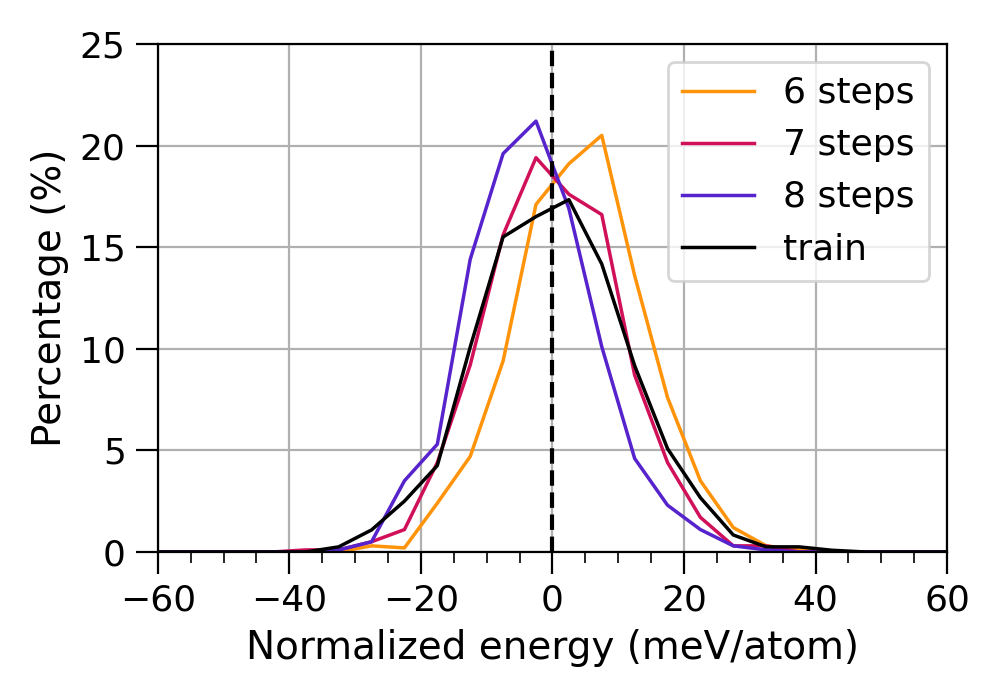"}
\caption{Normalized energy distributions of (relaxed) CDVAE-generated amorphous Si structures, compared to the training structures. The structures were generated using 6, 7, or 8 steps per noise level.}
\label{fig_cdvae_a-Si_energy}
\end{figure}

\clearpage

\begin{figure}[h!]
\centering
\includegraphics[width=4.5in]{"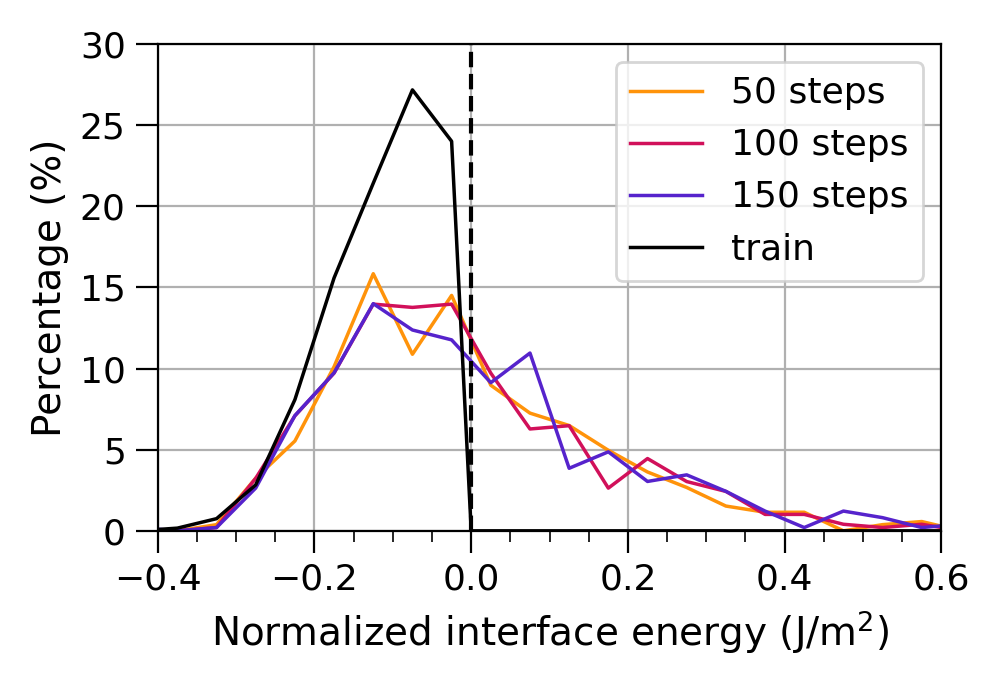"}
\caption{Normalized interface energy distributions of (relaxed) CrysTens-generated interface structures, compared to the training structures. The structures were generated using 50, 100, or 150 time steps.}
\label{fig_crystens_int_energy}
\end{figure}

\begin{table}[h!]
\centering
\caption{Hyperparameters used for the UniMat 3D U-Net model. The hyperparameter names correspond to those defined in ref. \citenum{wang2024imagen}. Default values were used for all other hyperparameters.}
\begin{tabularx}{0.5\textwidth}{XX}
\toprule
    Hyperparameter & Value \\
\midrule
    dim & 64 \\
    dim\_mults & (1, 2, 4) \\
    num\_resnet\_blocks & 3 \\
    layer\_attns & (False, True, True) \\
    layer\_cross\_attns & (False, True, True) \\
\bottomrule
\end{tabularx}
\label{table_unimat_hyperparameters}
\end{table}

\clearpage

\begin{figure}[h!]
\centering
\includegraphics[width=4.5in]{"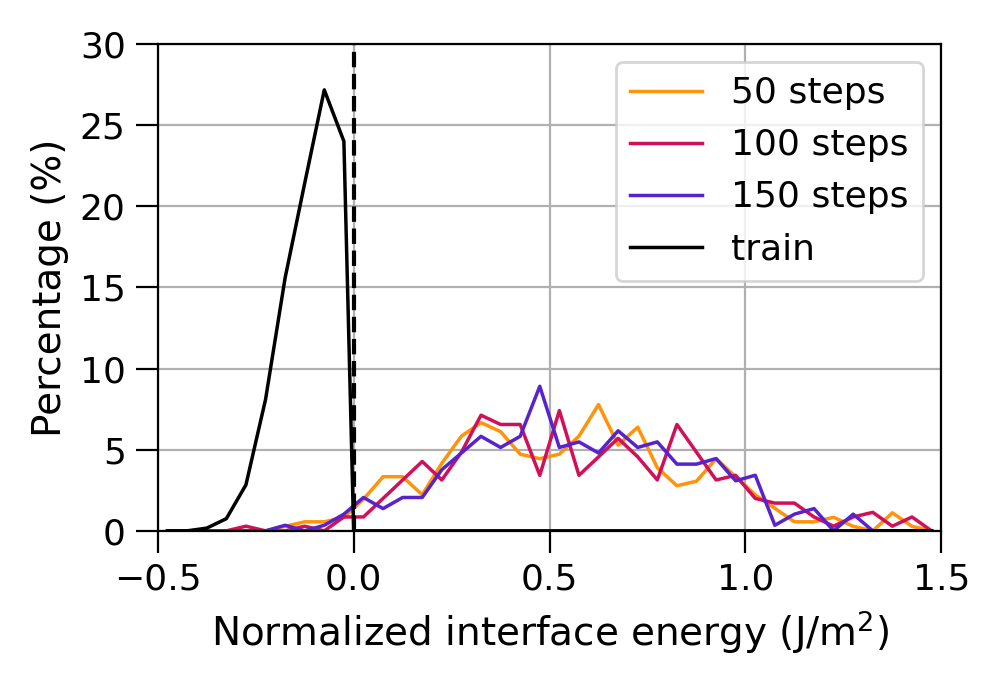"}
\caption{Normalized interface energy distributions of (relaxed) UniMat-generated interface structures, compared to the training structures. The structures were generated using 50, 100, or 150 time steps.}
\label{fig_unimat_int_energy}
\end{figure}

\begin{figure}[h!]
\centering
\includegraphics[width=4.0in]{"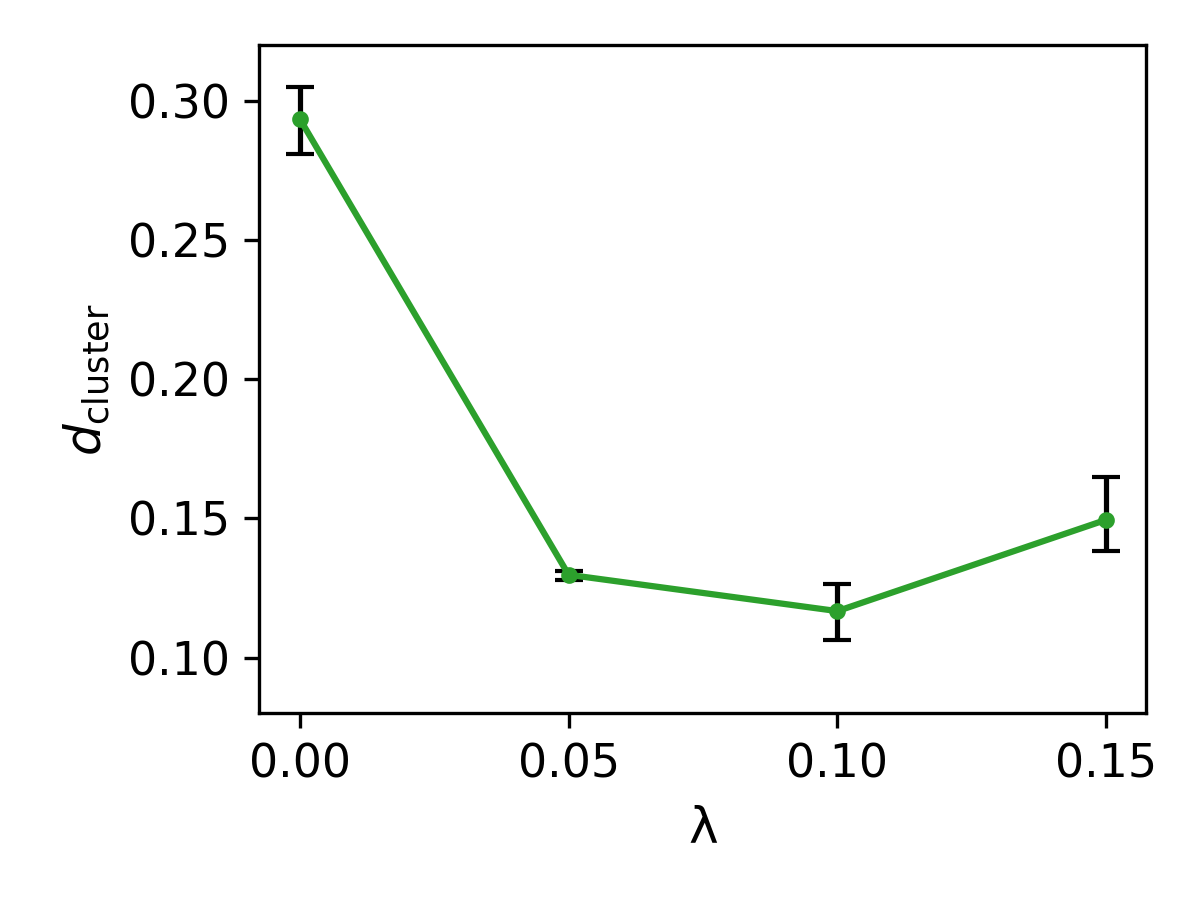"}
\caption{Euclidean distance of the cluster probability fingerprint for alloy structures (300 K, narrow SRO) generated using CryinGAN trained with different $\lambda$ values.}
\label{fig_lambda_d_cluster}
\end{figure}

\clearpage

\begin{figure}[t!]
\centering
\includegraphics[width=\textwidth]{"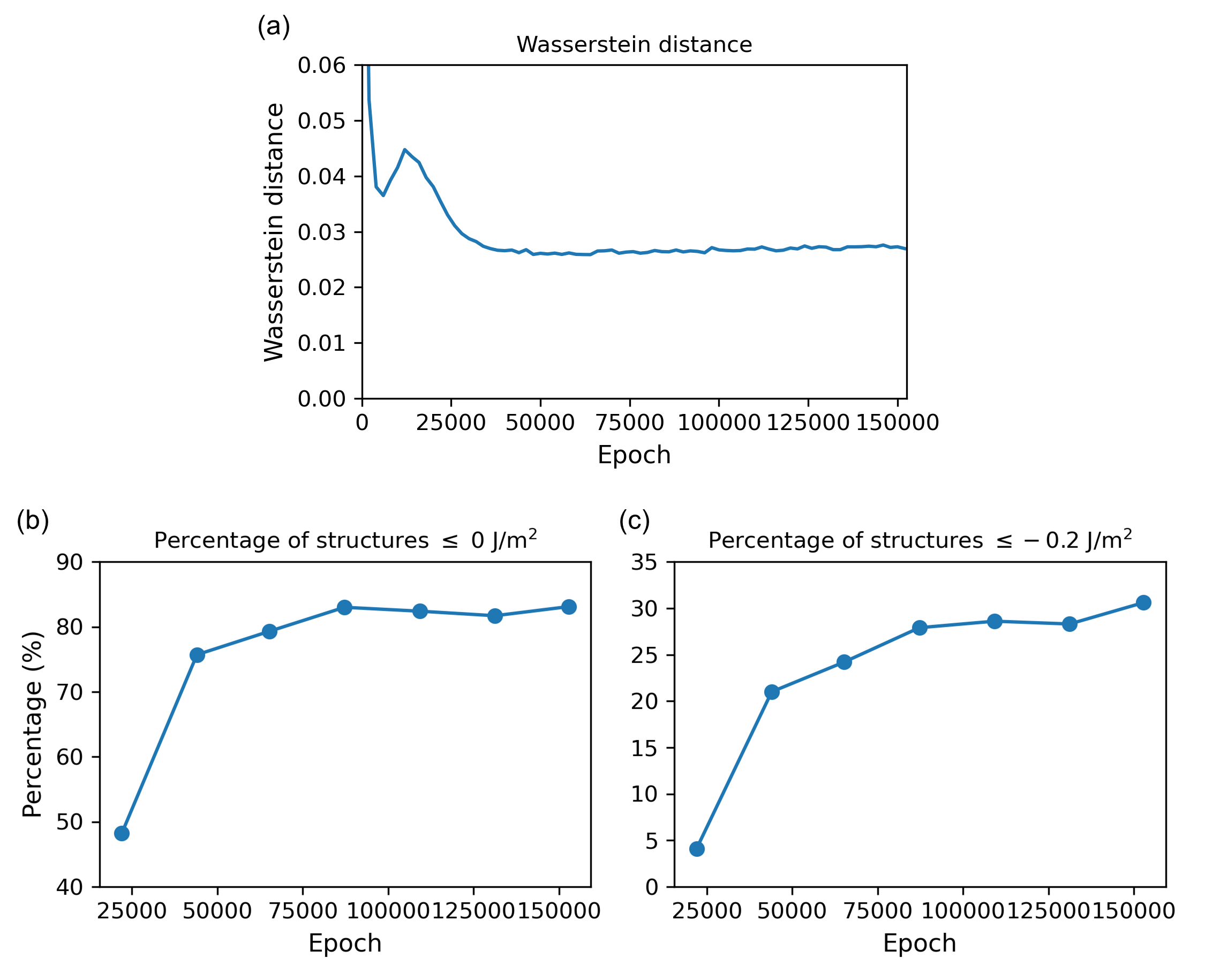"}
\caption{(a) Wasserstein distance as a function of epoch for CryinGAN trained with $\lambda = 0.05$. Note that the Wasserstein distance shown here does not include the gradient penalty term of the Wasserstein loss function. (b) Percentage of relaxed structures with normalized interface energy $\leq$ 0 J/m$^2$ as a function of epoch. (c) Percentage of relaxed structures with normalized interface energy $\leq -$0.2 J/m$^2$ as a function of epoch. Structures were generated using a CryinGAN model (trained for a certain number of epochs) and relaxed using the M3GNet interatomic potential. Each data point was calculated using 1,000 relaxed structures.}
\label{fig_best_GAN_training}
\end{figure}

\begin{table}[h]
\centering
\caption{Validation set mean absolute errors (MAEs) and losses of M3GNet models trained with different learning rates and batch sizes. The loss, $L$, is as defined in the main text. For each model, the epoch with the smallest $L$ is shown. The model with the smallest $L$ is highlighted in bold font. Note that we also trained a model with a learning rate of 0.005 and a batch size of 4, but we found the training to be unstable so the results are omitted here.}
\begin{tabular}{@{}lllllll@{}}
\toprule
Learning rate & Batch size & \makecell[l]{Energy MAE \\ (meV/atom)} & \makecell[l]{Force MAE \\ (meV/Å)} & \makecell[l]{Stress MAE \\ (GPA)} & Loss, $L$\\
\midrule
\textbf{0.001} & \textbf{4} & \textbf{2.75} & \textbf{20.9} & \textbf{0.0151} & \textbf{0.0251}\\
0.0005 & 4 & 2.82 & 22.2 & 0.0179 & 0.0268 \\
0.001 & 2 & 3.45 & 21.9 & 0.0203 & 0.0273 \\
0.001 & 6 & 2.88 & 21.3 & 0.0156 & 0.0257 \\
\bottomrule
\end{tabular}
\label{table_m3gnet_loss}
\end{table}

\clearpage

\begin{figure}[h!]
\centering
\includegraphics[width=\columnwidth]{"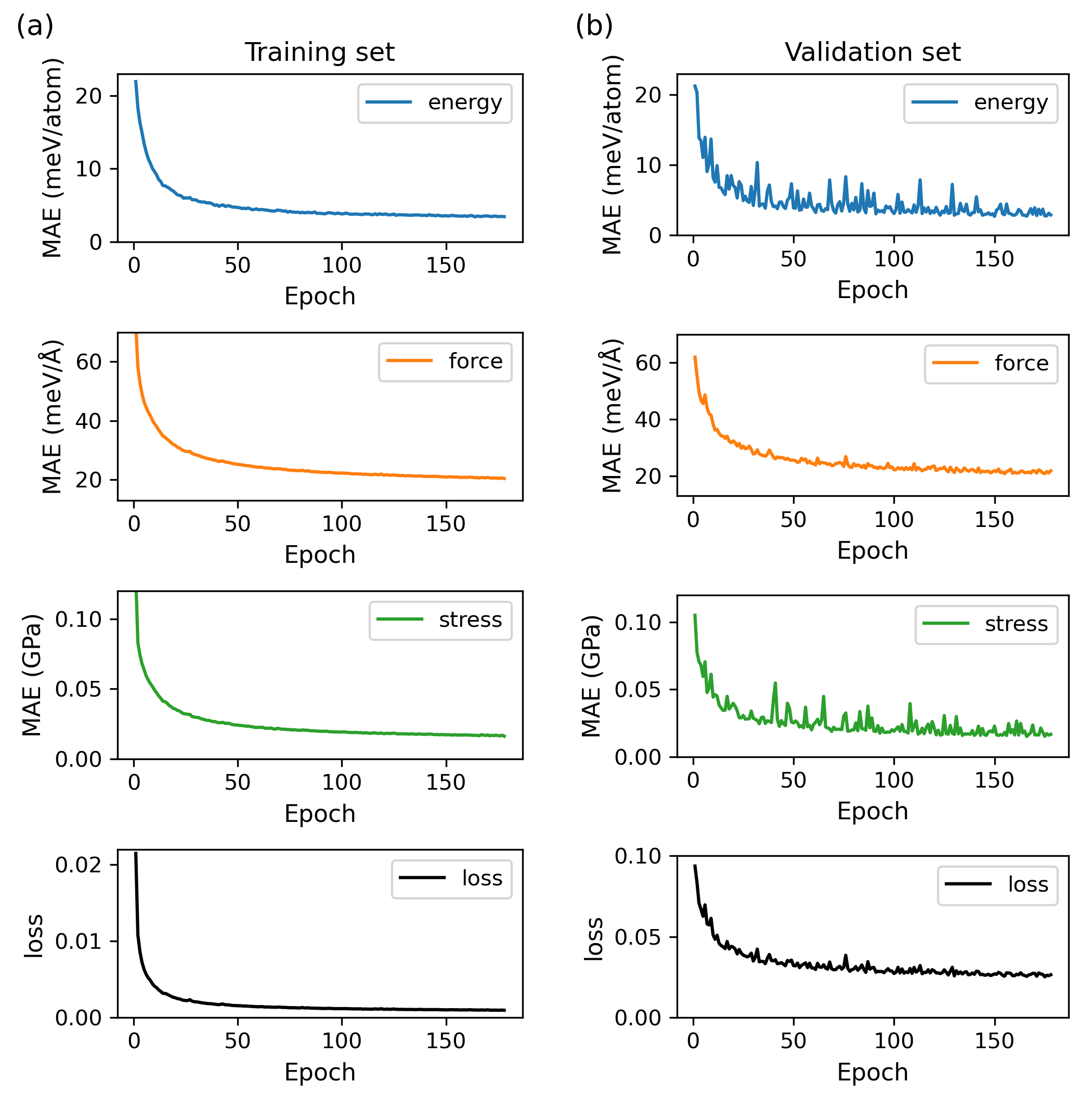"}
\caption{M3GNet training curves of the (a) training set and (b) validation set for the model trained with a learning rate of 0.001 and a batch size of 4. The mean absolute errors (MAEs) for energy, force, and stress, as well as the loss function are plotted against training epoch.}
\label{fig_m3gnet}
\end{figure}

\clearpage

\textbf{Supplementary Note 1: Oxide-chloride interface disorder mechanism}

To investigate the origin of the interfacial disorder observed for oxide-chloride interfaces, interfaces of binary materials (Li$_2$O-Li, Li$_2$O-LiCl, and Li$_2$O-MgS) were studied. All DFT calculations were performed using the same procedure as described in the main text. The Li ($1s^2\ 2s^1$), Cl ($3s^2\ 3p^5$), O ($2s^2\ 2p^4$), Mg ($2p^6\ 3s^2$), and S ($3s^2\ 3p^4$) electrons were treated as valence electrons in the pseudopotentials. Structural relaxations were first performed on unit cells of Li$_2$O ($Fm\overline{3}m$), Li ($Im\overline{3}m$), LiCl ($Fm\overline{3}m$), and MgS ($Fm\overline{3}m$). The cell shapes, cell volumes, and atom positions were allowed to relax, until the force on each atom was below 0.001 eV/Å. The Brillouin zone was sampled using a (8×8×8) Monkhorst-Pack k-point grid for Li, and a (6×6×6) Monkhorst-Pack k-point grid for Li$_2$O, LiCl, and MgS. The interface structures were constructed using the MPInterfaces package~\cite{RN128}, which implements the lattice matching algorithm proposed by Zur et al.~\cite{RN129}. The configurations of interfaces constructed and their lattice mismatches are listed in Table S4. All interfaces were constructed with vacuum spacings of at least 14 Å. We chose the (100) orientation as a representative plane for Li, LiCl, and MgS.

\vspace{12pt}

To study the effect of different terminations of Li$_2$O surfaces on the interfacial structure with LiCl, we chose the Li$_2$O(100) orientation which can either be Li-terminated of O-terminated. The surfaces of the Li$_2$O(100) slabs are polar, so half of the Li/O atoms were moved from one surface to the other to neutralize the polarity (resulting in ‘Tasker Type 2b’ surfaces~\cite{RN132, RN131}). Structural relaxations were performed on Li$_2$O(100)-LiCl(100) interfaces for both terminations, allowing the  cell shapes, cell volumes, and atom positions to relax until the force on each atom was below 0.05 eV/Å. The Brillouin zone was sampled using a (3×2×1) gamma-centered k-point grid. To study the effect of mechanical stiffness on interfacial structure, we chose the Li$_2$O(110) orientation, which exposes both Li and O at its surface, and paired it with Li(100), LiCl(100), and MgS(100). Structural relaxations were performed on these interfaces using the same procedure. The Brillouin zone was sampled using a (8×2×1), (3×2×1), and (3×2×1) gamma-centered k-point grid for Li$_2$O(110)-Li(100), Li$_2$O(110)-LiCl(100), and Li$_2$O(110)-MgS(100) respectively.

\vspace{12pt}

\begin{figure}[h!]
\centering
\includegraphics[width=\textwidth]{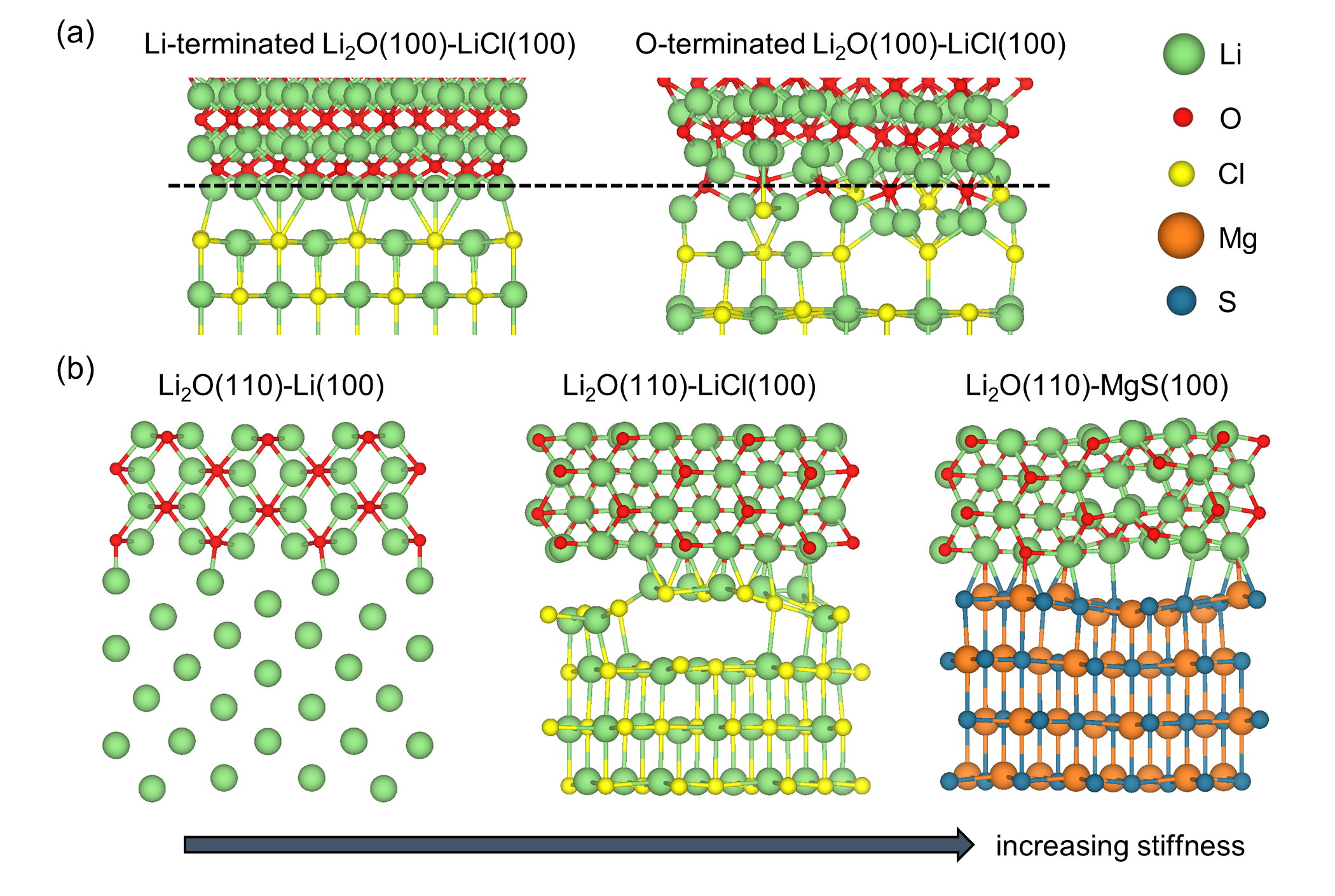}
\caption{(a) Interface structures of Li$_2$O(100)-LiCl(100) for Li-terminated and O-terminated Li$_2$O(100). The dashed line indicates the termination layer of Li$_2$O(100). (b) Interface structures for Li$_2$O(110)-Li(100), Li$_2$O(110)-LiCl(100), and Li$_2$O(110)-MgS(100). The mechanical stiffness of the bottom slab material increases from left to right.}
\label{fig_disorder}
\end{figure}

While many heterointerfaces can adopt regular epitaxial registries connecting two materials with well-defined crystalline orientations, other heterointerfaces show irregular, disordered interfacial patterns. We found that chlorides have an innate tendency to form disordered interfaces with oxides. As shown in Fig. \ref{fig_disorder}, using binary instead of ternary materials, disordered interfacial structures are obtained with only the combination of Li$_2$O and LiCl. We suggest two interdependent reasons for their occurrence. The first reason is the bond formation between O (in the oxide) and Li (in the chloride). Fig. \ref{fig_disorder}a shows the interface structure of Li$_2$O(100)-LiCl(100) for Li$_2$O(100) slabs that are either Li-terminated or O-terminated. When Li$_2$O(100) is Li-terminated, there is minimal rearrangement at the interface with only some Li-Cl bond formation. In contrast, when Li$_2$O(100) is O-terminated, significant atomic rearrangement is observed with a combination of Li-O and Li-Cl bond formation. The stronger interaction between O and Li is likely due to the higher charge density of O compared to Cl. However, Li-O bond formation alone does not lead to highly disordered interfaces, since such degree of disorder is not observed in other O-containing interfaces such as oxide-oxide and oxide-sulfide interfaces~\cite{RN331,RN332}.

\vspace{12pt}

The second reason that the chlorides can form  disordered interfaces is due to their intermediate mechanical stiffness. 
To form bonds with the O in the oxide slab, the other slab must distort its crystal structure to align its atoms with O, and the stiffness of the material determines the extent of the distortion allowed. 
Fig. \ref{fig_disorder}b shows the interface structure between Li$_2$O and materials in increasing order of mechanical stiffness (Li, LiCl, and MgS). MgS has the same rock salt crystal structure as LiCl but with higher stiffness~\cite{RN333,RN334}. Due to its higher stiffness, it is unable to distort significantly to bond with O. On the other hand, LiCl is soft enough that it can distort itself, whilst maintaining some of its structure, resulting in a more distorted interface structure. For a very soft material like Li metal~\cite{RN335}, the atoms can rearrange themselves relatively freely to bond with O, so large voids such as those observed for LiCl are less likely to form. The intermediate stiffness of the chlorides, combined with bond formation with O, results in the chloride solid electrolytes' ability to form highly disordered interface structures with the oxide cathode. The disordered nature of the Li$_3$ScCl$_6$(100)-LiCoO$_{2}$(110) interfaces is therefore understood from the two considerations above.

\clearpage

\textbf{Supplementary Note 2: Alternative CryinGAN architectures}

\begin{figure}[h!]
\centering
\includegraphics[width=\columnwidth]{"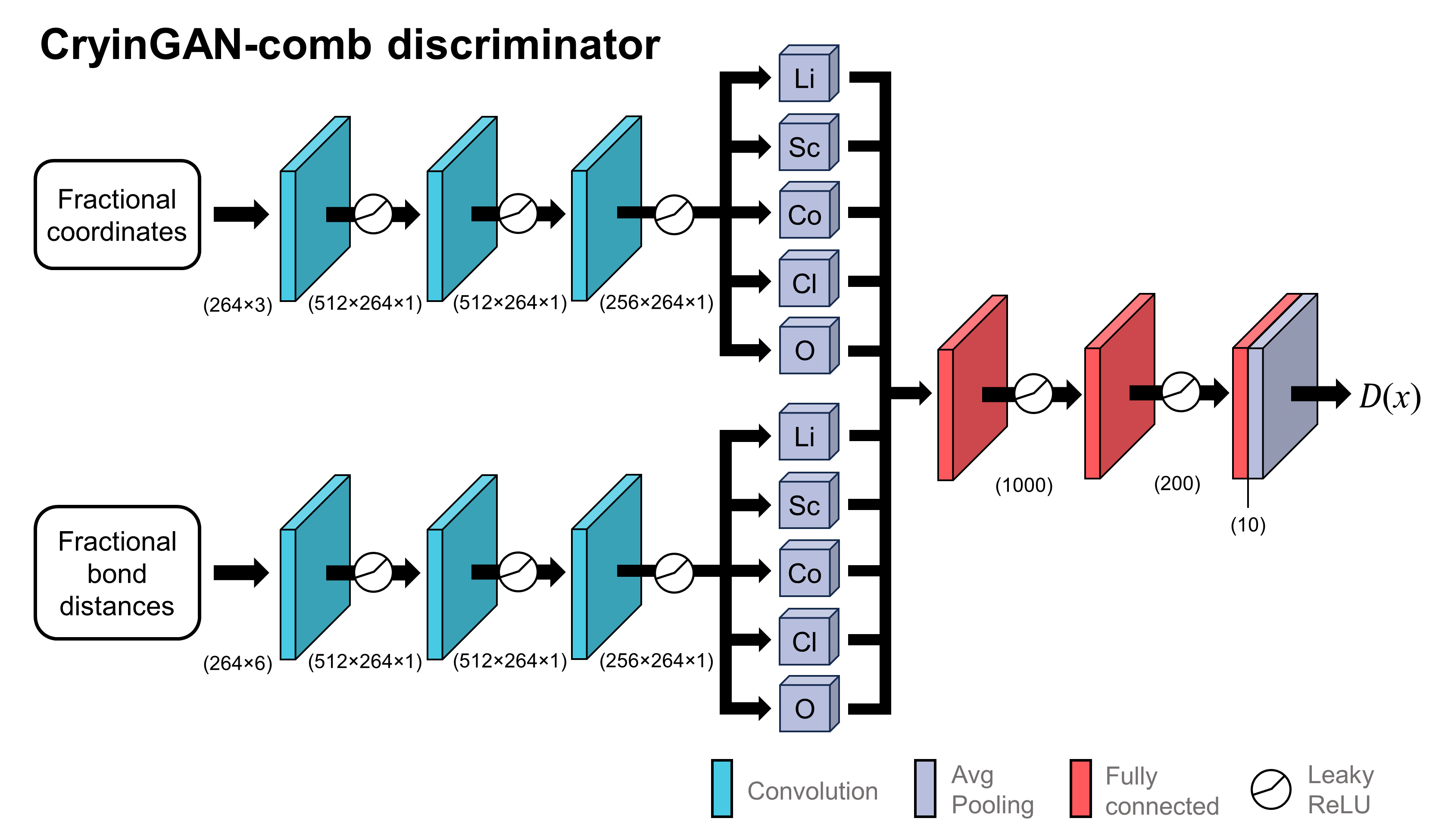"}
\caption{Discriminator architecture of CryinGAN-comb. The output of the two discriminators of CryinGAN are combined after pooling.}
\label{fig_CryinGAN-comb_arch}
\end{figure}

\begin{figure}[h!]
\centering
\includegraphics[width=\columnwidth]{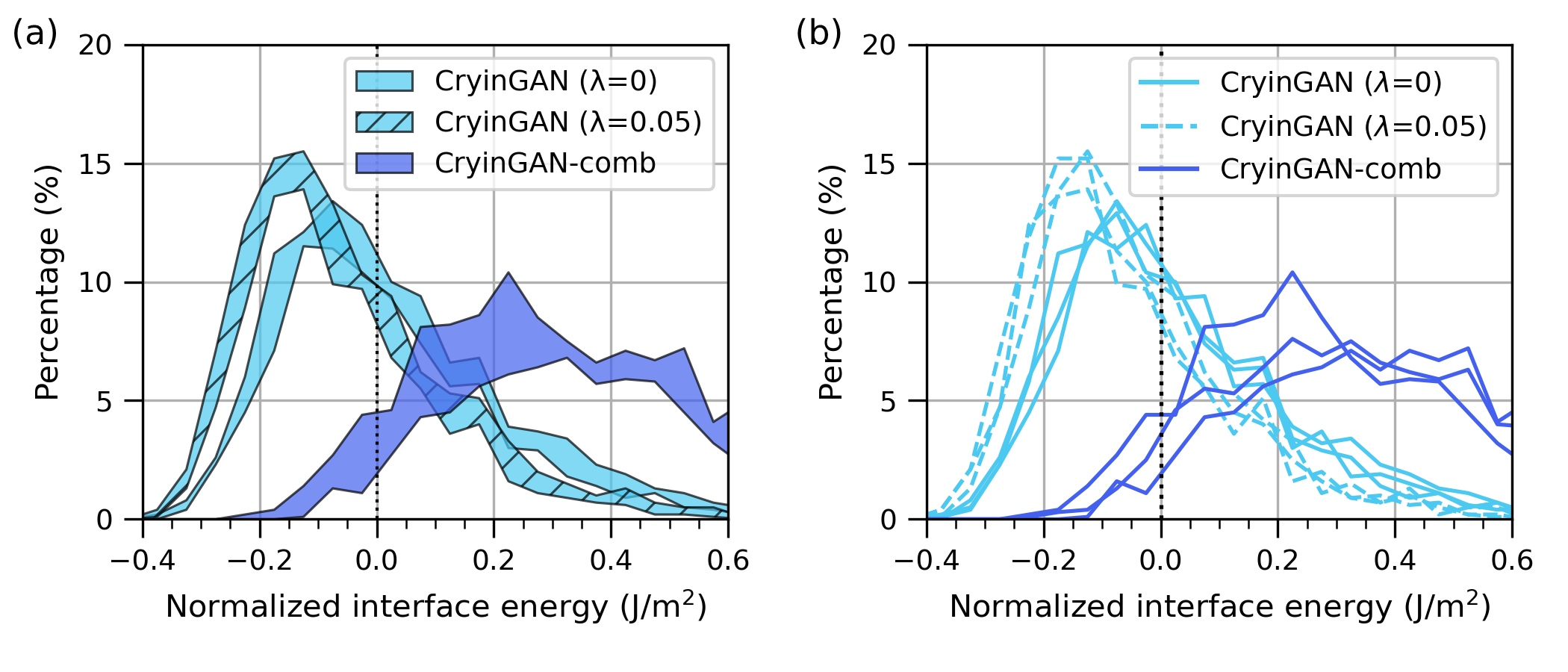}
\caption{Normalized interface energy distributions of structures generated using CryinGAN ($\lambda$ = 0, 0.05) and CryinGAN-comb. For each model configuration, 3 separate models were trained. The energy distribution are shown as shaded in (a) and unshaded in (b). All structures were relaxed using M3GNet, and the interface energies shown are based on M3GNet-calculated energies.}
\label{fig_CryinGAN-comb_energy}
\end{figure}

We considered a different architecture
that circumvents the need to tune $\lambda$ in CryinGAN, 
where both fractional coordinate and bond distance discriminators were combined into a single discriminator, which we refer to as CryinGAN-comb (see Fig. \ref{fig_CryinGAN-comb_arch}). 
The outputs were combined after the pooling layer, and the model was allowed to learn the relative importance of the coordinate and bond distance latent features on its own through fully connected layers. For CryinGAN-comb, the bond distances were directly obtained using fractional coordinates (instead of Cartesian coordinates), to provide a more direct link to the atomic coordinates which are also represented in fractional coordinates.
Structures were generated using CryinGAN-comb models and relaxed using M3GNet. 
Compared to CryinGAN trained with $\lambda = 0$, 
CryinGAN-comb generates structures with fewer atoms too close together (around 2 times fewer pairs of atoms $\leq$ 1.5 \r{A}). 
However, as shown in Fig. \ref{fig_CryinGAN-comb_energy}, the interface energy distribution of (relaxed) structures generated from CryinGAN-comb is significantly higher than CryinGAN.
These results show that combining the discriminators in CryinGAN-comb leads to less useful discriminator gradients to train the generator, resulting in poor quality of structures being generated. In contrast, separating the discriminators as in CryinGAN helps the GAN to learn more effectively.

\clearpage

\vspace{6pt}

\begin{figure}[t!]
\centering
\includegraphics[width=\columnwidth]{"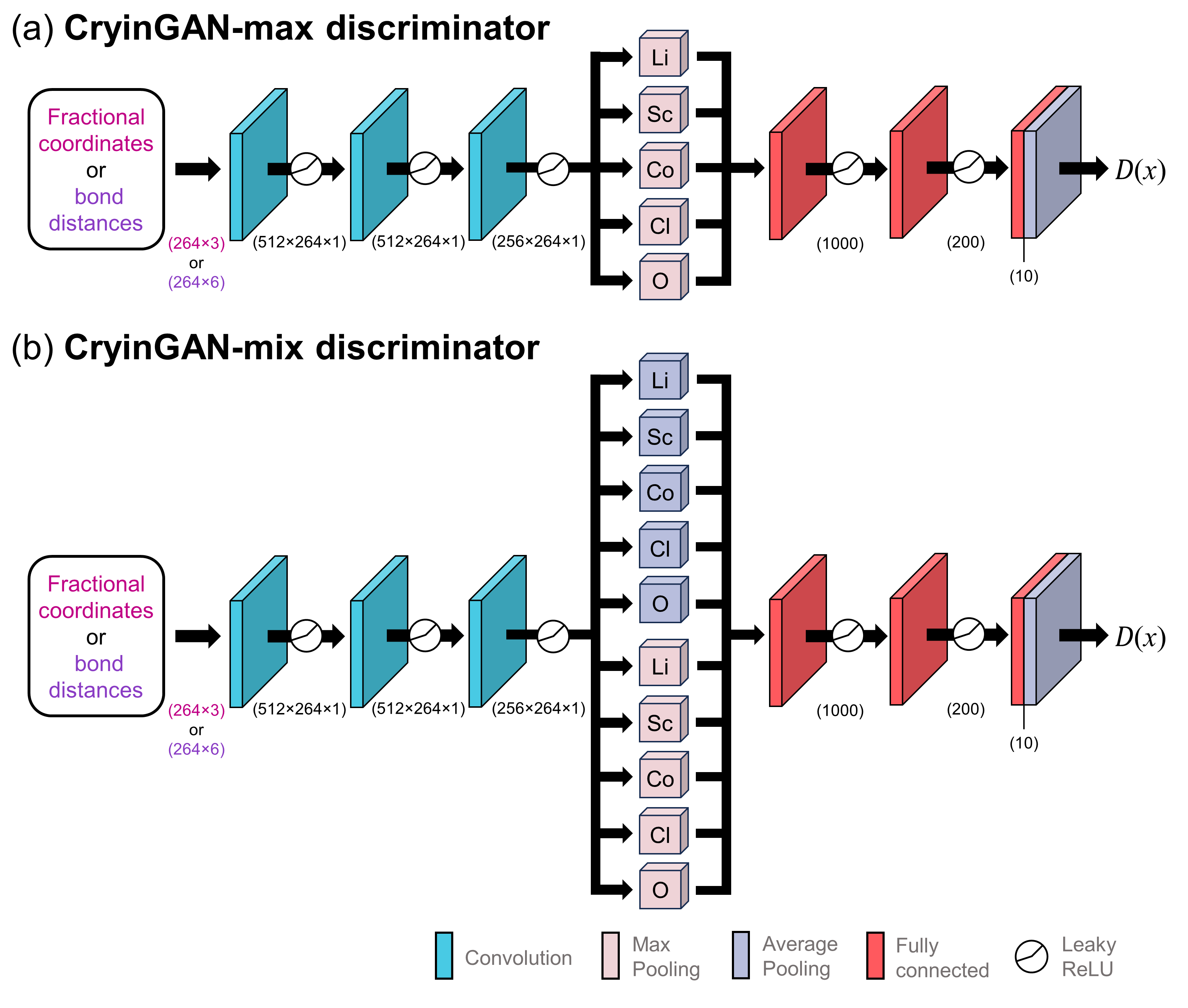"}
\caption{Discriminator architectures of (a) CryinGAN-max and (b) CryinGAN-mix. When the bond distance discriminator is not used, max/mix pooling is applied to the fractional coordinate discriminator. When the bond distance discriminator is used, max/mix pooling is applied to the bond distance discriminator, and average pooling is applied to the fractional coordinate discriminator.}
\label{fig_CryinGAN-max-mix_arch}
\end{figure}

Next, we considered different choices of the pooling layer for the discriminator. 
Wang et al.~\cite{wang2020sampling} found that the type of pooling operation affected the sampling sensitivity of the discriminator and the overall performance of the GAN. 
The sampling sensitivity describes how sensitive the discriminator is to changes in point density or the sampling pattern of the input point cloud. 
Their results suggest that max pooling produces a discriminator with lower sampling sensitivity than average pooling. 
The CryinGAN discriminators use average pooling, and we tested an architecture which uses max pooling instead, referred to as CryinGAN-max (see Fig. \ref{fig_CryinGAN-max-mix_arch}a). 
We also tested an architecture named CryinGAN-mix that uses the mix pooling operation proposed by Wang et al.~\cite{wang2020sampling}, where both max and average pooling operations are used together (see Fig. \ref{fig_CryinGAN-max-mix_arch}b). CryinGAN-max and CryinGAN-mix were trained using $\lambda$ = 0 and 0.05. For $\lambda$ = 0.05, only the pooling layer of the bond distance discriminator was changed, and we kept the average pooling for the fractional coordinate discriminator. This choice allows us to study the effect of the pooling choice on the two discriminators independently. 

\vspace{12pt}

\begin{figure}[t!]
\centering
\includegraphics[width=\columnwidth]{"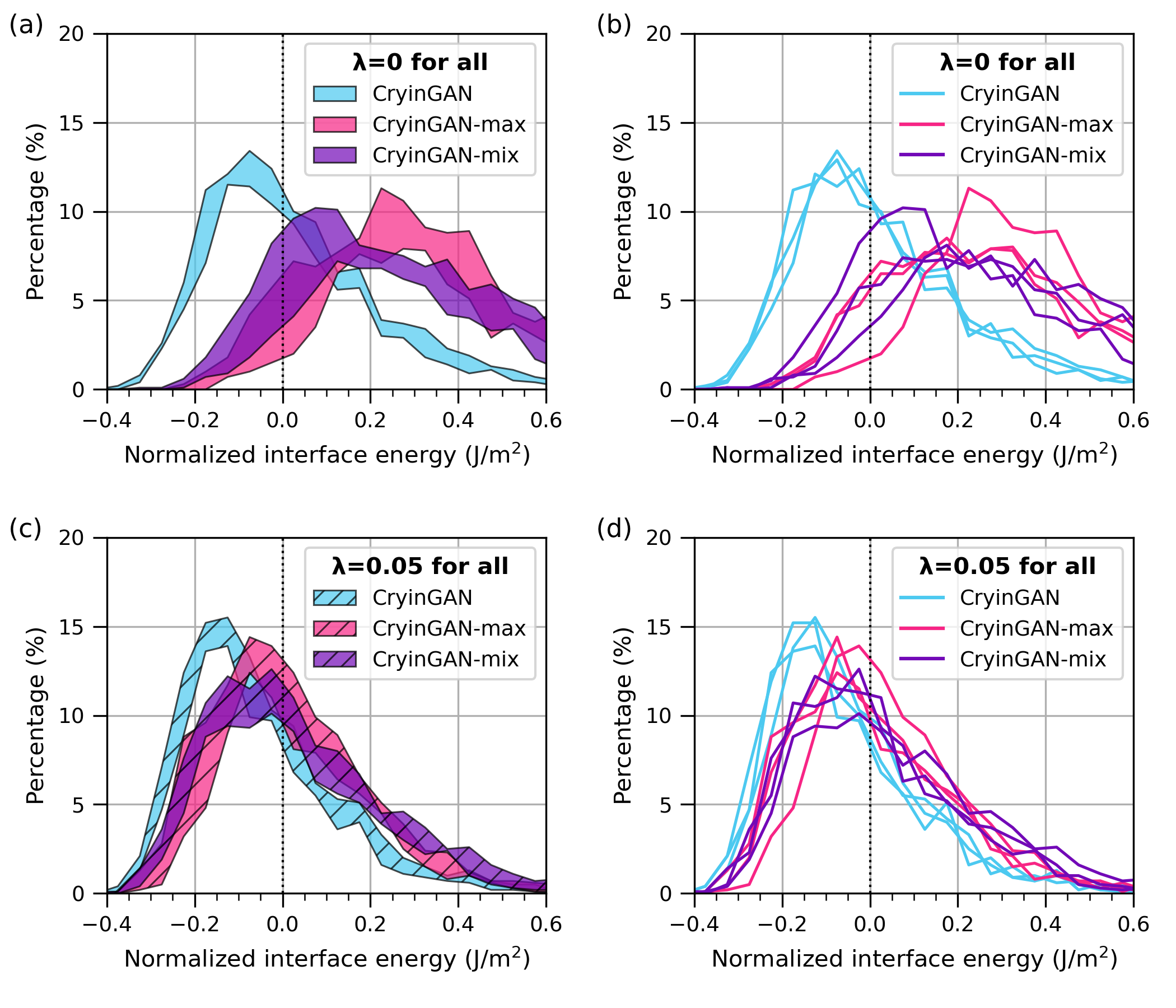"}
\caption{Normalized interface energy distributions of structures generated using CryinGAN (average pooling), CryinGAN-max (max pooling), and CryinGAN-mix (mix pooling). (a) and (b) show models without the bond distance discriminator ($\lambda$ = 0), where pooling is varied for the fractional coordinate discriminator. (c) and (d) show models with the bond distance discriminator ($\lambda$ = 0.05), where pooling is varied for the bond distance discriminator only (average pooling is used for the fractional coordinate discriminator). For each model configuration, 3 separate models were trained. The energy distribution are shown as shaded on the left, and unshaded on the right. All structures were relaxed using M3GNet, and the interface energies shown are based on M3GNet-calculated energies.}
\label{fig_CryinGAN-max-mix_energy}
\end{figure}

The interface energy distributions of relaxed structures generated using CryinGAN, CryinGAN-max, and CryinGAN-mix are shown in Fig. \ref{fig_CryinGAN-max-mix_energy}. 
For $\lambda$ = 0 (Fig. \ref{fig_CryinGAN-max-mix_energy}a-b), we observe that CryinGAN (average pooling) significantly outperforms CryinGAN-max and CryinGAN-mix. 
Although the use of max pooling was beneficial for the generation of 3D objects where the object shape is the most important aspect to capture~\cite{wang2020sampling}, it appears that a higher sampling sensitivity is needed for atomic configurations, for which the local coordination environment is the important aspect to capture. 
For $\lambda$ = 0.05 (Fig. \ref{fig_CryinGAN-max-mix_energy}c-d), we observe that the distributions are more similar across the three architectures, but CryinGAN still has the lowest energy distribution. 
The pooling choice of the bond distance discriminator does not relate to the sampling sensitivity like the fractional coordinate discriminator, and the choice appears to have a smaller effect on model performance. 
Overall, we find that average pooling works best for both discriminators.

\clearpage

\makeatletter
\setlength{\@fptop}{0pt}
\makeatother

\clearpage

\clearpage

\textbf{Supplementary Note 3: Interface Li coordination motif analysis of CryinGAN-generated structures}

\begin{figure}[h!]
\centering
\includegraphics[width=4in]{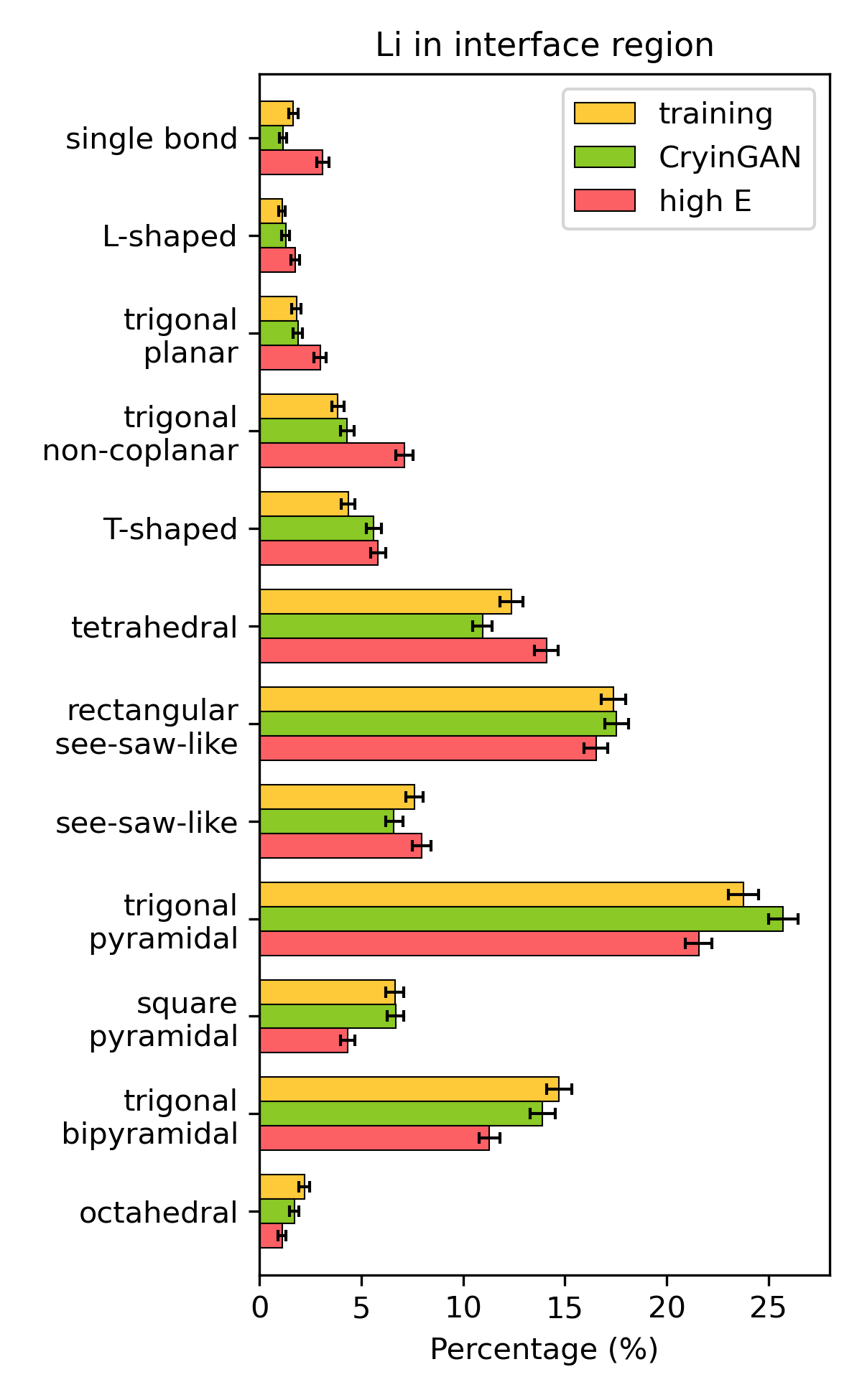}
\caption{Coordination motif distributions of Li in the interface region for three datasets: (1) training structures with low interface energy, (2) CryinGAN-generated structures, and (3) structures with high interface energy. All structures were relaxed using M3GNet followed by DFT calculations. Error bars represent 95 \% bootstrap confidence intervals. The coordination motifs are ordered in increasing coordination number from top to bottom.}
\label{fig_Li_motifs}
\end{figure}

Table \ref{table_fingerprint_Li} shows the Euclidean distance and cosine similarity between the average interface Li fingerprint of the CryinGAN/high energy dataset and the training dataset. 
Compared to the high energy dataset, the CryinGAN dataset has a slightly lower Euclidean distance to the low energy dataset, and similar cosine similarity. 
This small difference is better understood by examining the distribution of the most likely Li coordination motif as shown in Fig. \ref{fig_Li_motifs}.
Compared to the coordination motif distribution of the training dataset, the distribution of the high energy dataset is shifted upwards towards motifs with lower coordination number, indicating that fewer bonds are leading to higher energies. 
On the other hand, the coordination motif distribution of the CryinGAN-generated structures shows a higher similarity to the training structures, reflecting the smaller motif fingerprint Euclidean distance seen in Table \ref{table_fingerprint_Li}.

\begin{table}[h!]
\centering
\caption{Euclidean distance and cosine similarity between the average interface Li site fingerprint of the training structures and the CryinGAN/high-interface-energy structures. The 95 \% bootstrap confidence intervals are shown in brackets.}
    \begin{tabularx}{0.65\textwidth}{lXX}
    \toprule
        Dataset &  Euclidean distance \newline (95 \% CI) &  Cosine similarity \newline (95 \% CI) \\
    \midrule
         CryinGAN &  0.1297 \newline (0.1106 to 0.1455) &  0.9974 
         \newline (0.9968 to 0.9982)\\
         High energy & 0.1361 \newline (0.1226 to 0.1471) & 0.9973 
         \newline (0.9969 to 0.9979) \\
    \bottomrule
    \end{tabularx}
    \label{table_fingerprint_Li}
\end{table}

\clearpage

\textbf{Supplementary Note 4: RDF analysis of CryinGAN-generated interfaces}

\begin{figure}[h]
\centering
\includegraphics[width=3.6in]{"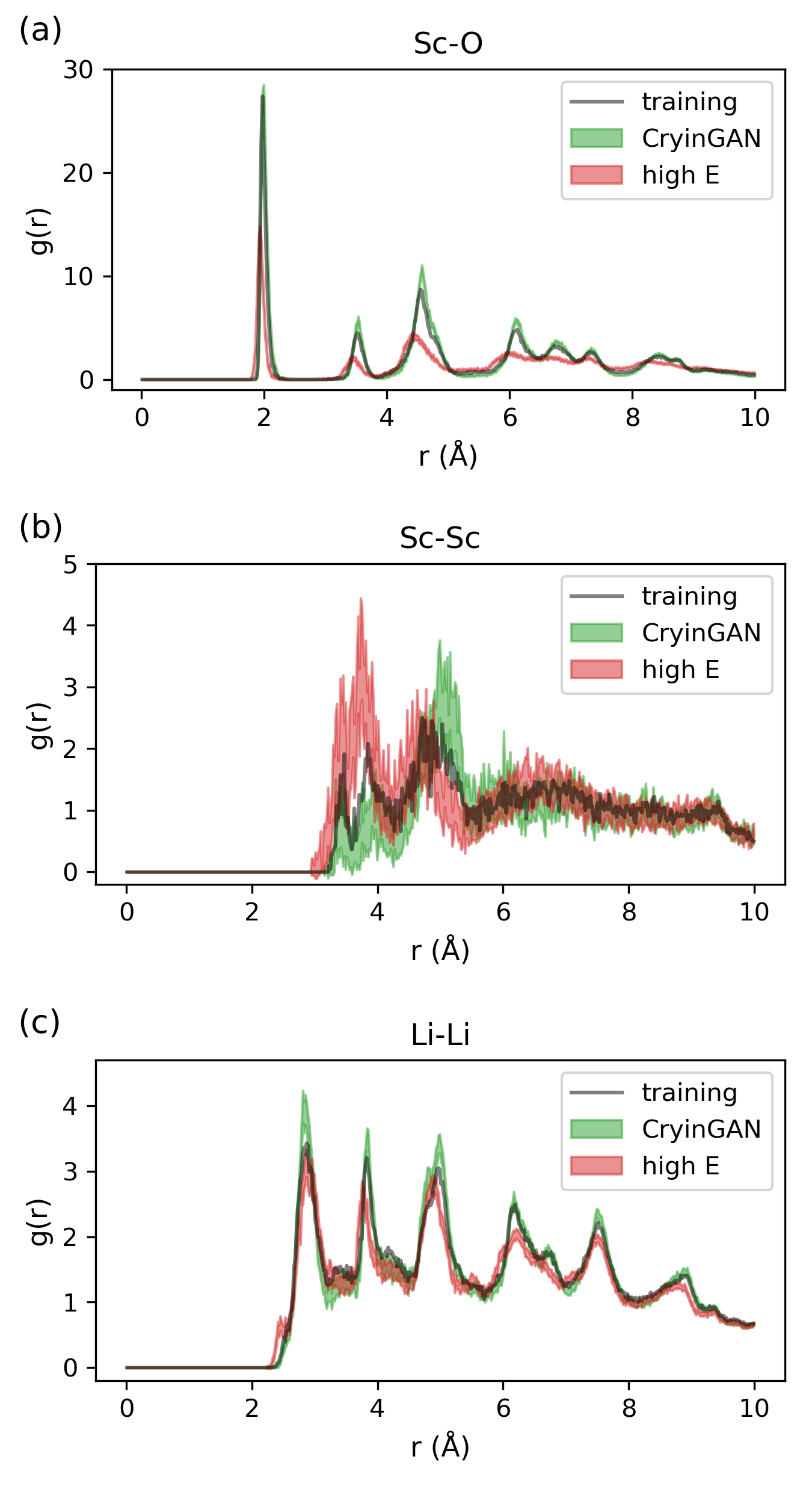"}
\caption{Radial distribution functions of atoms in the interface region for (a) Sc-O, (b) Sc-Sc, and (c) Li-Li. The thickness of each curve represents the 95 \% bootstrap confidence interval.}
\label{fig_rdf}
\end{figure}

The Sc-O, Sc-Sc, and Li-Li RDFs are shown in Fig. \ref{fig_rdf}.
For the Sc-O RDF (Fig. \ref{fig_rdf}a), the high energy dataset exhibits peaks with lower magnitude, reflecting its reduced degree of Sc-O bonding. 
On the other hand, the Sc-Sc RDF (Fig. \ref{fig_rdf}b) shows a higher peak around 4 \r{A}, indicating that the high energy structures have a higher proportion of Sc cations closer together. 
The associated Sc-Sc repulsion raises the energy of the structures. 
In the Li-Li RDF (Fig. \ref{fig_rdf}c), a new peak appears around 2.5 \r{A}. 
This short Li-Li interatomic distance again raises the energy of the structures due to Li-Li repulsion. 
In contrast, the CryinGAN dataset does not show any of these high-energy features, and its RDFs are similar to the training dataset. 

\clearpage

\bibliography{rsc.bib}